\documentclass[useAMS,usenatbib,uselongtable]{mn2e}
\usepackage{epstopdf}
\usepackage[pdftex]{graphicx}

\title[Young Stellar Objects in NGC 6823]{Young Stellar Objects in NGC 6823}
\author[Riaz et al.]
{B. Riaz,$^{1}$ E. L. Mart\'{i}n,$^{2}$
 R. Tata,$^{3}$ J.-L. Monin,$^{4}$, 
 N. Phan-Bao,$^{5}$ H. Bouy$^{6}$ \\
$^{1}$Centre for Astrophysics Research, Science \& Technology Research Institute, University of Hertfordshire, Hatfield, AL10 9AB, UK\\
$^{2}$Centro de Astrobiolog\'{i}a (CSIC/INTA), 28850 Torrej\'{o}n de Ardoz, Madrid, Spain\\
$^{3}$Instituto de Astrofis\'{i}ca de Canarias, E-38205 La Laguna, Tenerife, Spain\\
$^{4}$Universit\'{e} Joseph Fourier/CNRS Laboratoire d'Astrophysique de Grenoble, UMR 5571, BP 53, 38041 Grenoble C\'{e}dex 09, France\\
$^{5}$Department of Physics, HCMIU, Vietnam National University Administrative Building, Block 6, Linh Trung Ward, \\ Thu Duc District, HCM, Vietnam\\
$^{6}$European Space Agency (ESAC), PO Box 78, 28691 Villanueva de la Ca\~{n}ada (Madrid), Spain }

\begin{document}

\date{}

\pagerange{\pageref{firstpage}--\pageref{lastpage}} \pubyear{2002}

\maketitle

\label{firstpage}

\begin{abstract}

NGC 6823 is a young open cluster that lies at a distance of $\sim$2 kpc in the Vulpecula OB1 association. Previous studies using CCD photometry and spectroscopy have identified a Trapezium system of bright O- and B-type stars at its center, along with several massive O- B- and A-type stars in the cluster. We present optical {\it VRI} and near-infrared {\it JHK} photometric observations, complemented with {\it Spitzer}/IRAC archival data, with an aim to identify the young low-mass population and the disk candidates in this region. Our survey reaches down to {\it I}$\sim$22 mag and $K_{\rm s}$$\sim$18 mag. There is significant differential reddening within the cluster. We find a bimodal distribution for $A_{V}$, with a peak at $\sim$3 mag and a broader peak at $\sim$10 mag. We have classified the sources based on the [4.5]-[8] color, which is least affected by extinction. We find a $\sim$20\% fraction of Class I/Class II young stellar objects (YSOs) in the cluster, while a large 80\% fraction of the sources have a Class III classification. We have made use of the INT Photometric H$\alpha$ Survey (IPHAS) in order to probe the strength in H$\alpha$ emission for this large population of Class III sources. Nearly all of the Class III objects have photospheric ($r^{\prime}$-H$\alpha$) colors, implying an absence of H$\alpha$ in emission. This large population of Class III sources is thus likely the extincted field star population rather than the diskless YSOs in the cluster. There is a higher concentration of the Class I/II systems in the eastern region of the cluster and close to the central Trapezium. The western part of the cluster mostly contains Class III/field stars and seems devoid of disk sources. We find evidence of a pre-main sequence population in NGC 6823, in addition to an upper main-sequence population. The pre-main sequence population mainly consists of young disk sources with ages between $\sim$1-5 Myr, and at lower masses of $\sim$0.1-0.4 $M_{\sun}$. There may be a possible mass dependent age spread in the cluster, with the older stars being more massive than the younger ones. The presence of young disk sources in NGC 6823 indicates similar star formation properties in the outer regions of the Galaxy as observed for young clusters in the solar neighborhood. 



\end{abstract}


\section{Introduction}

The open cluster NGC 6823 is considered to be the core of the Vulpecula OB1 association. The cluster is surrounded by the reflection nebula NGC 6820, and is situated in a bright H II region, known as Sharpless 86 (Sharpless 1959), with associated dark clouds and pillar-like structures (e.g., Turner 1979; Chapin et al. 2008; Billot et al. 2010). The cluster has a Trapezium system, BD+22$\degr$3782, of bright O- and B-type stars at its center. Erickson (1971) conducted a proper-motion membership study of the cluster in a 30$\arcmin$$\times$30$\arcmin$ region centered on the central Trapezium system, and determined membership probabilities for 92 stars brighter than {\it V}=13 magnitude. Erickson (1971) noted the presence of an inner and outer cluster region, such that over one-half of the members were found to lie in a region extending to distances larger than 4$\arcmin$ from the cluster center. Stone (1988) conducted a follow-up {\it UBV} photometric study using photographic plates of the regions studied by Erickson (1971). The H-R diagrams for both the inner and outer cluster regions showed a well-defined upper main-sequence, with an estimated age of $\sim$2Myr. In addition, Stone (1988) noted a pre-main sequence population with ages between 0.2 and 0.5 Myr in the outer region of the cluster, extending out to 15$\arcmin$ from the cluster center. Based on the observed age dispersion, Stone suggested that NGC 6823 may have experienced a recent burst of stellar formation, possibly caused by a supernova explosion of one of the massive O-stars at the center, with a second generation of stars being formed in the outer region some $\sim$0.5 Myr ago. The presence of a pre-main sequence population in NGC 6823 has also been reported by Guetter (1992), Massey et al. (1995), Pigulski et al. (2000) and more recently by Bica et al. (2008), with an age spread estimated to be as large as $\sim$2-11 Myr for the cluster stars. There have been a few spectroscopic surveys of NGC 6823. The earliest type classified yet is O7, and the latest is K4 (Turner 1979; Massey et al. 1995; Shi \& Hu 1999). 

The cluster is located at a distance of 2-2.5 kpc (e.g., Turner 1979; Guetter 1992; Massey et al. 1995). The reddening across the cluster is found to be large and non-uniform. The {\it E(B-V)} color excesses range between 0.7 and 1.1 mag, with a mean value of $\sim$0.9 (e.g., Erickson 1971; Guetter 1992; Massey et al. 1995; Pigulski et al. 2000). The extinction is highest in the eastern part of the central Trapezium, in the direction of the reflection nebula NGC 6820 (e.g., Pigulski et al. 2000). The reddening for the eleven central Trapezium stars is found to be very uniform, with {\it E(B-V)} = 0.84$\pm$0.01 mag (Guetter 1992; Pigulski et al. 2000). There is evidence of foreground reddening toward NGC 6823, and a high contamination by foreground field stars (e.g., Guetter 1992; Pigulski et al. 2000; Bica et al. 2008). Guetter (1992) estimated that the foreground reddening toward the cluster is at least 0.6 mag. The presence of a nearby interstellar cloud in the direction of NGC 6823 has been confirmed by Fresneau \& Monier (1999), and Pigulski et al. (2000) suggest that more than half of the total absorption towards the cluster comes from this nearby absorbing matter, at a distance 0.2-0.5 kpc. 

Chapin et al. (2008) reported the detection of three pillar-like structures, {\it VulP-12-13-14}, located north-east of NGC 6823. Such pillars have been associated with recent episodes of star formation (e.g., Chen et al. 2006). The close proximity of the supernova remnant SNR G59.5+0.1 to NGC 6823 suggests a possible case of triggered star formation in this cluster (e.g., Billot et al. 2010). This is thus a young cluster that has likely seen a recent burst of star formation. Most of the past surveys in this region have been conducted at the {\it UBV} wavelengths, and have targeted the high mass population. We present deep optical {\it VRI} and near-infrared {\it JHK} photometric observations of NGC 6823, complemented with Spitzer/IRAC observations, with an aim to identify the low mass population and the disk candidates in this region. While searches for young very low mass stars have been conducted in several nearby star-forming regions, more such deep surveys need to be conducted in distant massive clusters. Our observations are presented in Section $\S$\ref{obs}. Section \S\ref{extinction} discusses the extent of reddening in this region. The disk population in NGC 6823 is discussed in Sections $\S$\ref{disks} and $\S$\ref{halpha}. Section $\S$\ref{spatial} presents the spatial distribution of the young stellar objects (YSOs) detected, and in Section $\S$\ref{trigger}, we investigate for any observational diagnostics of triggered star formation in NGC 6823. 

\section{Observations and Data Reduction}
\label{obs}

The optical {\it V}, {\it R}, and {\it I} band images were obtained using the Prime Focus camera (WHT/WFC detector) mounted on 4m William Herschel Telescope (WHT) in La Palma, Canary Islands, Spain. The camera  contains two EEV-42-80 thinned and AR coated CCDs butted along their long axis to provide a 4K x 4K pixel mosaic. The pixel scale for this detector is 0.24 arcsec/pixel with a field of view of 16.4 x 16.4 arcmin. Observations were performed in May 2005, obtaining one 3s exposure and one 600s exposure in each of the bands. The 3s exposure was to obtain photometry of the bright objects which would saturate in the 600s exposures. The near-infrared (NIR) {\it J}, {\it H}, $K_{s}$ band images were obtained using the Infrared Side Port Imager (ISPI) mounted on CTIO 4m Blanco Telescope in Cerro Tololo, Chile. ISPI is a 2K x 2K HgCdTe HAWAII-2 array with a pixel scale 0.3 arcsec/pixel in the near infrared with a 10.25 x 10.25 arcmin field of view. Observations were performed in March 2007. 

The optical and the NIR data were reduced using Image Reduction and Analysis Facility (IRAF) system routines. For the NIR images, a single stacked image in each band (J, H, $K_s$) was obtained by combining several short-exposure images. We conducted Point Spread Function (PSF) photometry on the bias subtracted and flat fileded images., using various routines under the IRAF/ DAOPHOT package (Stetson 1987). The routine {\it daofind} is first applied to detect all the stellar sources. The computation of the PSF to be fitted was performed by the routine {\it psf}. The PSF was modeled to be the sum of an analytic bivariate Gaussian function and empirical corrections from the best Gaussian of the true observed brightness values within the average profile of several stars in the image. PSF photometry was performed with the routine {\it allstar}, which after classifying the stars in groups, compiles a catalog of the most likely candidate stars, based on their PSF fitting and the physical conditions, and subtracts them from the original image. The NIR absolute photometry was calibrated using the 2MASS photometric system. The optical photometry was calibrated using sources from CADC photometric standard list for the NGC 6823 field. Astrometric solutions were obtained for the optical and NIR frames using {\it imwcs} command in WCSTools (Mink 1999). We then created an optical+NIR catalog using {\it gcompare}, a tool from the ELIXIR system. Separate catalogs were created for the bright and the faint sources, and any duplicates were removed. We then cross matched these catalogs with data from the {\it Spitzer} GLIMPSE survey. Our final catalog consists of 627 sources, with good photometric detections ($\sigma$ $<$0.1 mag) in all four IRAC bands. For 17 of these sources, we have {\it Spitzer} MIPSGAL 24$\micron$ matches obtained from Billot et al. (2010). None of the sources were detected at 70$\micron$. Billot et al. estimate an uncertainty of $<$0.2 mag for the 24$\micron$ photometry. The optical, NIR and {\it Spitzer}/IRAC photometry for these sources is listed in Table~\ref{phot}, and the MIPS 24$\micron$ photometry is listed in Table~\ref{mips}.  

\begin{figure*}
\includegraphics[width=\linewidth]{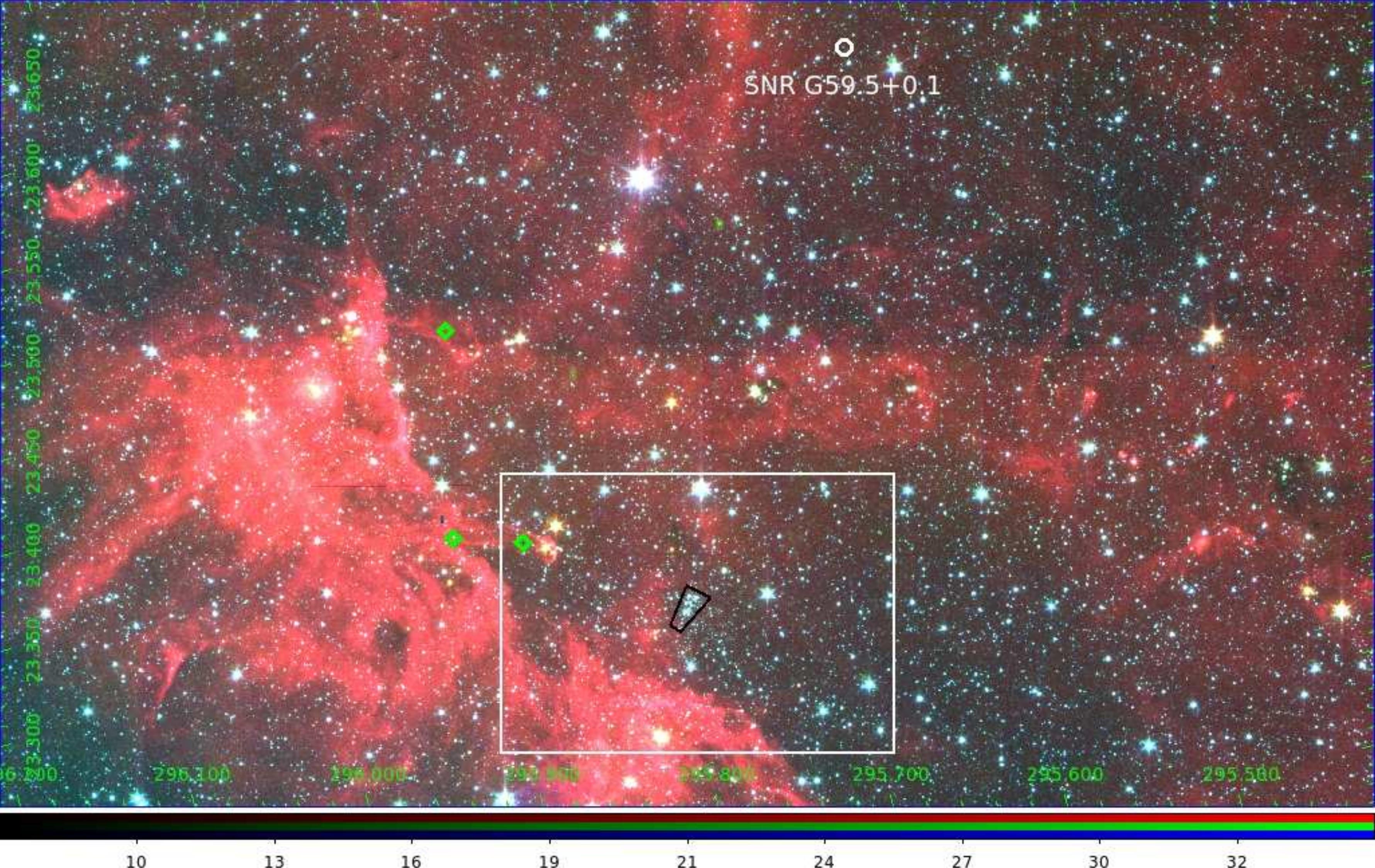}
    \caption{A spatial view of NGC 6823 and the Vulpecula association. This is an IRAC composite image of 3.6$\micron$ (blue), 4.5$\micron$ (green), and 8$\micron$ (red) observations. The color intensity bar in units of MJy sr$^{-1}$ is shown at the bottom of the image. North is up, east is to the left. The region in NGC 6823 that we have surveyed is marked by a white rectangle, and is centered on the Trapezium (marked in black). In the north of the Trapezium is the supernova remnant SNR G59.5+0.1. Also marked are the three pillar-like structures, {\it VulP-12-13-14} (green diamonds).  }
      \label{snr}
 \end{figure*}

Figure~\ref{snr} gives a spatial view of NGC 6823. This is an IRAC composite image of 3.6$\micron$ (blue), 4.5$\micron$ (green), and 8$\micron$ (red) observations. The survey area from our work is marked by white lines, the central Trapezium is marked in black. Taylor et al. (1992) detected a supernova remnant, SNR G59.5+0.1, in the direction of the Vulpecula OB1 association. The location of this SNR and the three pillars, {\it VulP-12-13-14}, detected by Chapin et al. (2008), are also marked. 

\section{Analysis}

\subsection{Extinction}
\label{extinction}

\begin{figure*}   
\includegraphics[width=120mm]{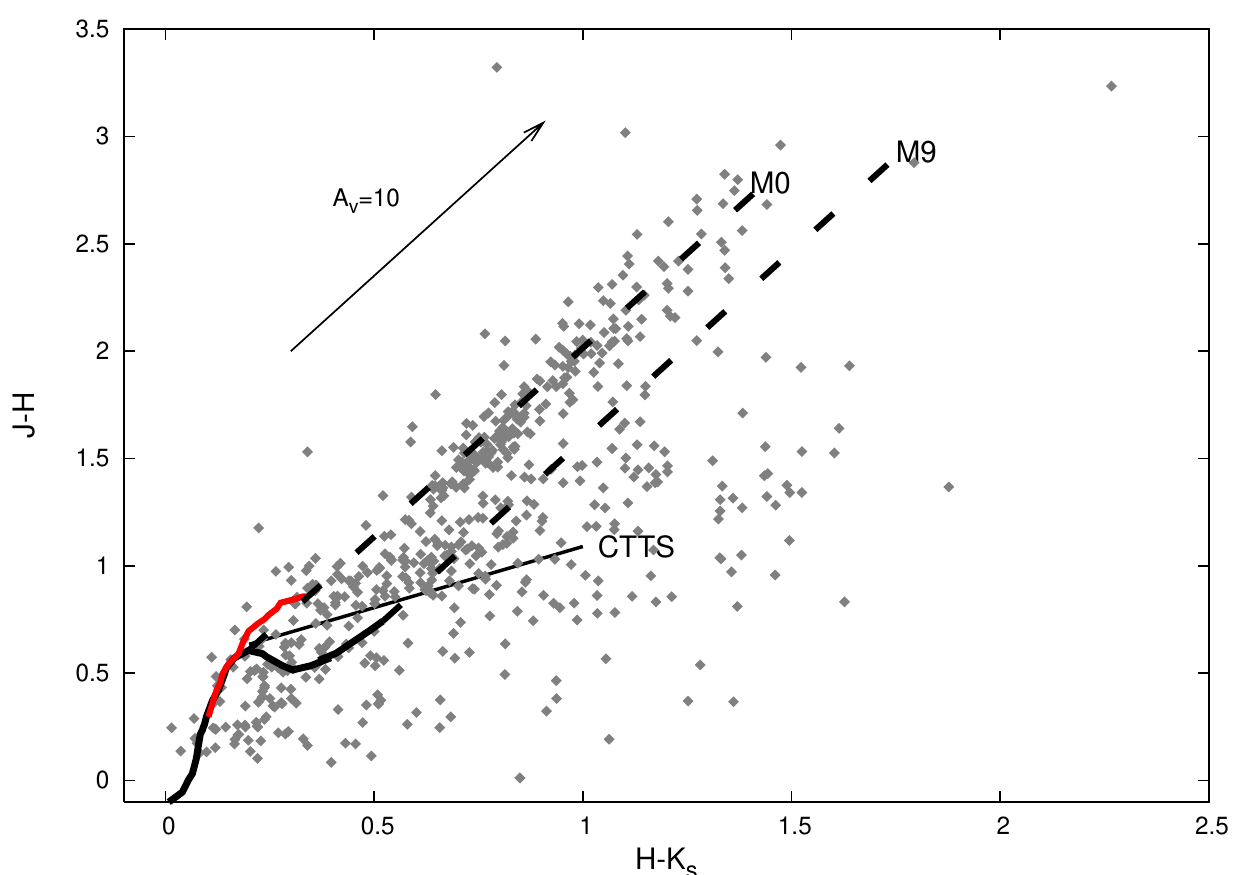}
    \caption{The near-infrared ccd for NGC 6823. The thick black and red curves indicate the zero-age main sequence for dwarfs and giants, respectively. The thin black line marks the CTTS locus from Meyer et al. (1997). The reddening vector shown is from the extinction law of Reike \& Lebofsky (1985). The black dashed lines are parallel to the reddening vector, and intersect the dwarf main-sequence at spectral types of M0 (left line) and M9 (right line). Typical 1-$\sigma$ errors in (J-H) and (H-K$_{s}$) are 0.06 mag and 0.02 mag, respectively.  }
      \label{Av}
 \end{figure*}

Figure~\ref{Av} shows the NIR color-color diagram (ccd) for the NGC 6823 cluster. The thick black and red curves indicate the zero-age main sequence for dwarfs and giants, respectively (covering spectral types O-M9). The zero-age main sequence was assembled from the work of Bessell \& Brett (1988), Leggett (1992), Leggett, Allard \& Hauschildt (1998), and Kirpatrick et al. (2000). The blue line marks the classical T Tauri star (CTTS) locus from Meyer et al. (1997). The dashed lines are parallel to the reddening vector, and intersect the dwarf main-sequence at spectral types of M0 and M9. The NIR ccd shows a highly extincted population of early-type main-sequence dwarfs concentrated near ($H-K_{s}$) $\sim$ 0.7, ($J - H$) $\sim$ 1.5. These stars form a sequence parallel to the reddening vector, and comparing with the length of the $A_{V}$ vector, we can infer that these are located at $A_{V}$ $\geq$ 10. There is another group of stars that lies close to the CTTS locus near ($H-K_{s}$) $\sim$ 0.6, ($J - H$) $\sim$ 1. This is likely the young disk population in the cluster. We also find a more dispersed population that have modest $K_{s}$-band excesses, with ($H-K_{s}$) $\geq$ 1. This `red' population lies to the right of the principal sequence of reddened field stars, and has a different slope than the reddening vector shown. 

The median ($H-K_{s}$) color for YSOs is only about $\sim$0.2 magnitude (e.g., Wilking et al. 1999), which makes it difficult to distinguish young stars with modest $K_{s}$-band excesses from the extincted field stars. Such a distinction can be made more clearly if a combination of NIR and mid-infrared (MIR) colors is used. In order to estimate the extent of reddening in NGC 6823, we have considered the Rayleigh-Jeans Color Excess (RJCE) method described in Majewski et al. (2011). This method uses a combination of NIR and MIR measured colors, and extinction to a star is determined by dereddening the observed colors along color excess vectors to the intrinsic stellar locus. Majewski et al. have based the RJCE method on the ({\it H} - [4.5]) color, since the spread in the intrinsic ({\it H} - [4.5])$_{0}$ color is found to be very small ($\leq$ 0.4 mag) over a wide range of B through M spectral types. Due to this near constancy, the interstellar reddening can be estimated using only the observed ({\it H} - [4.5]) color. The RJCE conversion relation is, $A_{K}$ = 0.918 ({\it H} - [4.5] - 0.08), where Majewski et al. (2011) have considered an intrinsic ({\it H} - [4.5])$_{0}$ color of 0.08, and the reddening law from Indebetouw et al. (2007) has been used to derive the conversion relation. 





Figure~\ref{Av2}a shows the $A_{V}$ histogram for NGC 6823 ($A_{K}$ = 0.112 $A_{V}$; Rieke \& Lebofsky 1985). We find a bimodal shape for the $A_{V}$ distribution, with a peak at $A_{V}$$\sim$3 mag and a broader peak at $A_{V}$$\sim$10 mag. To demonstrate the bimodality, we applied a simple $\chi^{2}$ goodness-of-fit test. Our null hypothesis is that a unimodal distribution cannot adequately represent the observed $A_{V}$ distribution. We fitted the $A_{V}$ histogram by a Gaussian distribution, using the MPFITPEAK program in IDL. The Gaussian fit is shown in Fig.~\ref{Av2}a (bottom panel). The peak value of the Gaussian is 42.64 and the Gaussian sigma is 7.07. The number of degrees of freedom (dof) is computed as the difference of the number of elements and the number of free parameters used for the fit. The Gaussian peak value (height), sigma (width), and the centroid were used as the free parameters for fitting. The resulting $\chi^{2}$ (summed squared residuals) value of the fit is found to be 1745.06, and the reduced-$\chi^{2}$ value ($\chi^{2}$/dof) is 32.3. For an ideal fit, we would expect the reduced-$\chi^{2}$ value to be less than or near one. The large value obtained indicates that the fit has missed too many data points and is a poor fit. The associated $\chi^{2}$ probability, {\it Q}, which estimates the significance of the $\chi^{2}$ test, is found to be 0.0. A high value of {\it Q} indicates that the resulting $\chi^{2}$ value will be different if the test is repeated again. If the observed $A_{V}$ distribution were unimodal with a single broad peak observed around $A_{V}$$\sim$10 mag, we would have obtained a good fit with the Gaussian distribution and the resulting reduced-$\chi^{2}$ value would be close to one. Thus our null hypothesis is correct and the bimodality in the $A_{V}$ distribution is likely real. 


A large fraction of the sources are part of the more extincted population with $A_{V}$$>$5 mag. Fig.~\ref{Av2}b compares the spatial distribution of these two groups of sources, with $A_{V}$ $\leq$ 5 mag (blue points) and $A_{V}$ $>$5 mag (green points). The central Trapezium in NGC 6823 consists of 11 bright O- and B-type stars, which show nearly uniform reddening of $A_{V}$ = 2.57$\pm$0.031 mag (Guetter 1992; Pigulski et al. 2000). The Trapezium stars are thus part of the smaller group of sources with a peak at $A_{V}$$\sim$3 mag. Overall, we do not find any prominent high- or low- extinction regions in the cluster. Pigulski et al. (2000) had noted higher extinction in the eastern part of the central Trapezium, and had explained it by the presence of the reflection nebula NGC 6820 in that direction. In Fig.~\ref{Av2}c (top panel), we have compared the extent of extinction in the eastern and western parts of the cluster. We have plotted here the separation in RA for each source from the central Trapezium. We have considered the star E81 Sh-d [identification numbers from Erickson (1971) and Sharpless (1959)] located at ($\alpha$, $\delta$) = (295.8$\degr$, 23.3$\degr$) to be the center of the Trapezium system. This is a B1.5V star, and has a 98\% cluster membership probability from Erickson (1971). In Fig.~\ref{Av2}c, positive values for $\Delta$RA imply sources located to the east of the central Trapezium. A wider range in $A_{V}$ between 0 and 50 mag is observed in the eastern region, while sources in the western region show two confined distributions at $\sim$0 and $\sim$10 mag. The bottom panel in Fig.~\ref{Av2}c compares the cumulative distribution functions, obtained from a Kolmogorov-Smirnov (K-S) statistical test, for the high-extincted ($A_{V}$$>$5 mag) and the low-extincted ($A_{V}$$\leq$5 mag) sources located to the east (solid line) and to the west (dashed line) of the central Trapezium. The 'X' variable in these distributions implies the values for $A_{V}$. For the low-extincted sources ({\it bottom panel, left}), the east and the west distributions have similar shapes. According to the K-S test, the maximum difference between these cumulative distributions is 0.11, and the corresponding probability that the two distributions are similar is {\it P}=0.7. In comparison to this, the cumulative fraction of the high-extincted sources with $A_{V}$ between $\sim$10 and 20 mag is slightly higher in the western part of the cluster than the east. This can be seen from the bottom, right panel in Fig.~\ref{Av2}c. The difference between the two cumulative distributions is not significant; according to the K-S test, the maximum difference is 0.24, and the probability {\it P} that the two distributions are similar is 0.6. There may be a slightly higher concentration of the high-extincted sources located west of the Trapezium, but no significant spatial concentration of the high- or low-extincted sources is evident. Nevertheless, there is significant differential reddening within the cluster, and we can infer that the cluster lies behind at least $\sim$3 magnitudes of visual extinction.

The main uncertainties associated with the RJCE method are due to the spread in the intrinsic stellar colors, the uncertainties in the extinction law used, and the uncertainties in the photometric measurements. Majewski et al. (2011) estimate an uncertainty on $A_{K}$ of $\leq$0.1 mag. We note that the intrinsic ({\it H} - [4.5])$_{0}$ color of 0.08 used in their relation is for F, G and K stars, and the intrinsic color could be as large as 0.4 mag if the later type stars are taken into account. This can lead to a difference of $\sim$3 magnitude in the $A_{V}$ values, which is small in the context of Fig.~\ref{Av}a. The intrinsic variability of a star makes it difficult to accurately estimate the extinction towards each star individually, but such methods are reliable in estimating the extent of reddening over the full survey area studied. A commonly used method of estimating $A_{V}$ is by dereddening each source in the ($H-K_{s}$)/($J - H$) ccd to the CTTS locus (e.g., Luhman et al. 2006). This method, however, could lead to large uncertainties if the sample consists of field stars and diskless weak-line T Tauri stars (WTTS). Also, the spread in the intrinsic NIR colors that include the {\it J}-band photometry is larger than the NIR+MIR color combinations (e.g., Majewski et al. 2011). The RJCE provides a more reliable method for estimating extinction, compared to previous dereddening methods based on NIR colors only.

\begin{figure*}
\includegraphics[width=150mm]{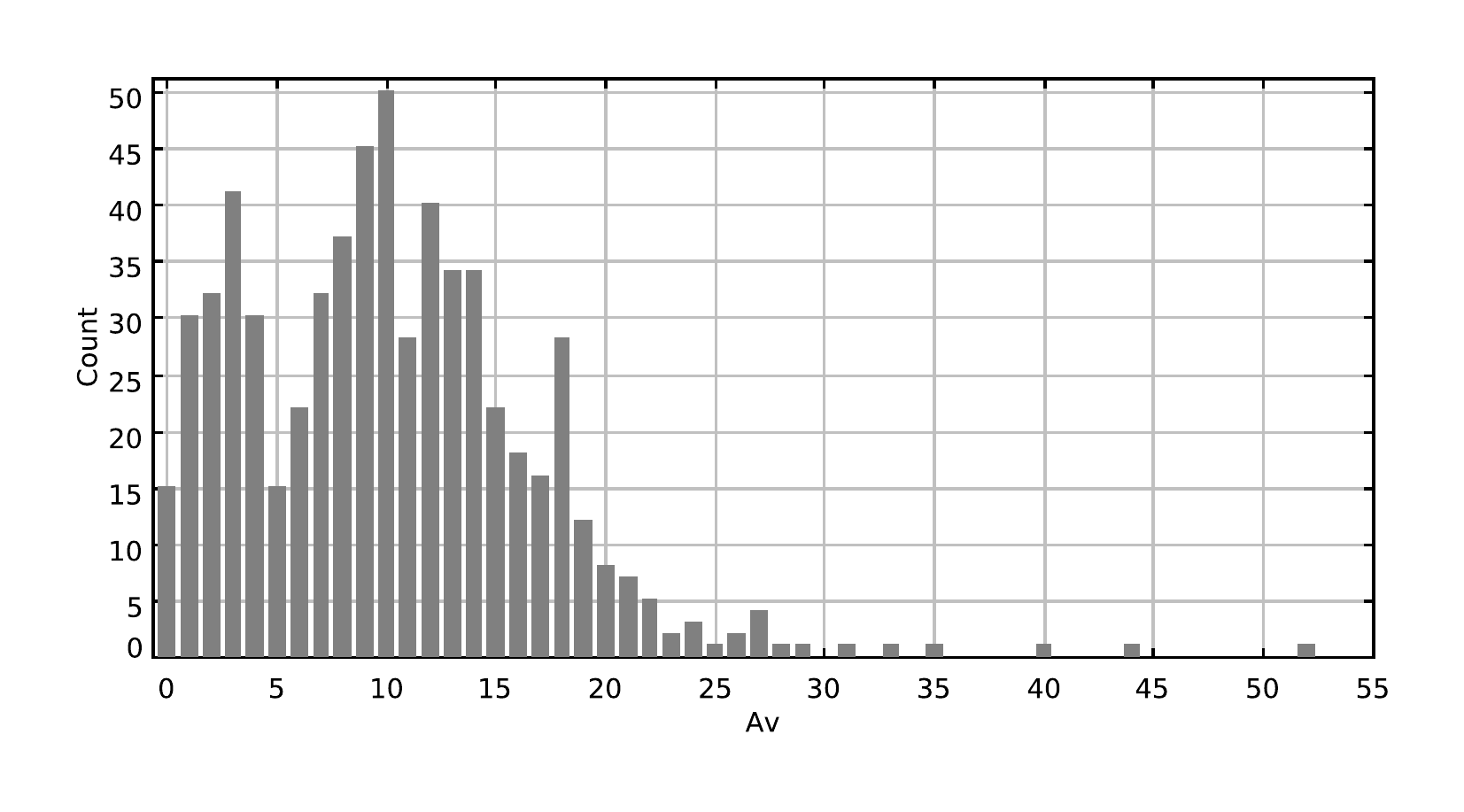} \\ \vspace{0.2in}
\includegraphics[width=95mm]{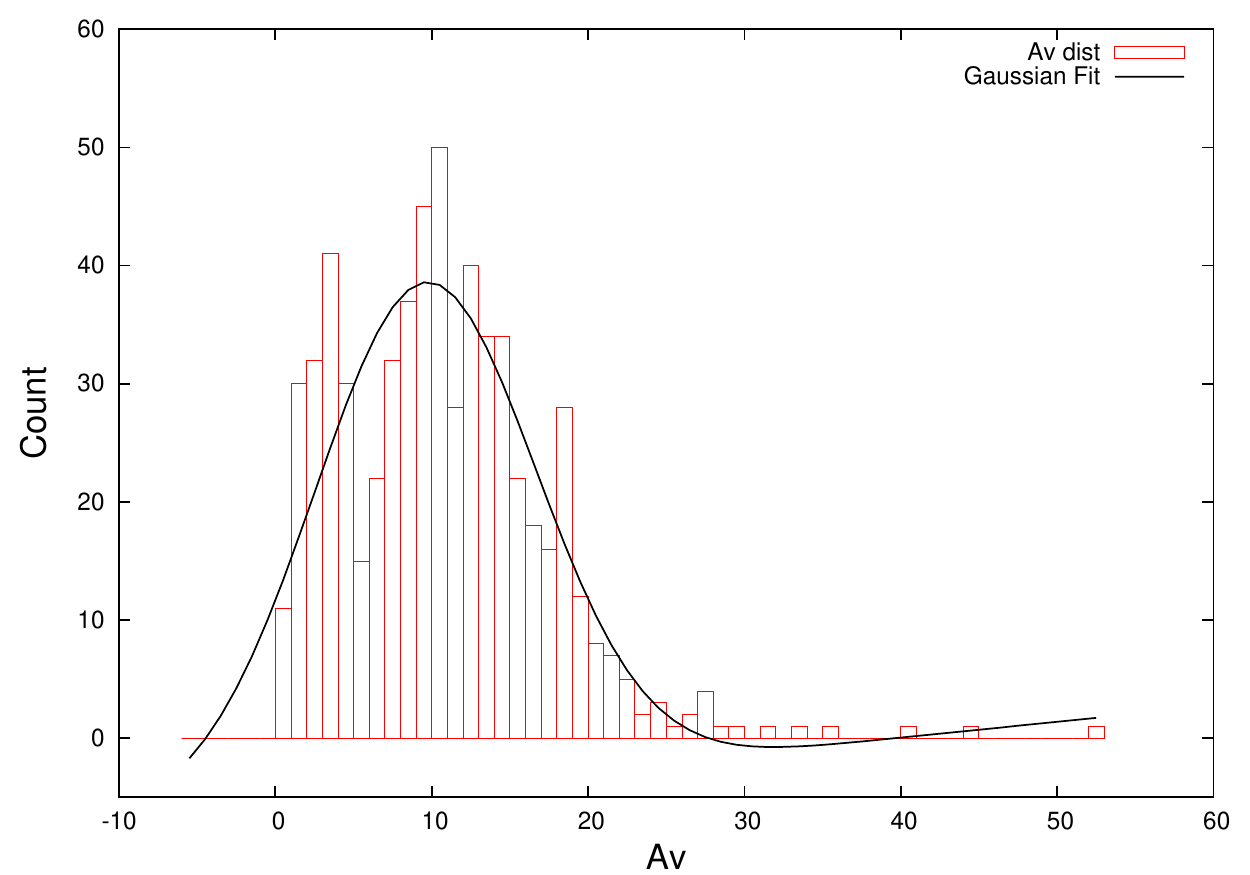} \\
    \caption{(a): ({\it Top}) The $A_{V}$ histogram for NGC 6823 sources, obtained using the RJCE method. A bimodal distribution can be seen, with peaks at $\sim$3 and $\sim$10 mag. {\it Bottom} panel shows a Gaussian fit to the observed $A_{V}$ distribution. More details are provided in the text.  }
      \label{Av2}
 \end{figure*}
 
 
 \begin{figure*}
\setcounter{figure}{2}
  \includegraphics[width=\linewidth]{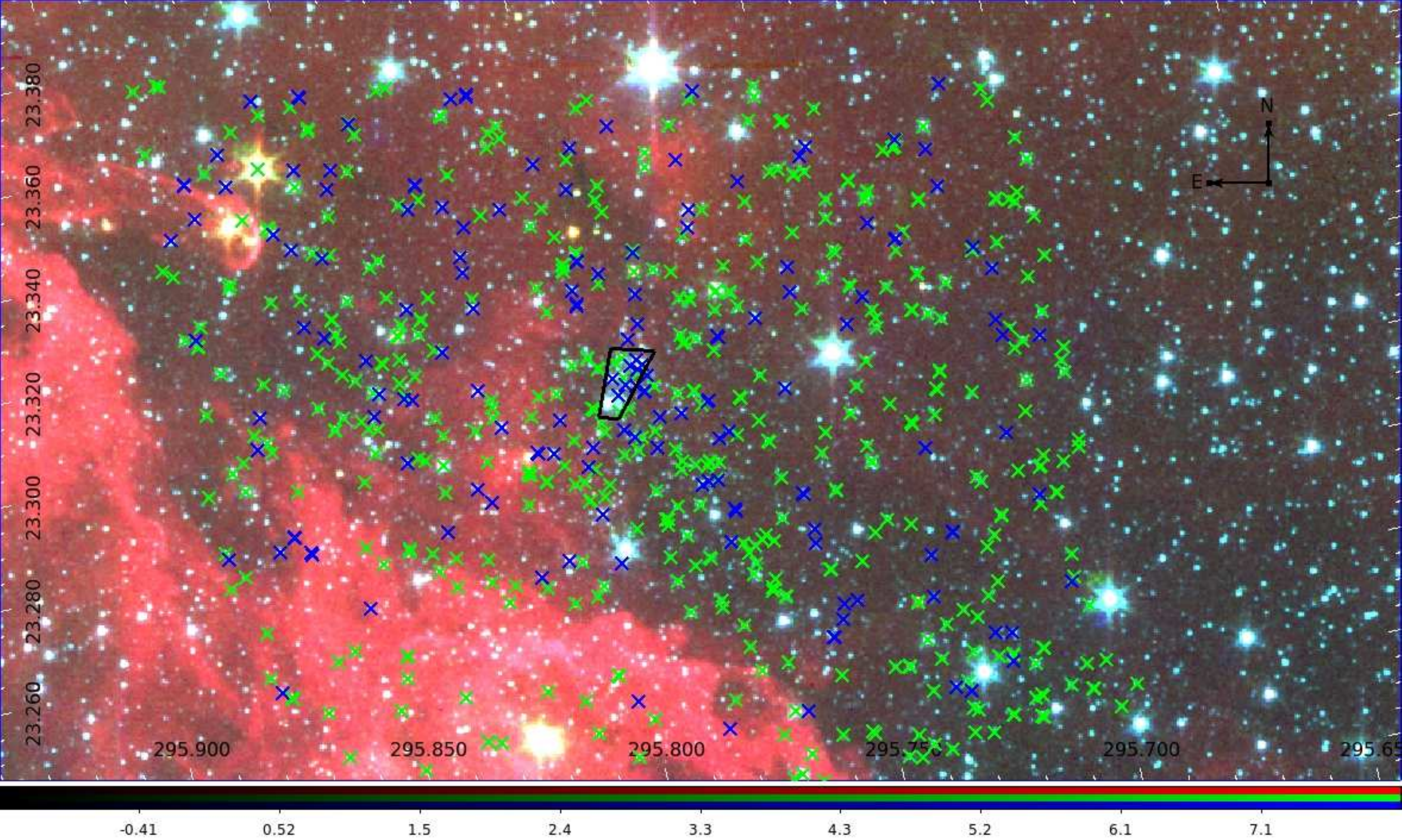}           
    \caption{(b): A comparison of the spatial distribution of the two groups of sources with $A_{V}$ $\leq$ 5 mag (blue points) and $A_{V}$ $>$5 mag (green points). The central Trapezium is marked by black lines. }
      \label{Av2}
 \end{figure*}
 
 
 \begin{figure*}
\setcounter{figure}{2}
\includegraphics[width=100mm]{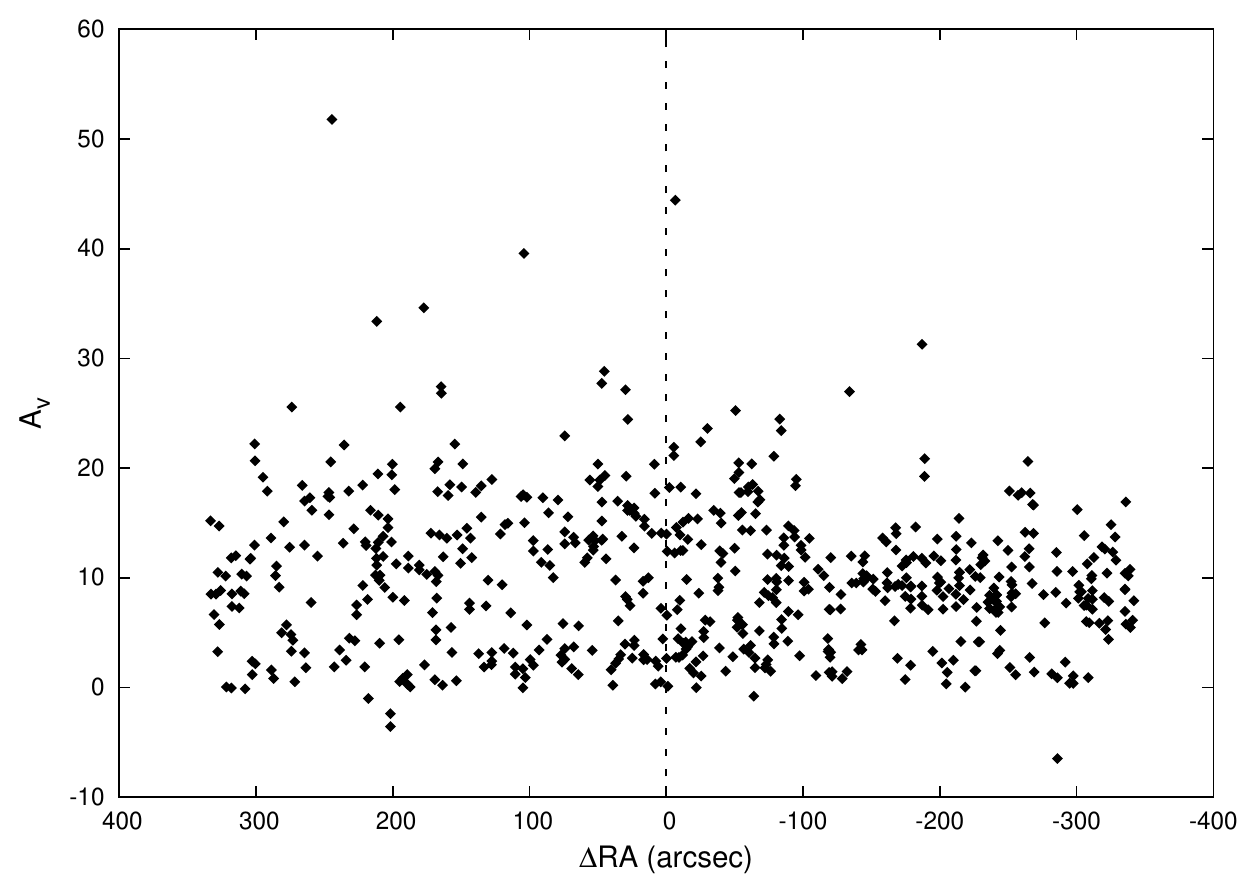} \\  \vspace{0.2in}
\includegraphics[width=70mm]{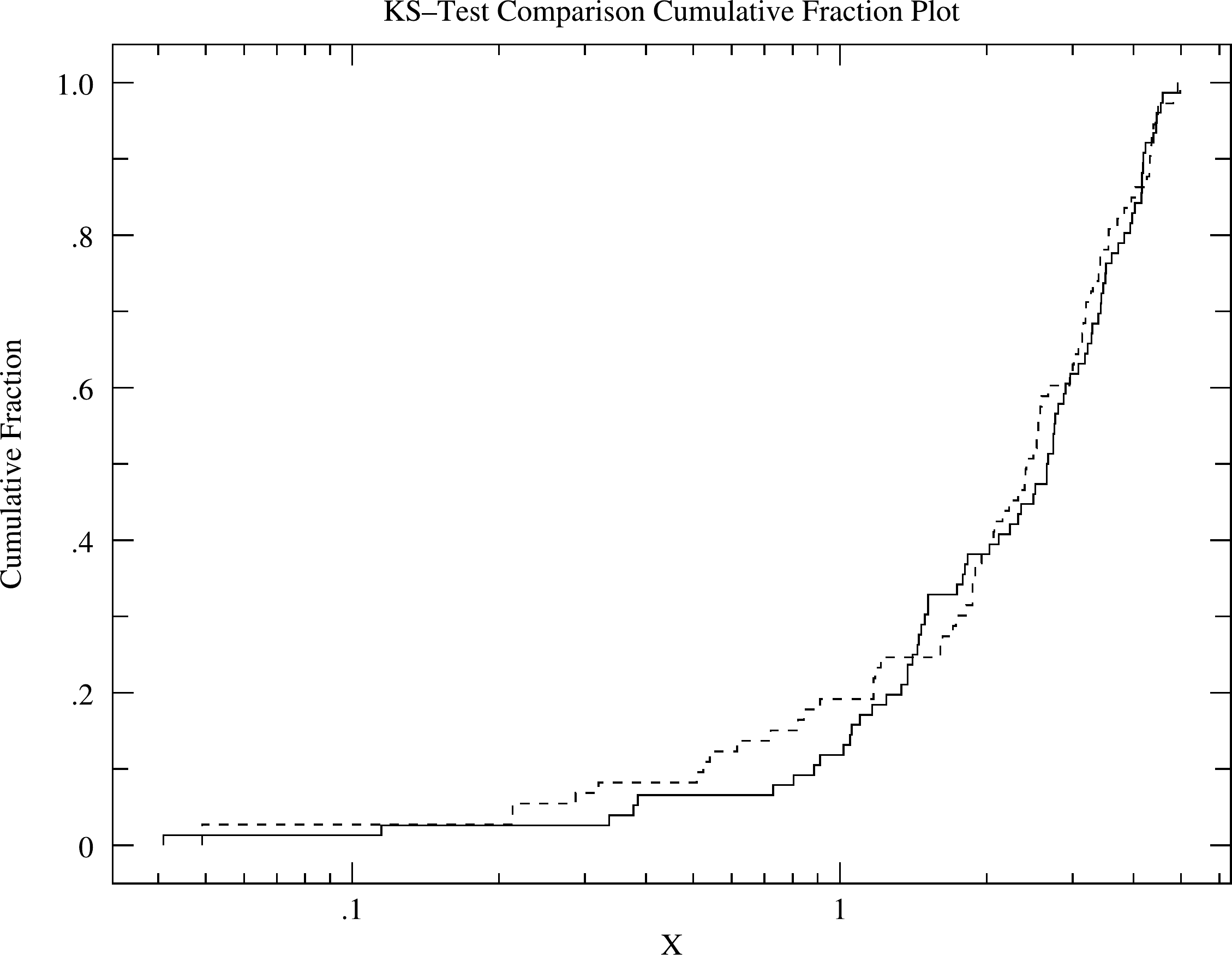} 
\includegraphics[width=70mm]{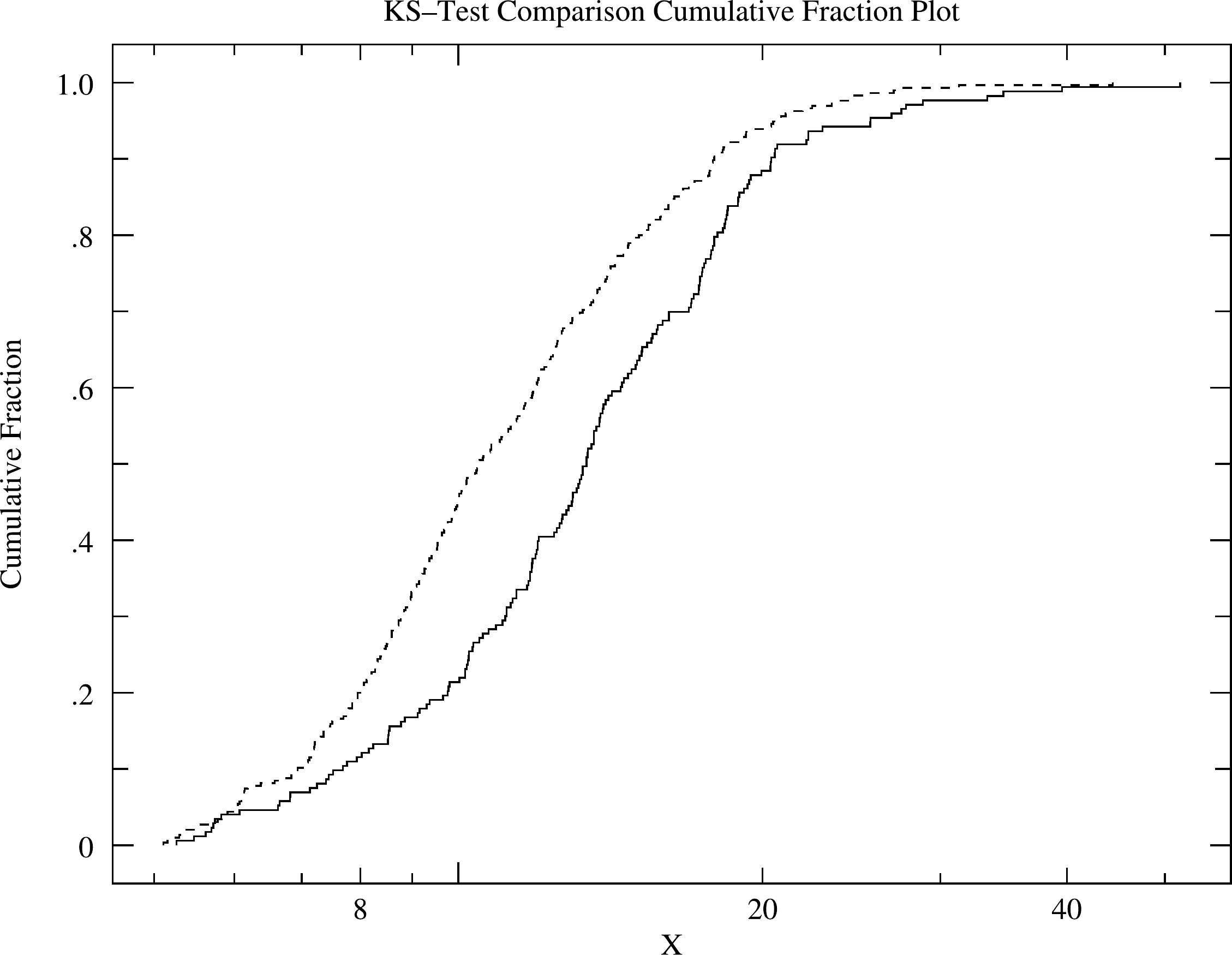} 
    \caption{(c): ({\it Top}) The separation in RA from the central Trapezium vs. $A_{V}$. $\Delta$RA = 0 implies the location of the central Trapezium, positive values for $\Delta$RA imply location east of the Trapezium. {\it Bottom} panel compares the cumulative distribution functions, obtained from a K-S test, for the low-extincted  ({\it left}) and the high-extincted ({\it right}) sources located to the east (solid line) and to the west (dashed line) of the central Trapezium. Here, 'X' implies the values for $A_{V}$. More details are provided in the text.   }
      \label{Av2}
 \end{figure*}

\subsection{IRAC Source Classification}
\label{disks}

We have used the empirically-derived methods described in Gutermuth et al. (2008) to classify the sources that show excess emission in the mid-infrared IRAC bands. There are a few well-known classification schemes that use the 2-8$\micron$ slope to classify YSOs into the different evolutionary classes (e.g., Allen et al. 2004; Lada et al. 2006). These schemes include the 3.6$\micron$ photometry, which is strongly affected by reddening from dust. From the extinction law by Indebetouw et al. (2005) and Flaherty et al. (2007), the extinction is nearly constant at 0.43$A_{K}$ in the 4.5$\micron$ through 8$\micron$ bands, but rises to 0.56$A_{K}$
in the 3.6$\micron$ band. Gutermuth et al. (2005) have made use of the flattening of the extinction law in the IRAC 4.5 to 8$\micron$ bands, and have developed a classification scheme based on the [4.5]-[5.8] color to distinguish between Class I and Class II YSOs. This color is much less affected by dust extinction than
any [3.6]-based color. 

The first phase in their classification method is to remove the various extragalactic contaminants. After applying the respective color cuts given in Gutermuth et al. (2008) for identifying possible extragalactic contaminants, we have found 137 sources which have colors consistent with being PAH, AGN, or shock emission sources. After removing these contaminants, the remaining 490 sources were classified into Class I or Class II by applying the appropriate color constraints defined in Gutermuth et al. (2008). In total, we find 94 Class II and 2 Class I systems. The remaining sources have colors consistent with photospheric emission, and were classified as Class III sources. We note that objects lying in this Class III category can be field stars and/or diskless YSOs. They can also be transition disks with significant inner disk clearing or large inner holes in the disk. Such sources exhibit photospheric emission in the IRAC bands, but could flare up at longer wavelengths. In the discussion that continues, we will refer to the Class III sources as Class III/field stars. In total, these Class III/field stars are 394 in number. We thus have a $\sim$20\% fraction of Class I and Class II YSOs, while about 80\% are Class III/field stars.

\begin{figure*}
\includegraphics[width=80mm]{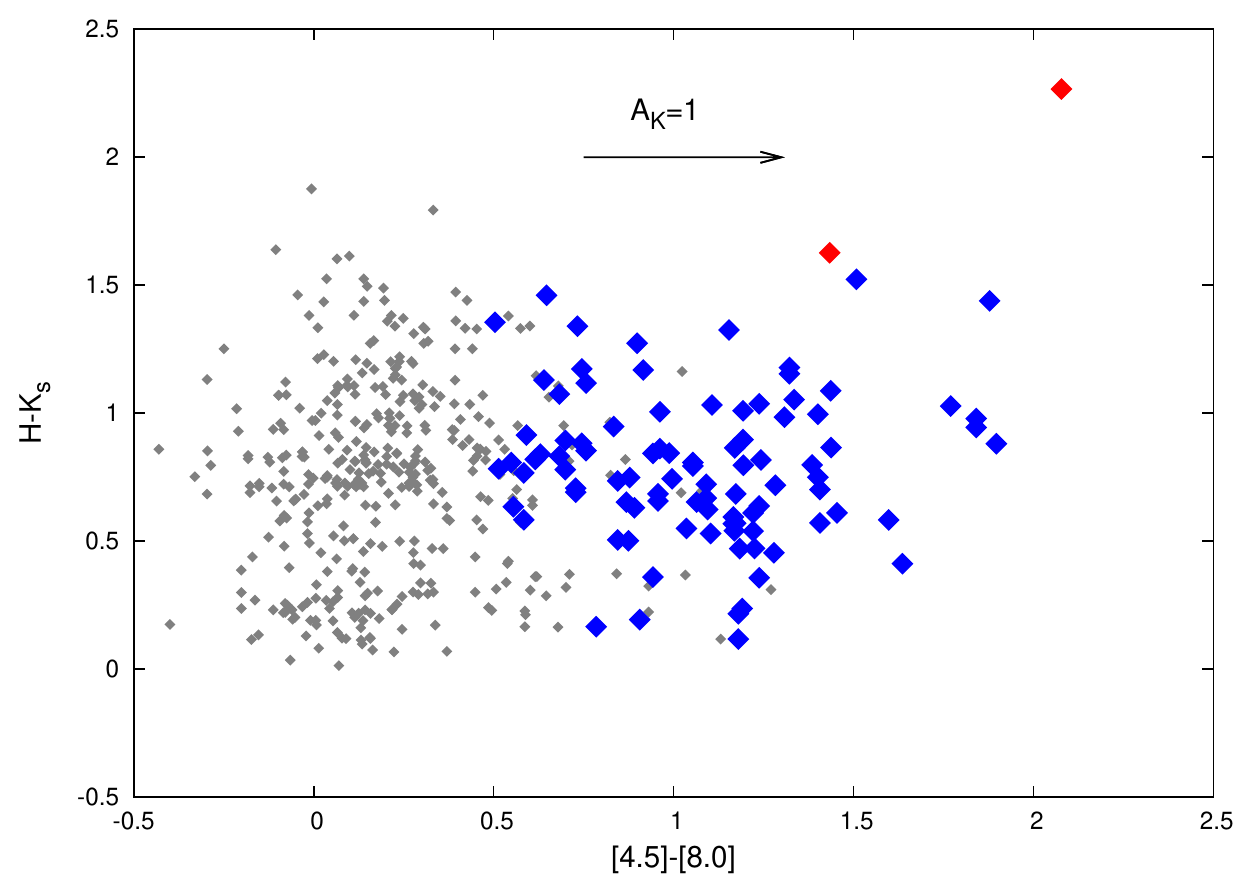} 
\includegraphics[width=80mm]{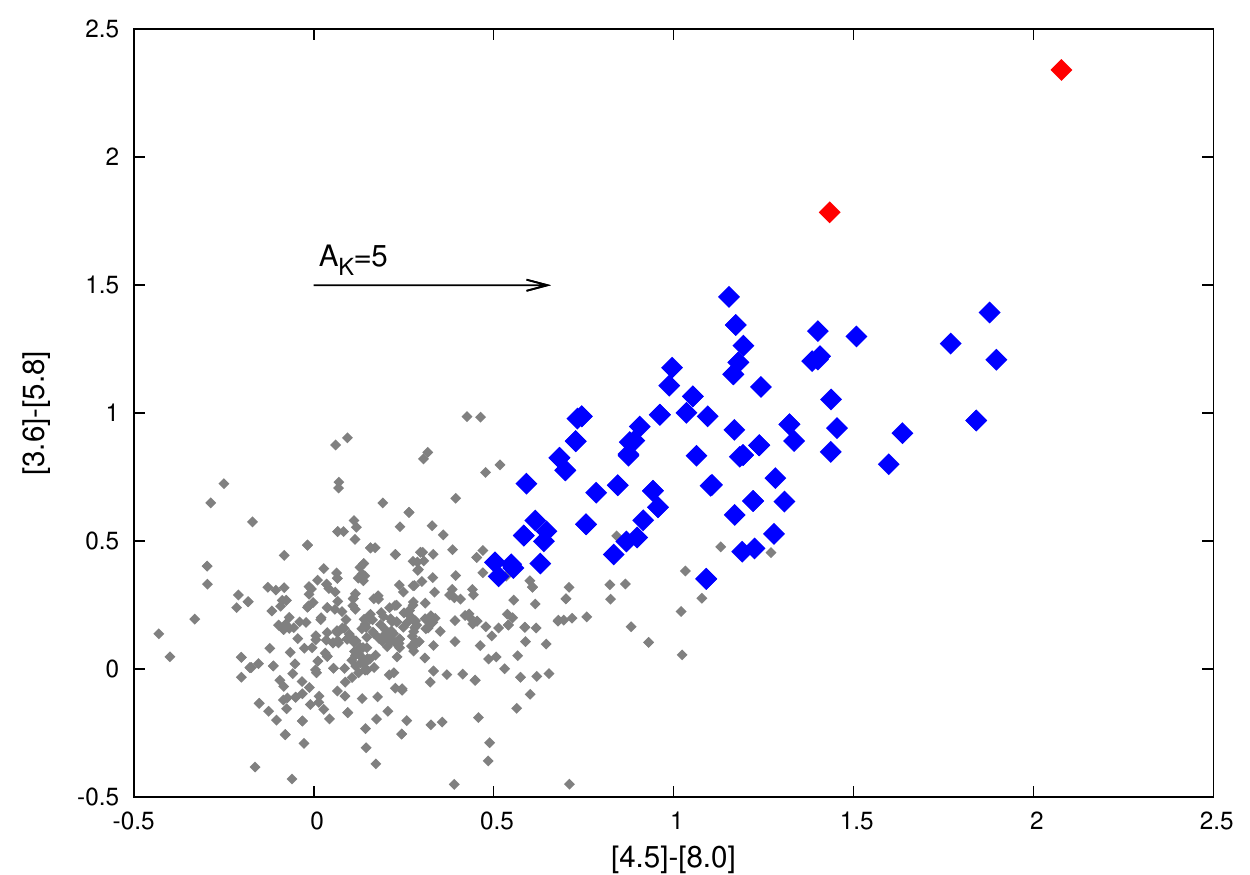} \\
\includegraphics[width=80mm]{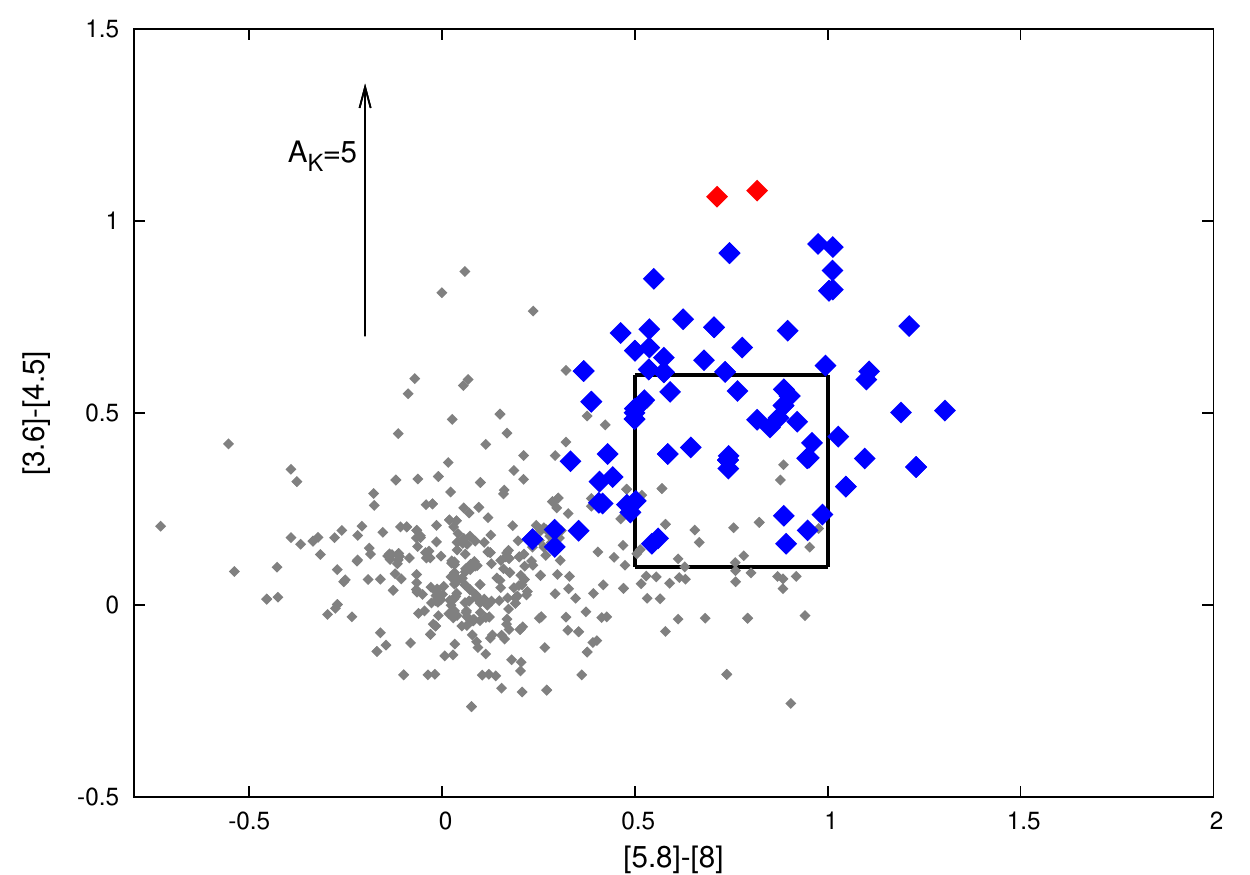}        
\includegraphics[width=80mm]{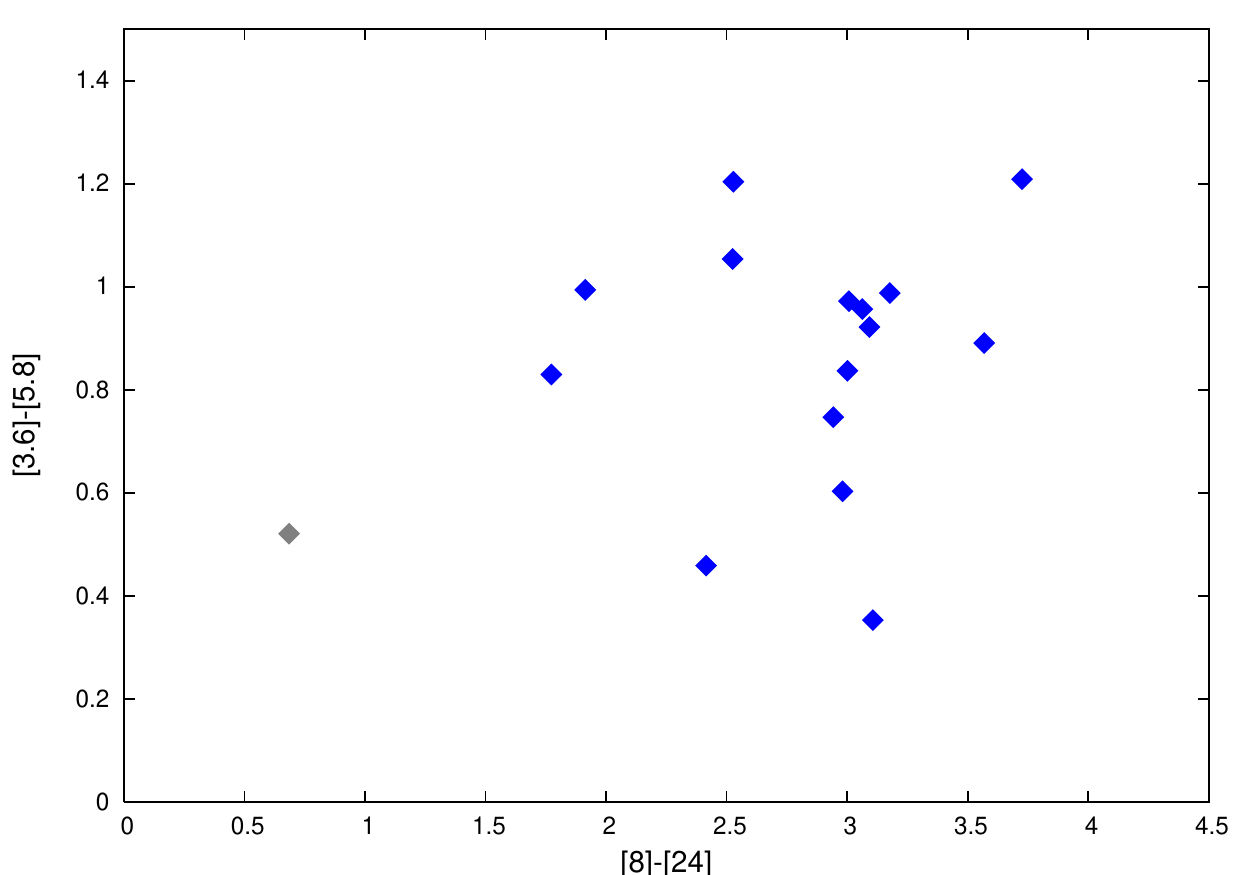}         
    \caption{IRAC ccds for NGC 6823 sources. Red points denote Class I systems, blue are Class II and grey are Class III/field stars. The reddening vectors are plotted using Indebetouw et al. (2005) extinction law. The boxed area in the bottom panel (left) denotes the median Taurus colors.  }
    \label{IRAC}
 \end{figure*}

Figure~\ref{IRAC}a shows the ($H-K_{s}$) versus [4.5]-[8] ccd for the 490 sources. The YSOs and the diskless sources show a similar wide range in the NIR colors, but the IRAC color provides a good separation between the photospheric and the disk population at [4.5]-[8]$\sim$0.4. There are a few Class III/field stars with IRAC colors redder than this value, and are more mixed with the Class II sources. These could be the Class III YSOs in the cluster, that show slightly redder IRAC colors than the main-sequence population. The IRAC ccd in Fig.~\ref{IRAC}b shows a nice separation between the different evolutionary classes. The protostars are the reddest in the [3.6]-[5.8] color and lie in the top right corner. The Class II systems can be identified at [4.5]-[8]$\geq$0.5, [3.6]-[5.8]$\geq$0.4, while the Class III/field stars form a locus near [4.5]-[8] and [3.6]-[5.8]$\sim$0.2-0.3. 
 
In Fig.~\ref{IRAC}c, the inner box indicates the IRAC colors observed for the T Tauri stars in Taurus, from the work of Hartmann et al. (2005). The median IRAC colors in Taurus are [3.6]-[4.5] = 0.4, [4.5]-[5.8] = 0.5, [5.8]-[8] = 0.8. For the Class II sources in NGC 6823, the median colors are [3.6]-[4.5] = 0.5, [4.5]-[5.8] = 0.3, [5.8]-[8] = 0.7. Most NGC 6823 Class II YSOs show stronger excess in the [3.6]-[4.5] color compared to Taurus, which can be explained by the higher extinction observed in this cluster. Taurus, in comparison, is relatively free of extinction ($A_{V}$$\leq$2 mag; e.g., Luhman et al. 2006). 

\begin{figure*}
\centering
 \includegraphics[width=4cm]{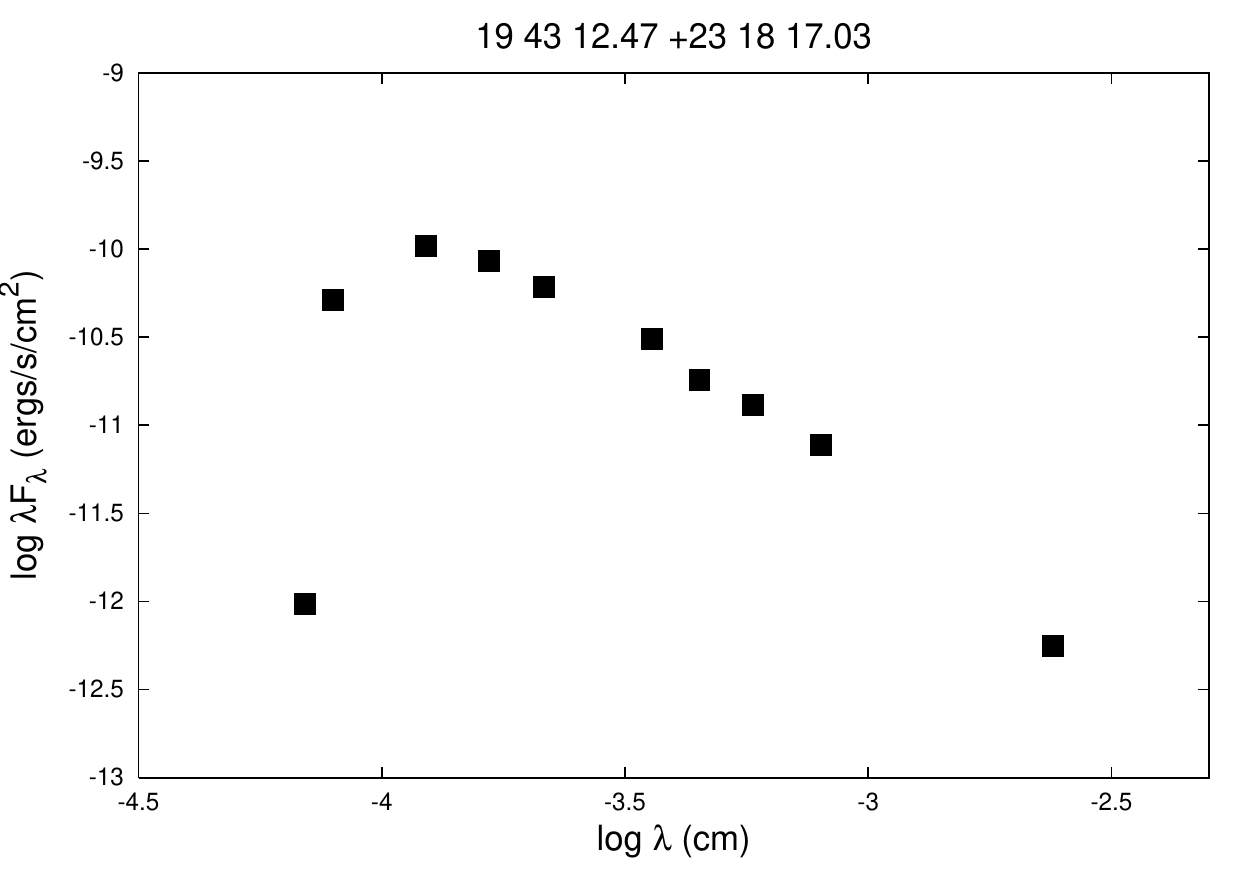} 
  \includegraphics[width=4cm]{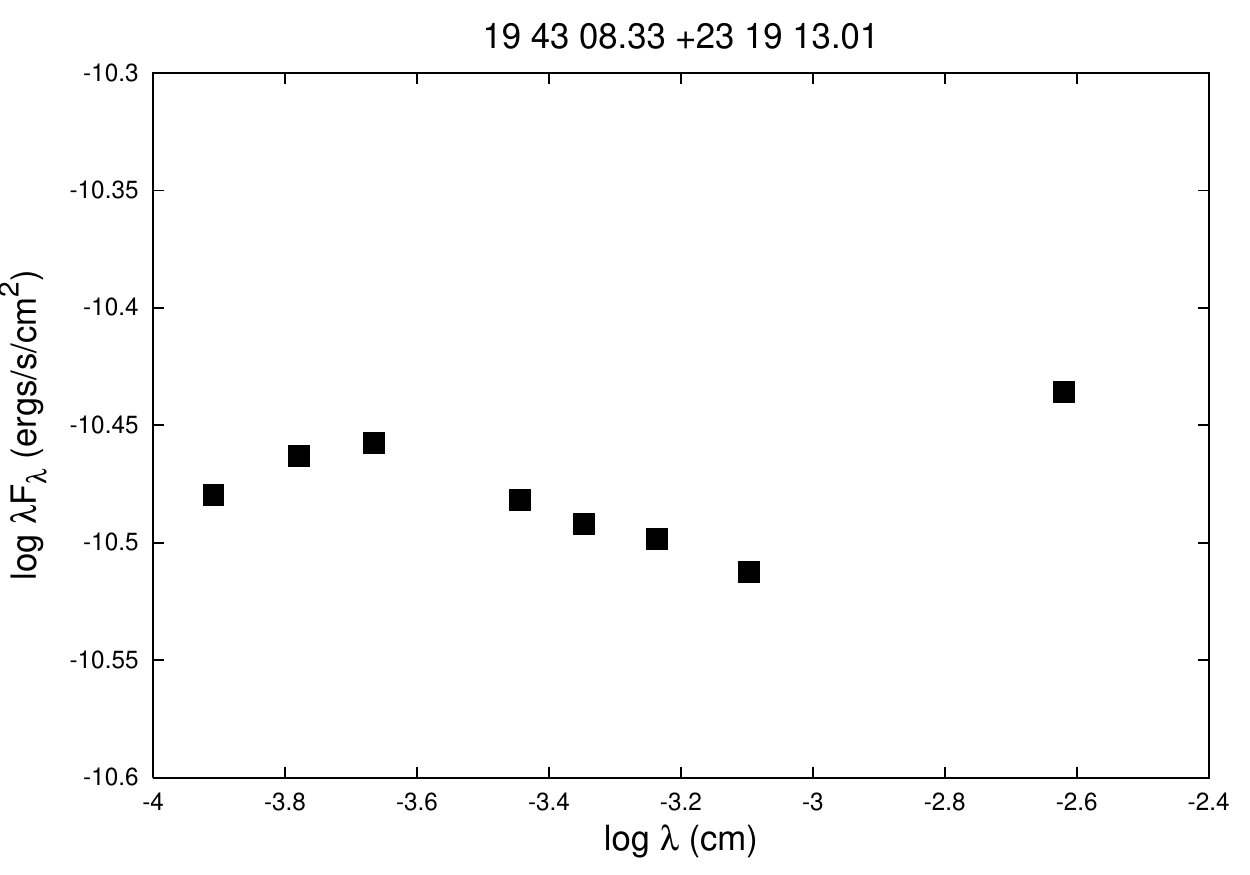} 
  \includegraphics[width=4cm]{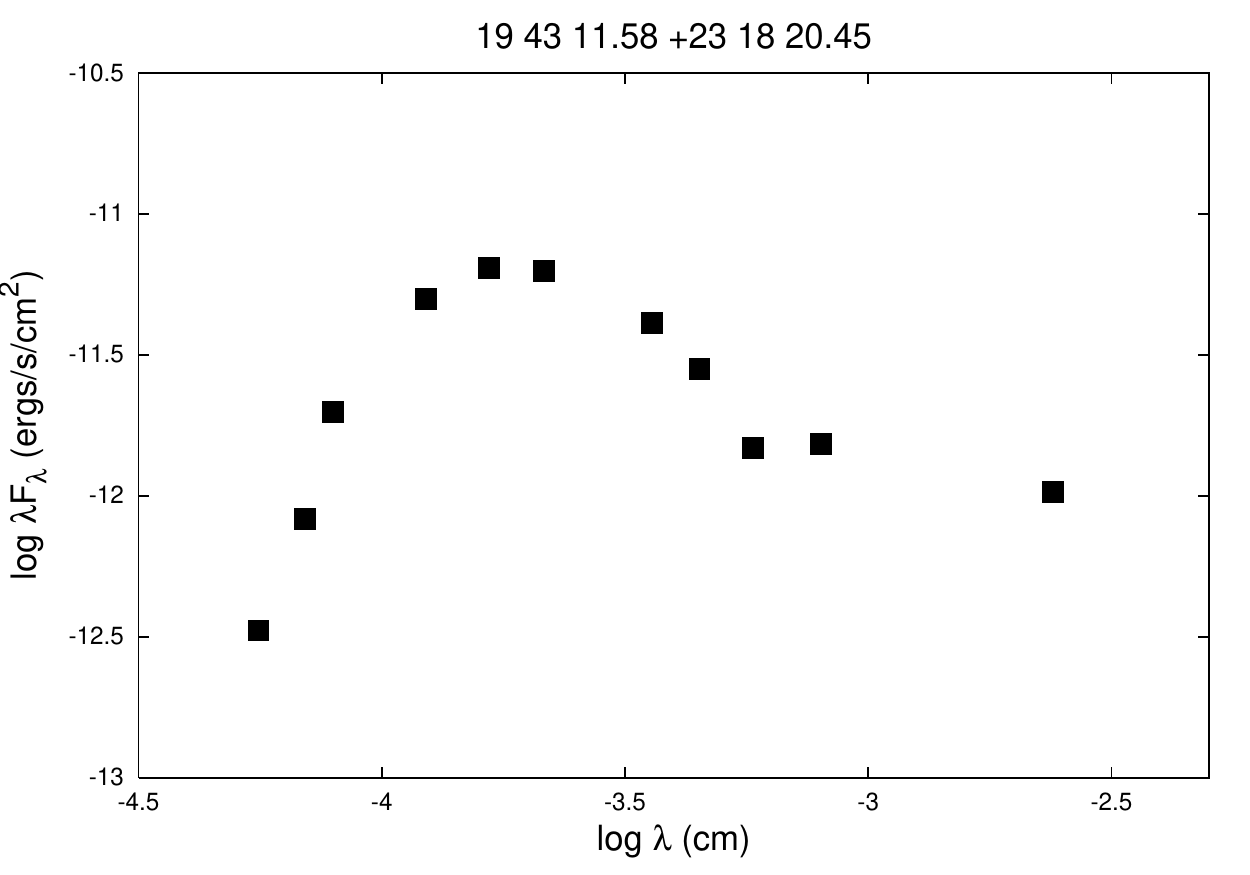}   
 \includegraphics[width=4cm]{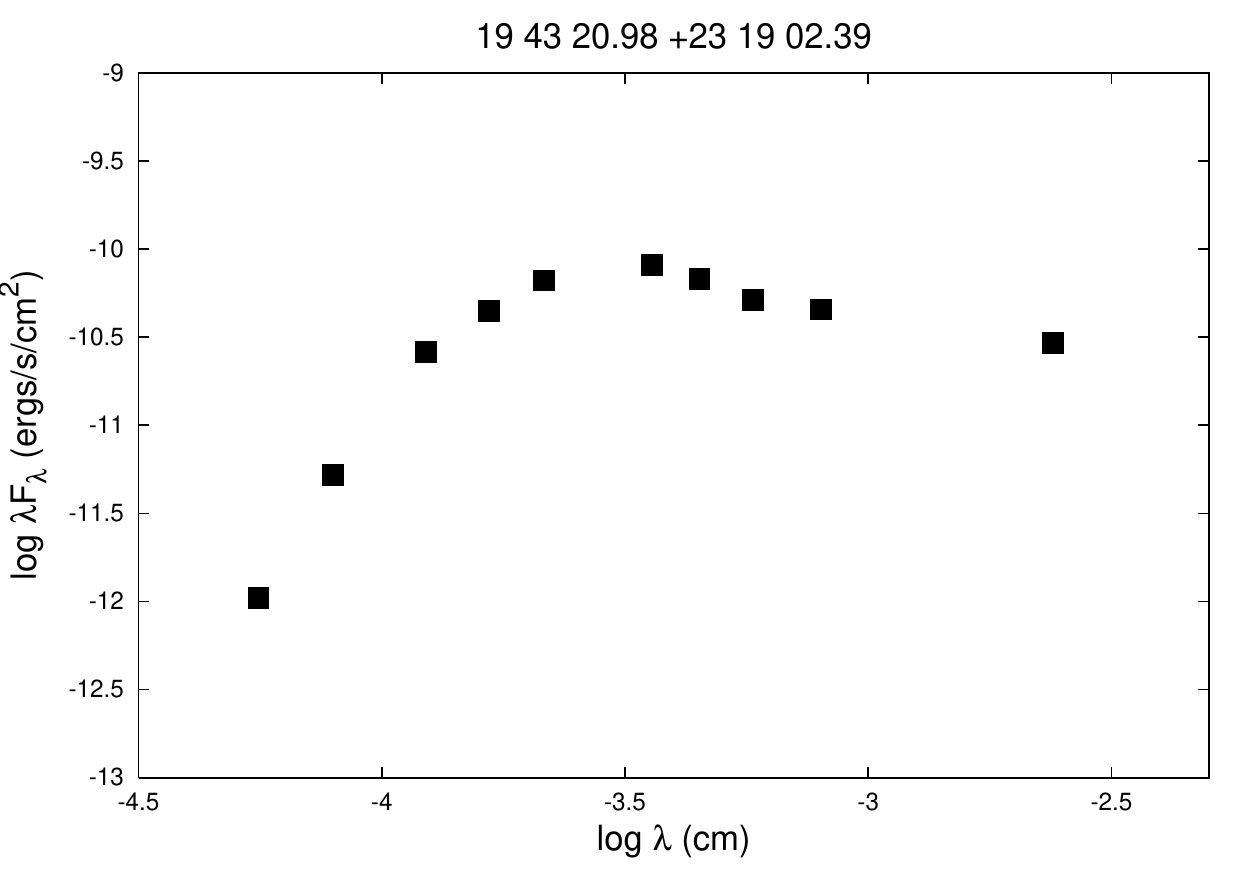}  \\  \vspace{0.2cm}    
  \includegraphics[width=4cm]{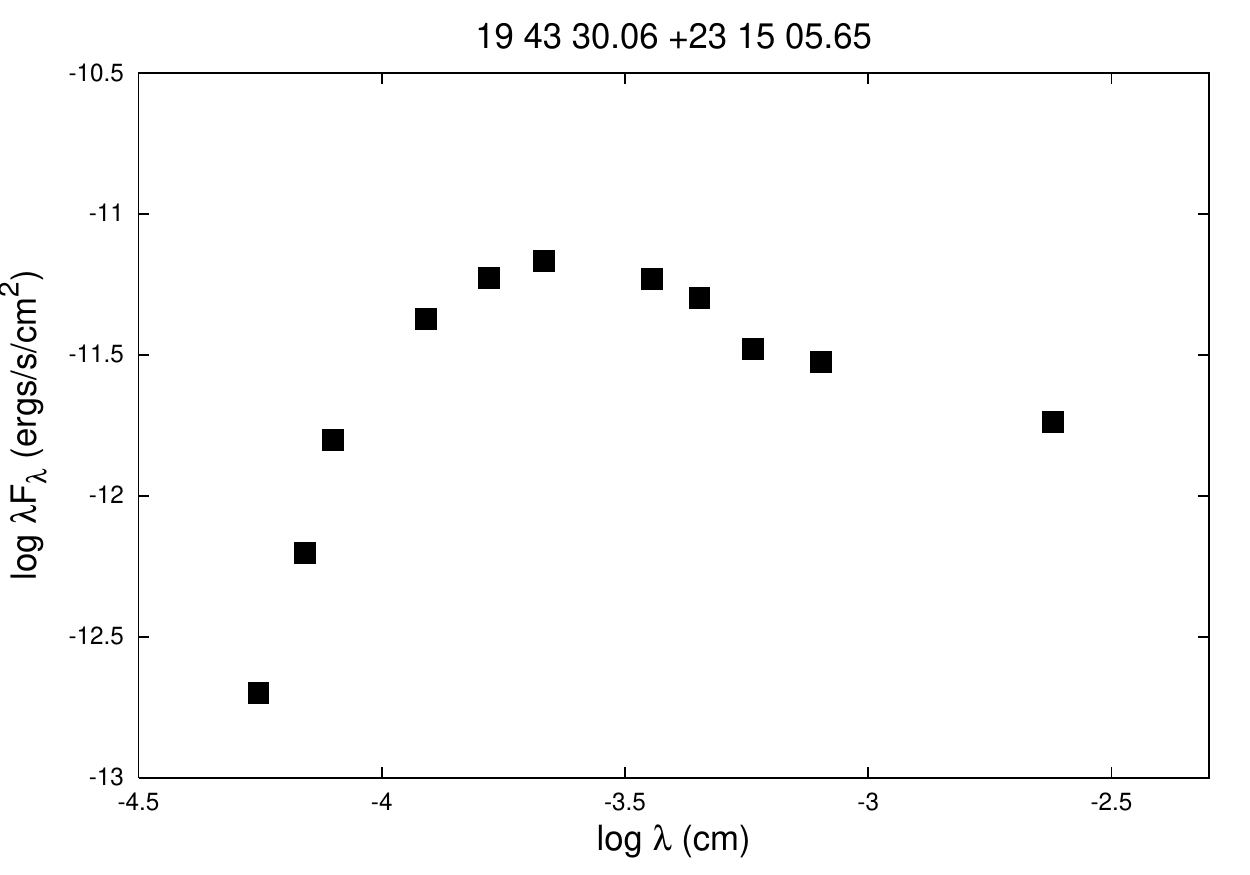} 
   \includegraphics[width=4cm]{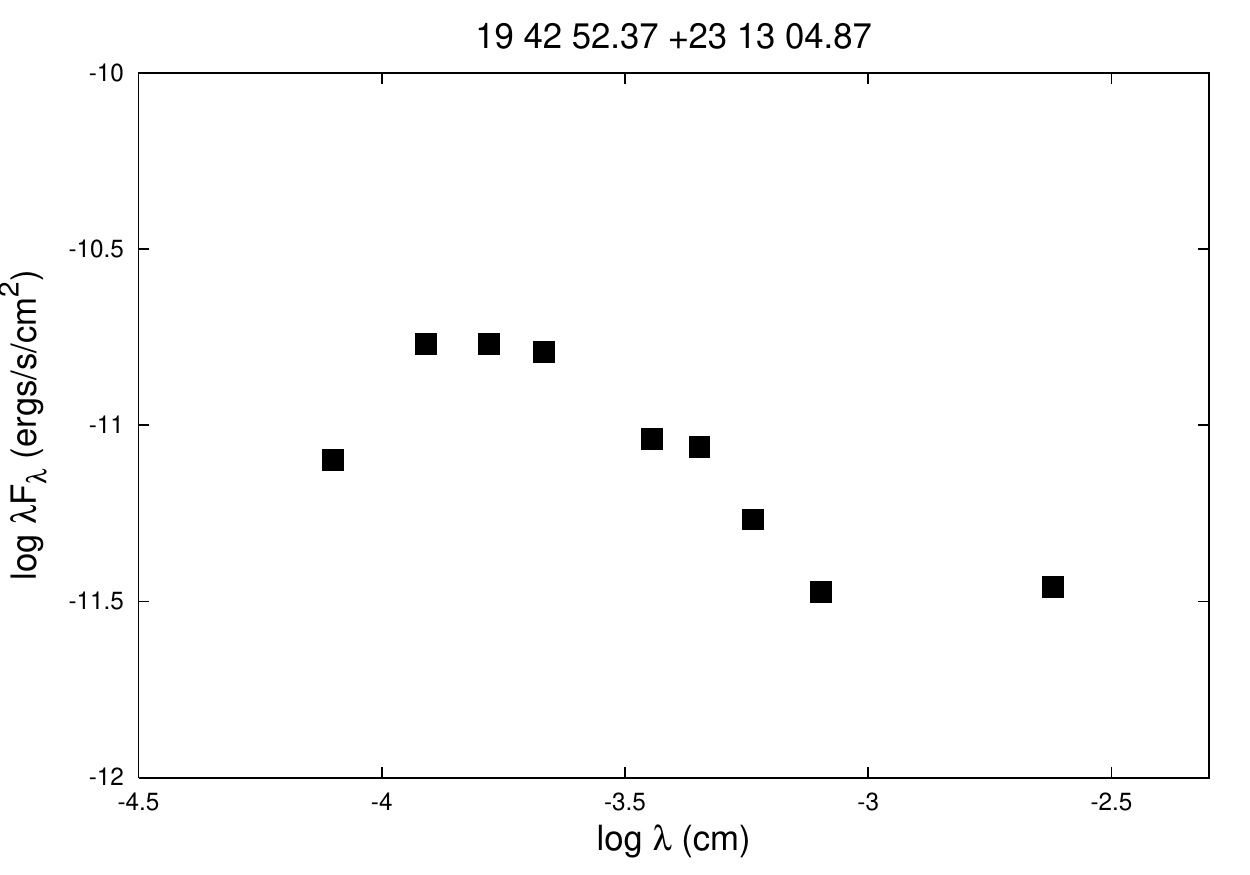} 
 \includegraphics[width=4cm]{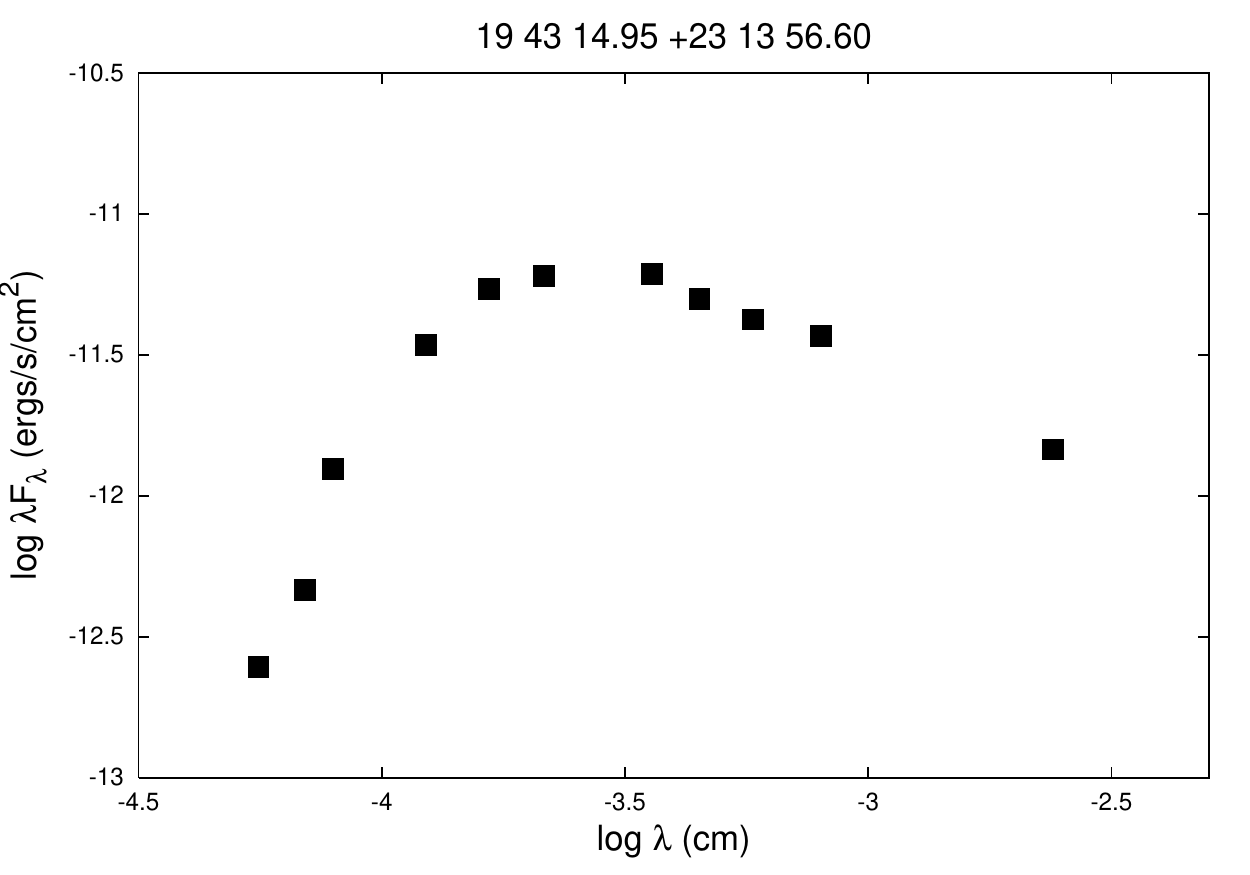} 
 \includegraphics[width=4cm]{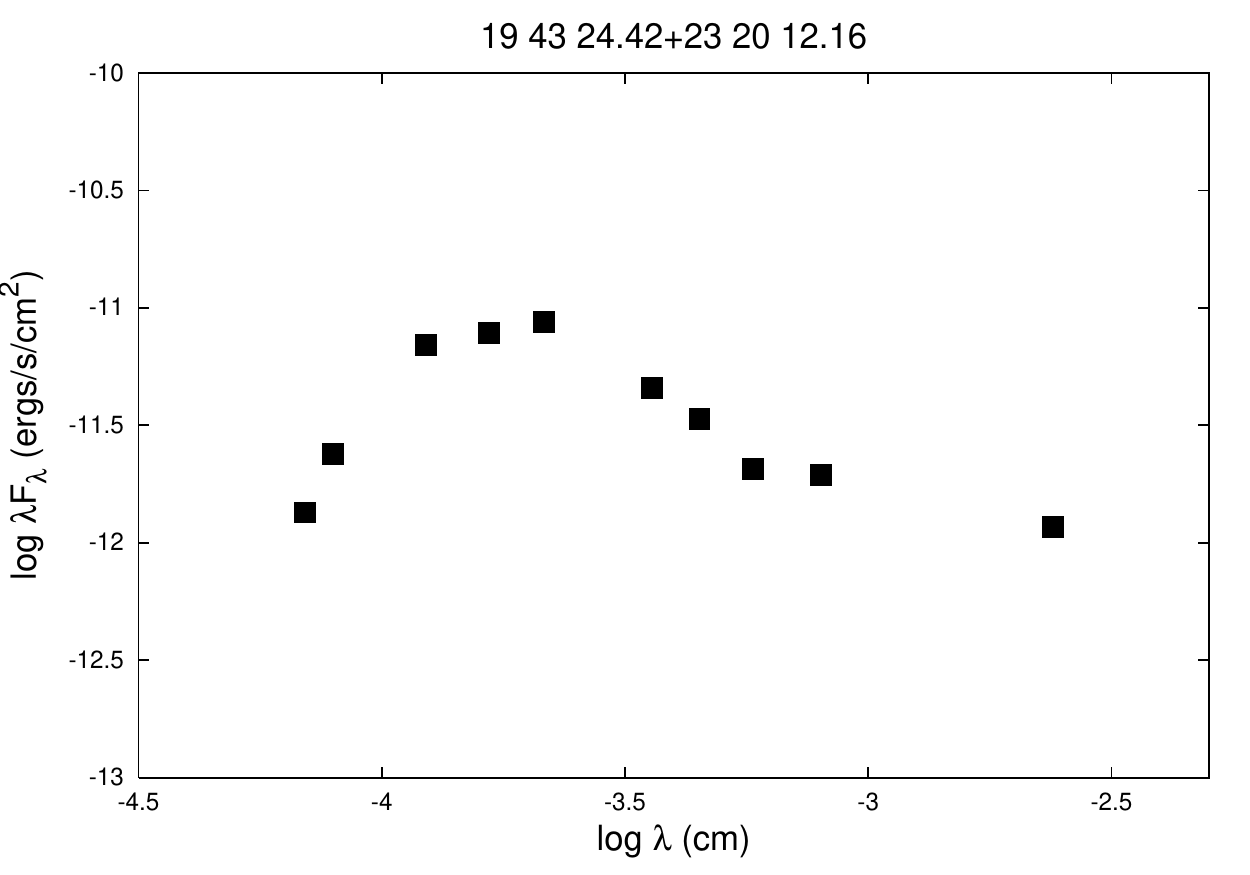} \\  \vspace{0.2cm}    
  \includegraphics[width=4cm]{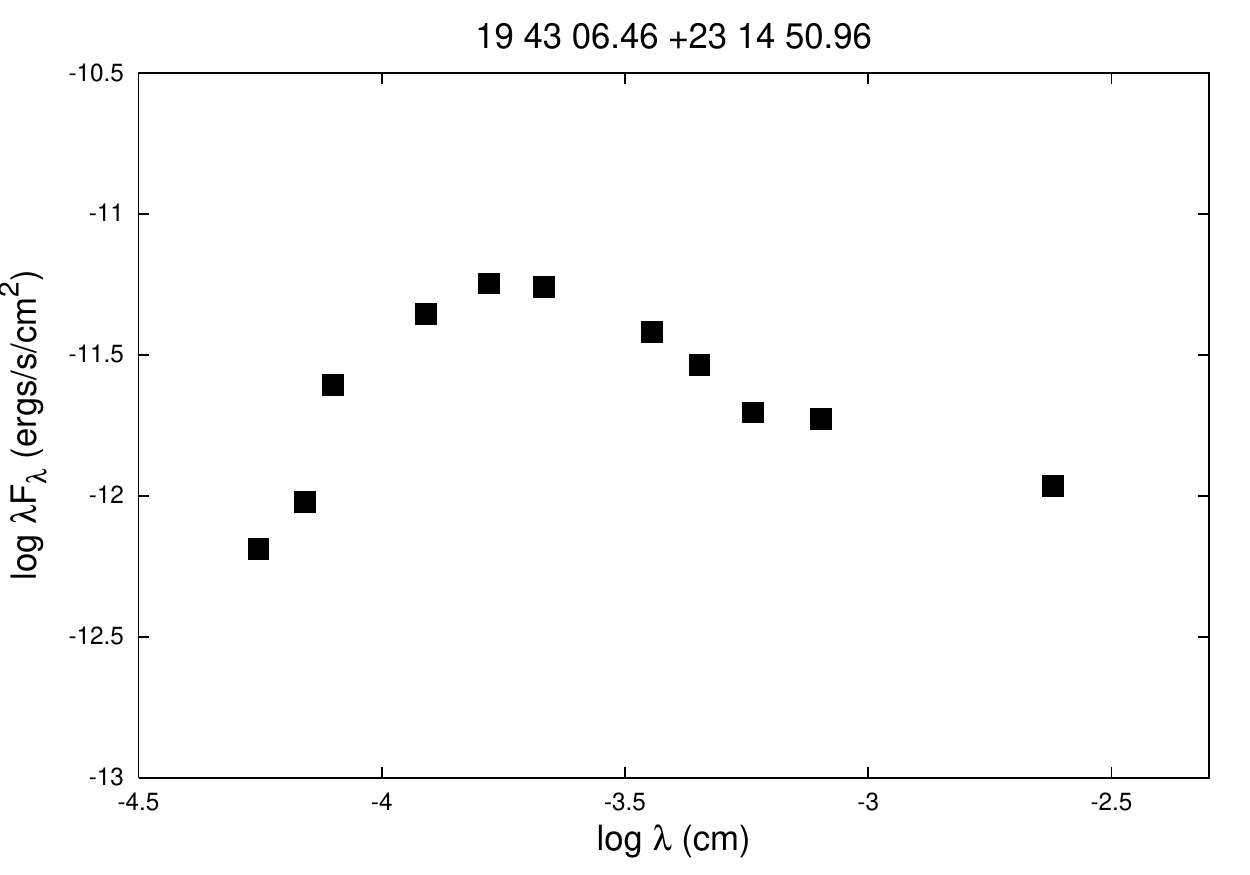} 
  \includegraphics[width=4cm]{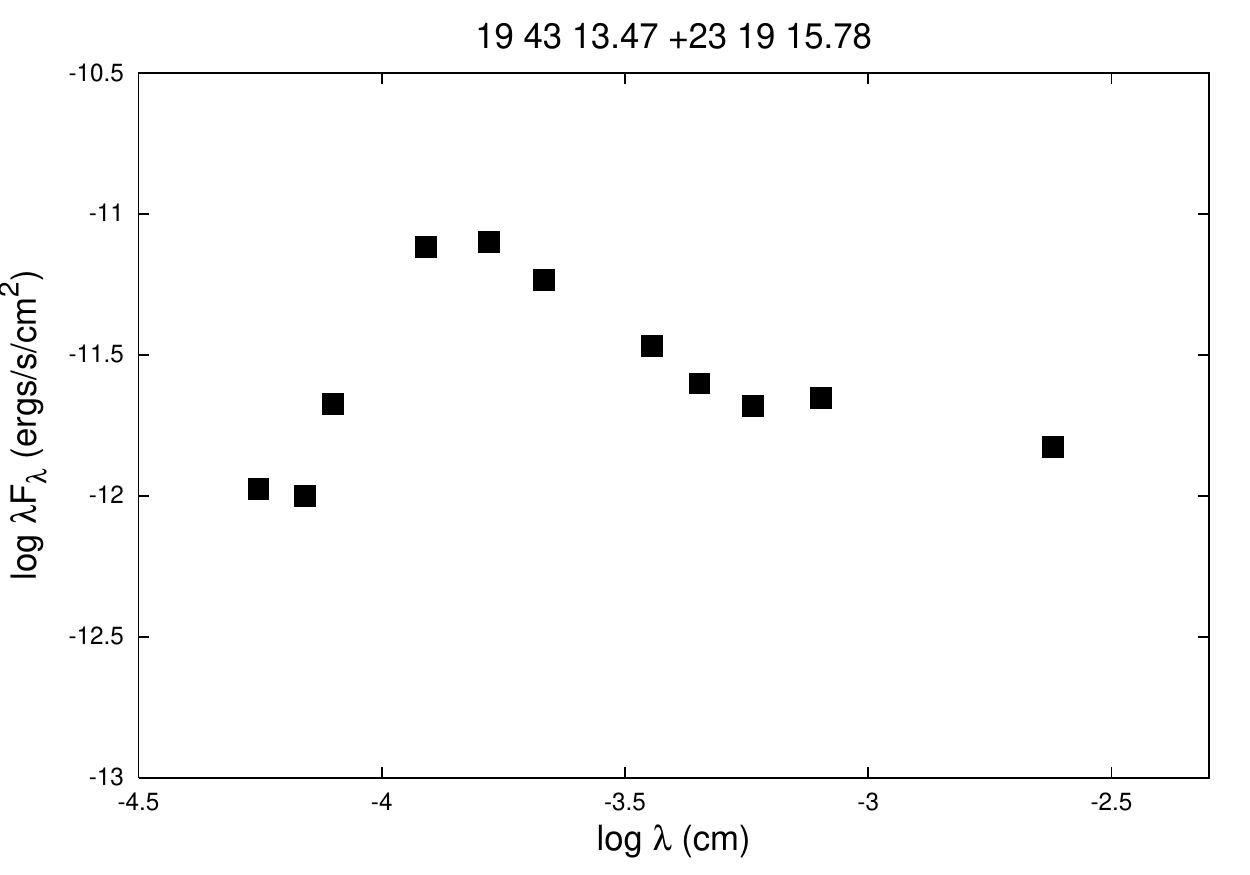}
  \includegraphics[width=4cm]{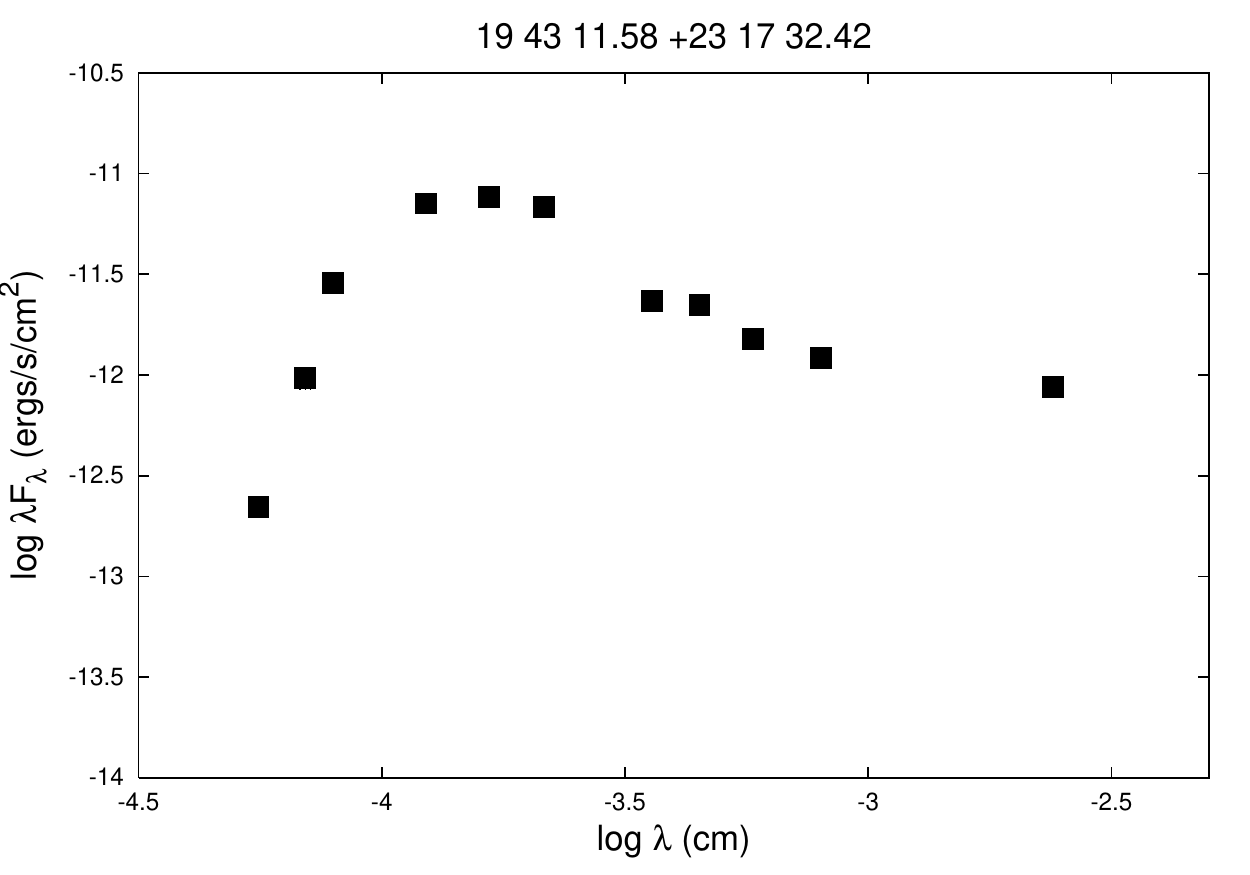} 
 \includegraphics[width=4cm]{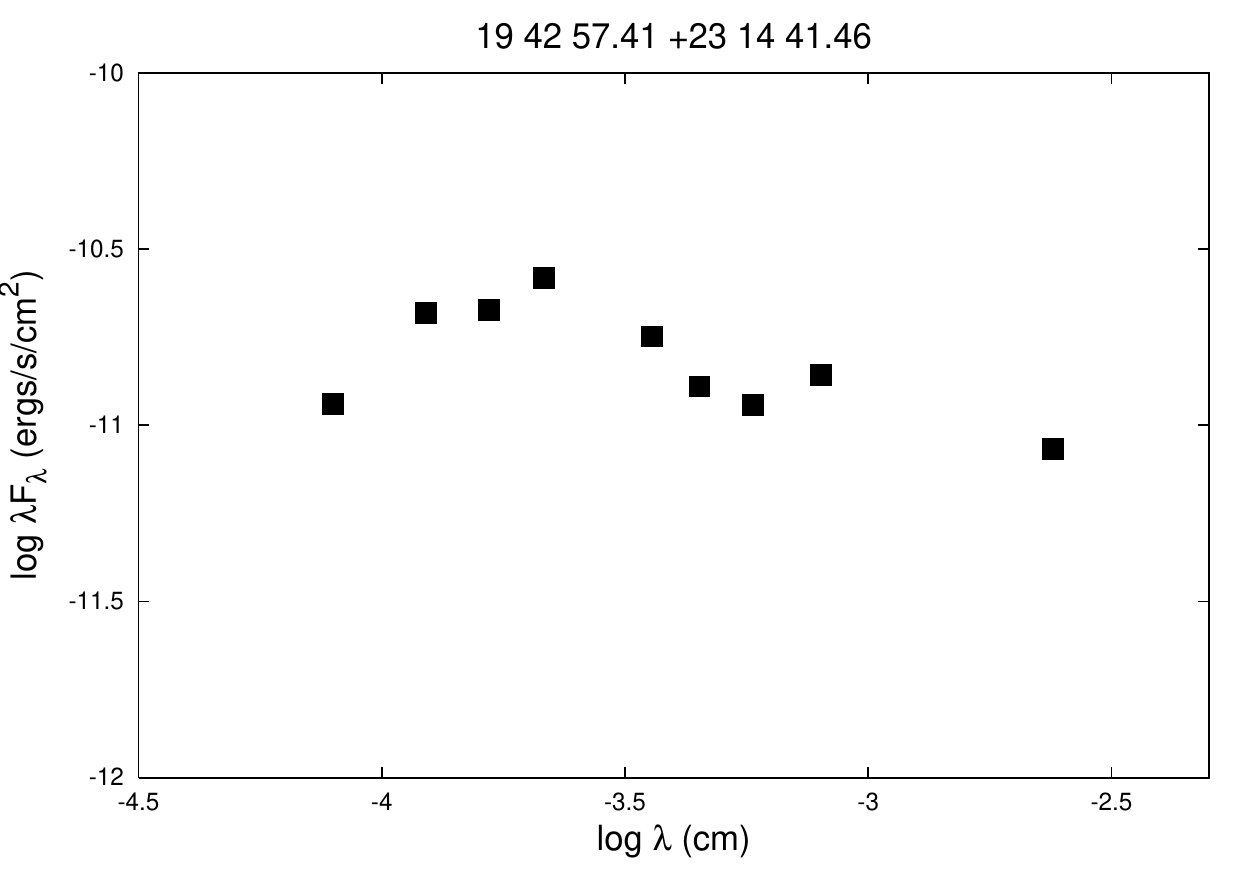} \\  \vspace{0.2cm}   
 \includegraphics[width=4cm]{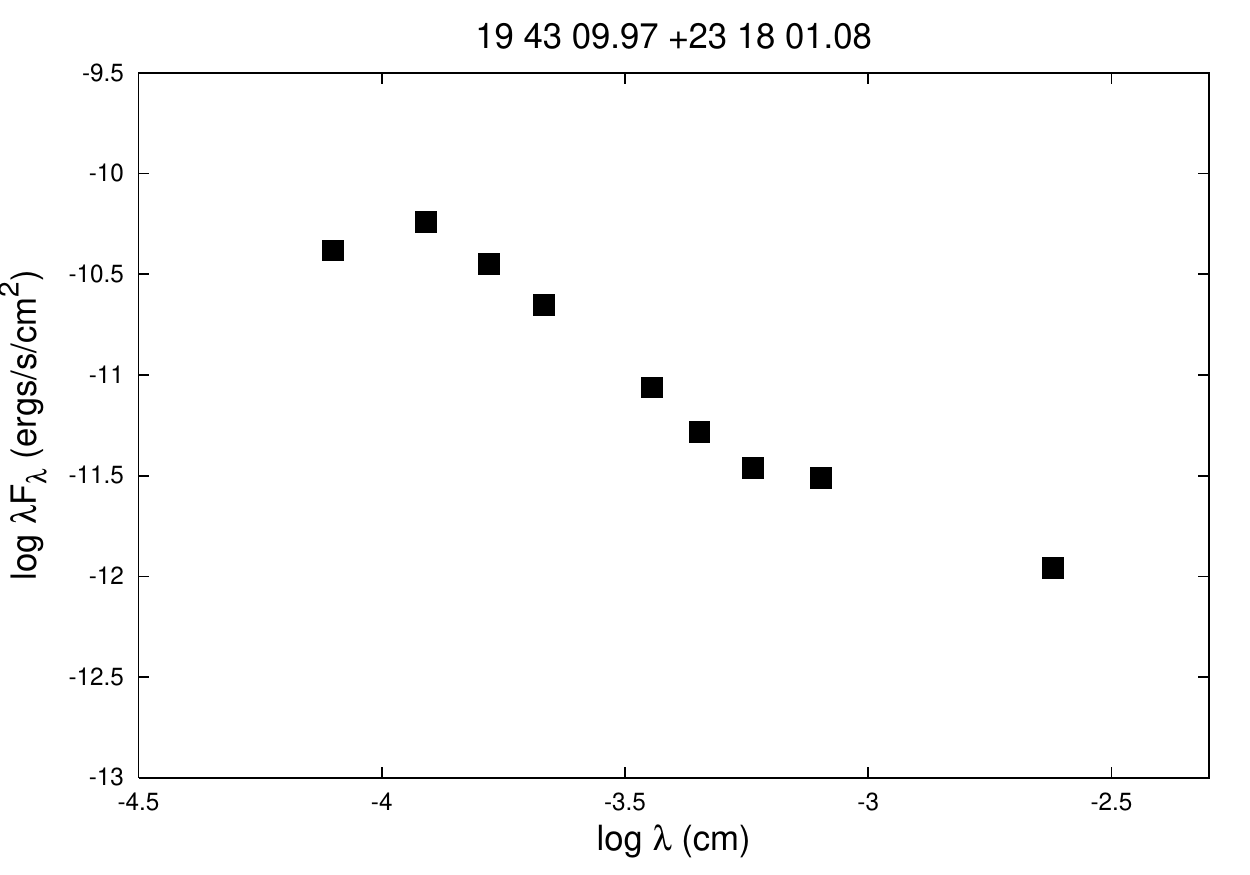} 
 \includegraphics[width=4cm]{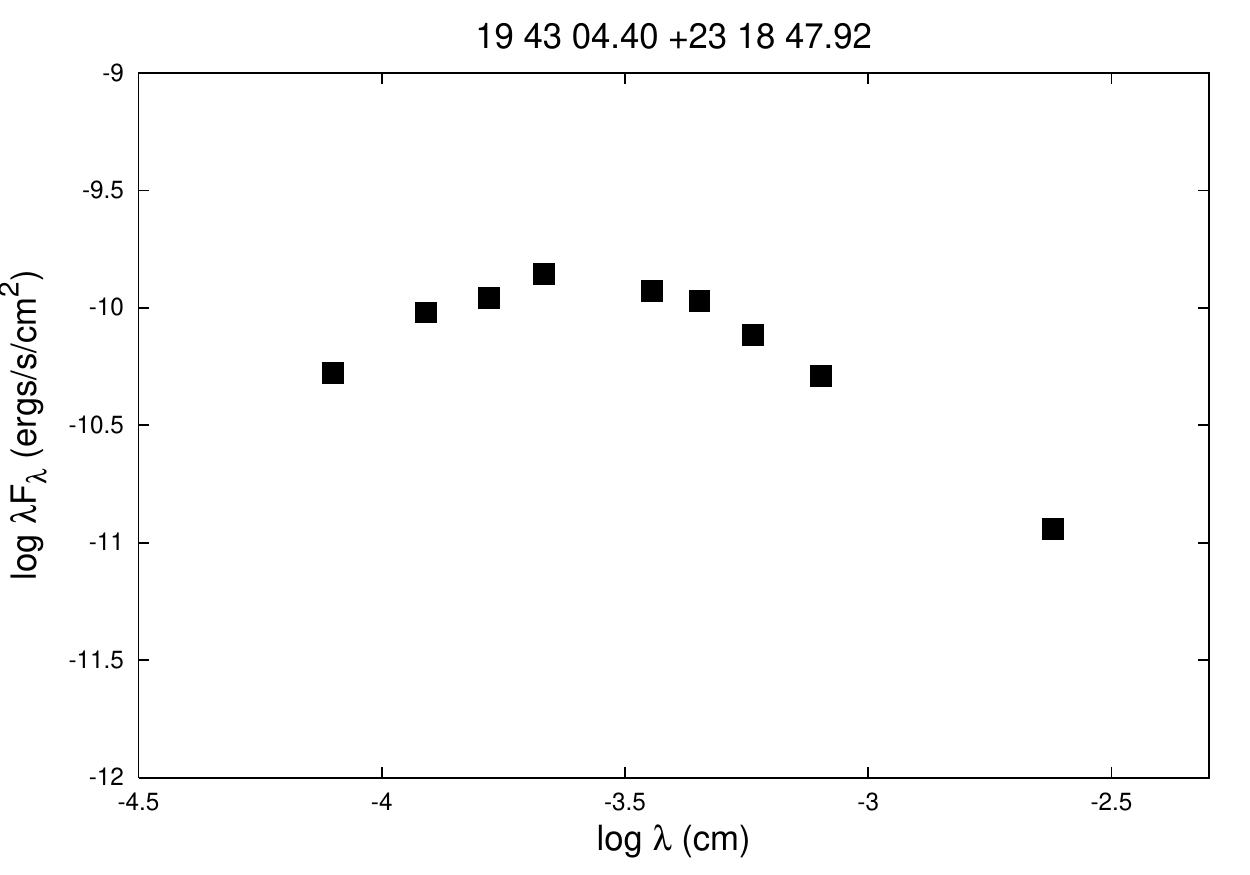}
 \includegraphics[width=4cm]{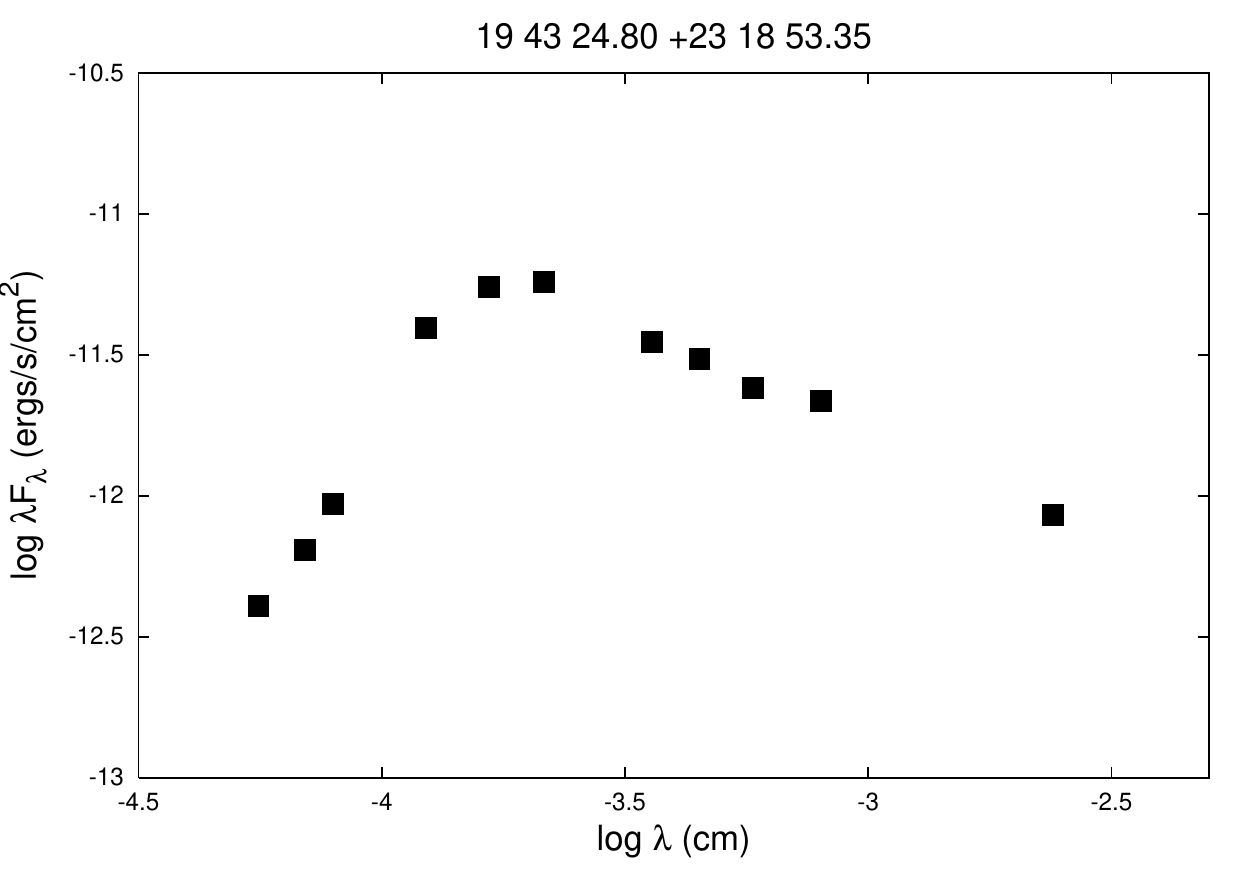} 
 \includegraphics[width=4cm]{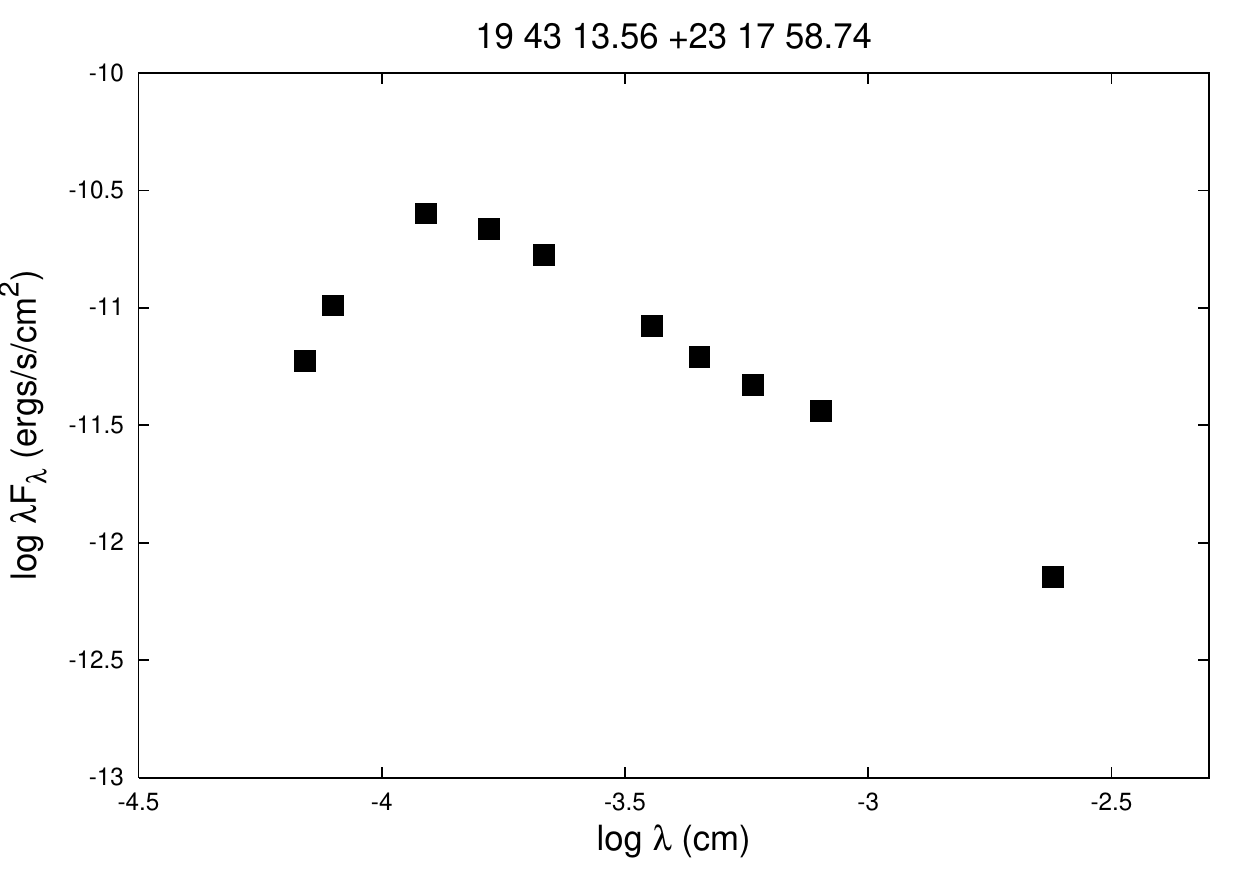} \\ 
 \caption{SEDs for the 16 sources with 24$\mu$m MIPSGAL matches.  }
  \label{sed}
\end{figure*}

 
We have MIPSGAL matches from Billot et al. (2010) at 24$\micron$ for 15 Class II YSOs, and one Class III/field star. Fig~\ref{IRAC}d shows the IRAC-MIPS ccd for these 16 disks. The Class II YSOs show larger excesses in the [8]-[24] color than [3.6]-[5.8], which indicates some flaring in the outer disks for these systems. The [8]-[24] color for the one Class III/field star is consistent with photospheric emission. Fig.~\ref{sed} shows the spectral energy distributions (SEDs) for these 16 sources. The top left SED is for the one Class III/field star (194312.47+231817.03). For most Class II sources, the shape of the SED has either flattened between 8 and 24$\mu$m, or shows some flaring between these two points. The YSO 194308.33+231913.01 shows strong flaring at 24$\micron$. The MIPSGAL survey has a 5-$\sigma$ rms point source sensitivity of 1.3 mJy at 24$\mu$m. Thus only the brightest disks could be detected by this survey. Deeper observations should result in a larger number of detections.



\subsection{H$\alpha$ Emission Strength}
\label{halpha}

\begin{figure*}  
\includegraphics[width=150mm]{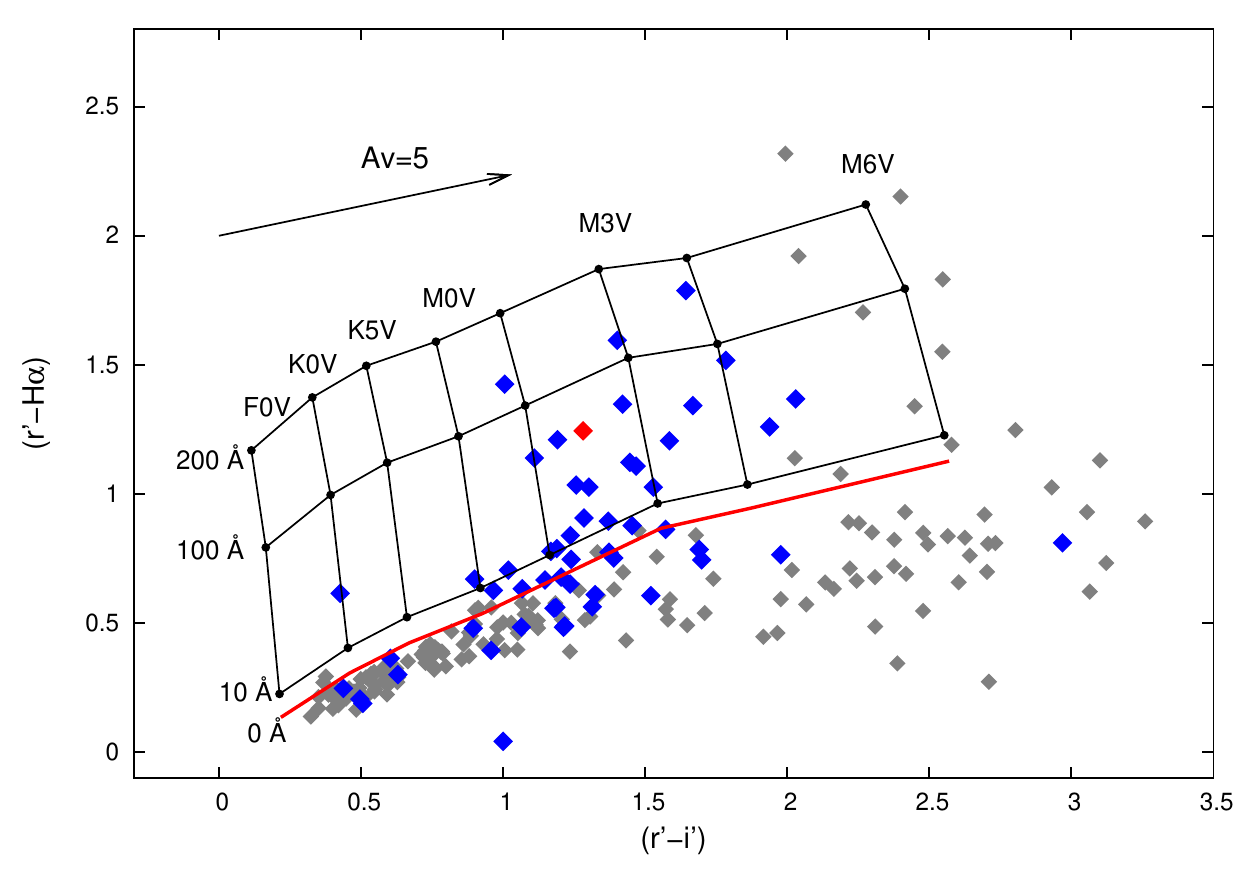}
    \caption{The ($r^{\prime}-i^{\prime}$) versus ($r^{\prime}$-H$\alpha$) ccd for NGC 6823. Also overplotted is the grid of simulated IPHAS colors from Barentsen et al. (2011) for spectral types of F0-M6 and H$\alpha$ EW of 0-200$\AA$. Red points denote Class I systems, blue are Class II and grey are Class III/field stars.  }
      \label{Halpha}
 \end{figure*}

One way to distinguish between Class III YSOs and field stars is the strength in the H$\alpha$ emission. We matched our NGC 6823 catalog with the INT Photometric H$\alpha$ Survey (IPHAS) catalog (Drew et al. 2005). We were able to obtain IPHAS matches for 213 sources. Fig.~\ref{Halpha} plots the IPHAS ($r^{\prime}-i^{\prime}$) versus ($r^{\prime}$-H$\alpha$) colors. Also overplotted is the grid of simulated IPHAS colors from Barentsen et al. (2011) for spectral types of F0-M6. These simulated colors track the increase in the H$\alpha$ emission for a given spectral type. We have not applied any extinction correction to the observed colors or the simulated grid. The  ($r^{\prime}-i^{\prime}$) color thus represents a combination of the interstellar reddening and the intrinsic color of the source. The ($r^{\prime}$-H$\alpha$) color gets redder with increasing emission or larger equivalent width (EW) in the H$\alpha$ spectral line. The reddening vector shown in Fig.~\ref{Halpha} is from the optical-IR extinction law of Howarth (1983) and Schlegel et al. (1998). 

Nearly all Class III/field stars lie below the 0$\AA$ EW track, implying an absence of H$\alpha$ in emission and photospheric ($r^{\prime}$-H$\alpha$) colors for these sources. We find a concentration of these photospheric sources at ($r^{\prime}-i^{\prime}$) of $\sim$0.3-1, and these are likely the FGK type field stars. There is also a dispersed population at ($r^{\prime}-i^{\prime}$) $\geq$2 that lies below the 0$\AA$ track. The red ($r^{\prime}-i^{\prime}$) colors for these sources are likely affected by high extinction, as these lie in cluster regions where $A_{V}$$>$7 mag (Section \S\ref{spatial}). We expect the diskless sources that lie below the 0$\AA$ track to be main-sequence stars.


Young disk sources are expected to be CTTS with ongoing accretion in the disk. A boundary of H$\alpha$ EW=10 $\AA$ is traditionally used to make a distinction between CTTS and the non-accreting WTTS (e.g., Barentsen et al. 2011). Most of the Class II systems in Fig.\ref{Halpha} lie above the 10$\AA$ track, and can be classified as accreting CTTS systems. We also have about 20 Class II sources that lie below this boundary, which indicates the presence of passive or non-accreting disks around these sources. Emission in H$\alpha$ arises from the hot spots on the stellar surface, where the accreting material from the circumstellar disk strikes onto the star at free-fall velocities (e.g., Hartmann et al. 2005). The origin of H$\alpha$ emission in disk bearing sources is from the inner disk regions close to the central source ($<$1 AU for a typical T Tauri disk). The H$\alpha$ emission line is thus a gas diagnostic of the warm inner part of the disk and is not indicative of the total gas content in the disk. The few Class II cases with an absence of H$\alpha$ emission may be more evolved systems with inner disk clearings. If the disk material has dissipated in the inner regions then we would not expect emission in the H$\alpha$ line. Fig.~\ref{Halpha} indicates that selecting YSOs based on the strength in H$\alpha$ emission could result in a rejection of young disk sources with inner disk clearings. 



Among the Class III sources, there are 8 that have H$\alpha$ EW larger than 10$\AA$ and may be classified as CTTS systems.  Emission in H$\alpha$, however, could also arise from an active chromosphere. Chromospheric activity in field stars is found to show a rise towards later types, with a peak at spectral type of M6-M7 (e.g., Riaz et al. 2006a). The 8 Class III sources with strong H$\alpha$ emission have ($r^{\prime}-i^{\prime}$) colors similar to the later spectral types of M4-M6. These could be chromospherically active field stars, rather than young CTTS objects. 

We could select all CTTS sources as candidate cluster members, and reject the sources without H$\alpha$ emission, particularly the Class III/field stars with photospheric ($r^{\prime}$-H$\alpha$) colors as background/foreground field stars. However, some of these photospheric sources may be diskless YSOs. We plan to conduct an X-ray imaging survey of NGC 6823 which will be important in identifying the genuine YSOs among the Class III/field star population. The IPHAS photometry for the 213 sources is listed in Table~\ref{iphas}, with the possible accretors marked by an asterisk. 


\subsection{Spatial Distribution}
\label{spatial}

\begin{figure*}
 \includegraphics[width=\linewidth]{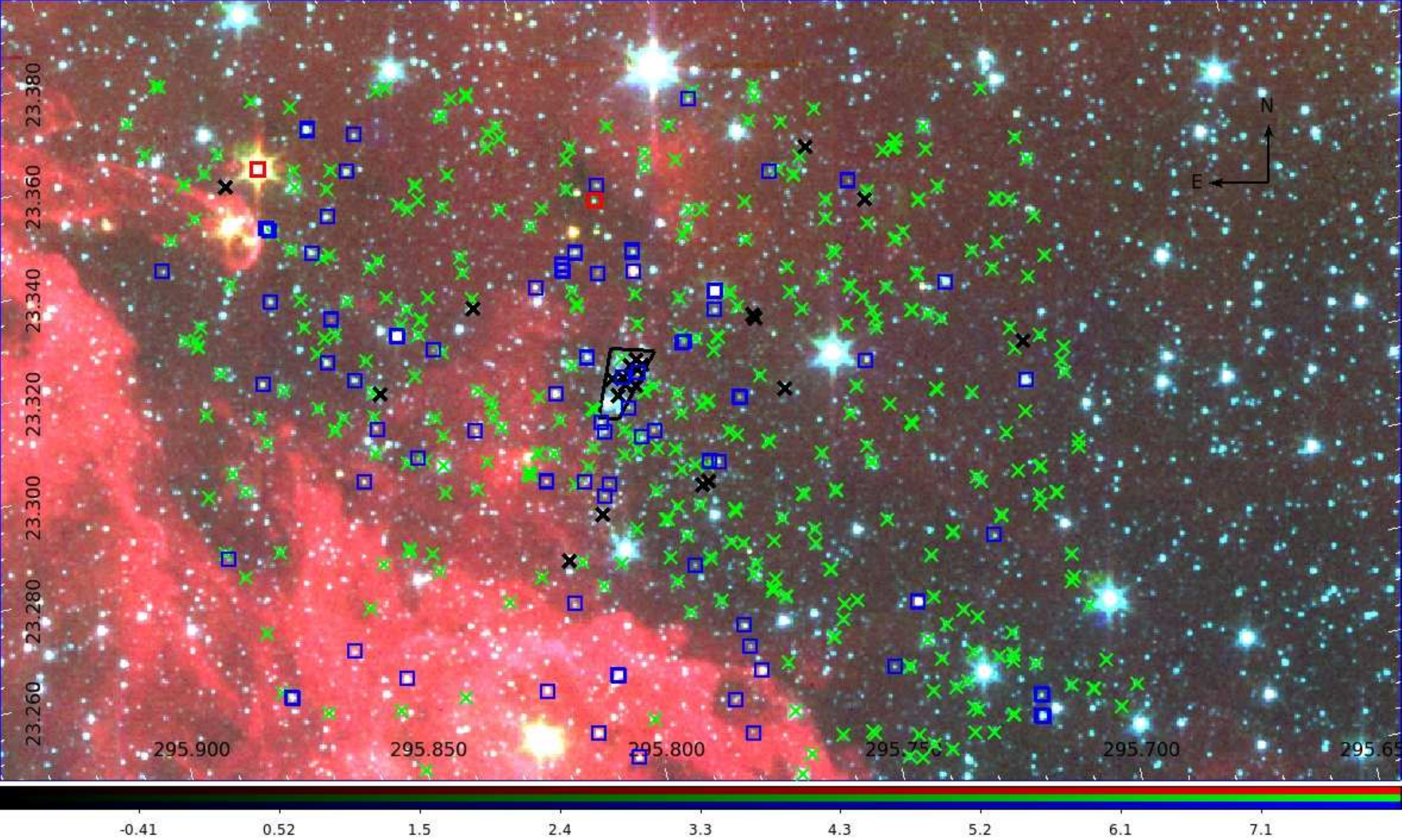}
    \caption{The spatial distribution for Class I (red), Class II (blue), and Class III/field stars (green). Black crosses are the known OB stars in the cluster.  }
      \label{class}
 \end{figure*}

\begin{figure*}
\includegraphics[width=80mm]{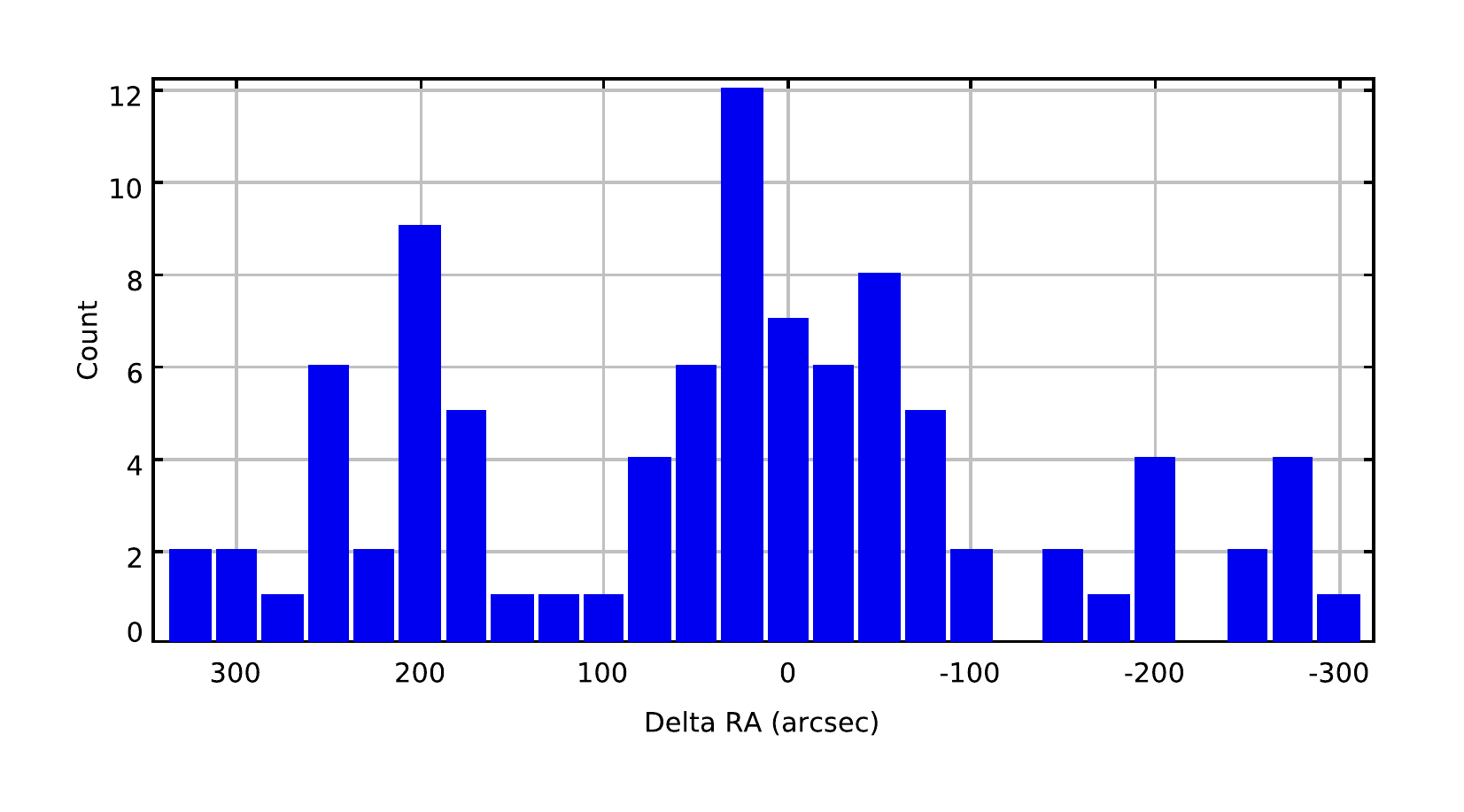} 
\includegraphics[width=80mm]{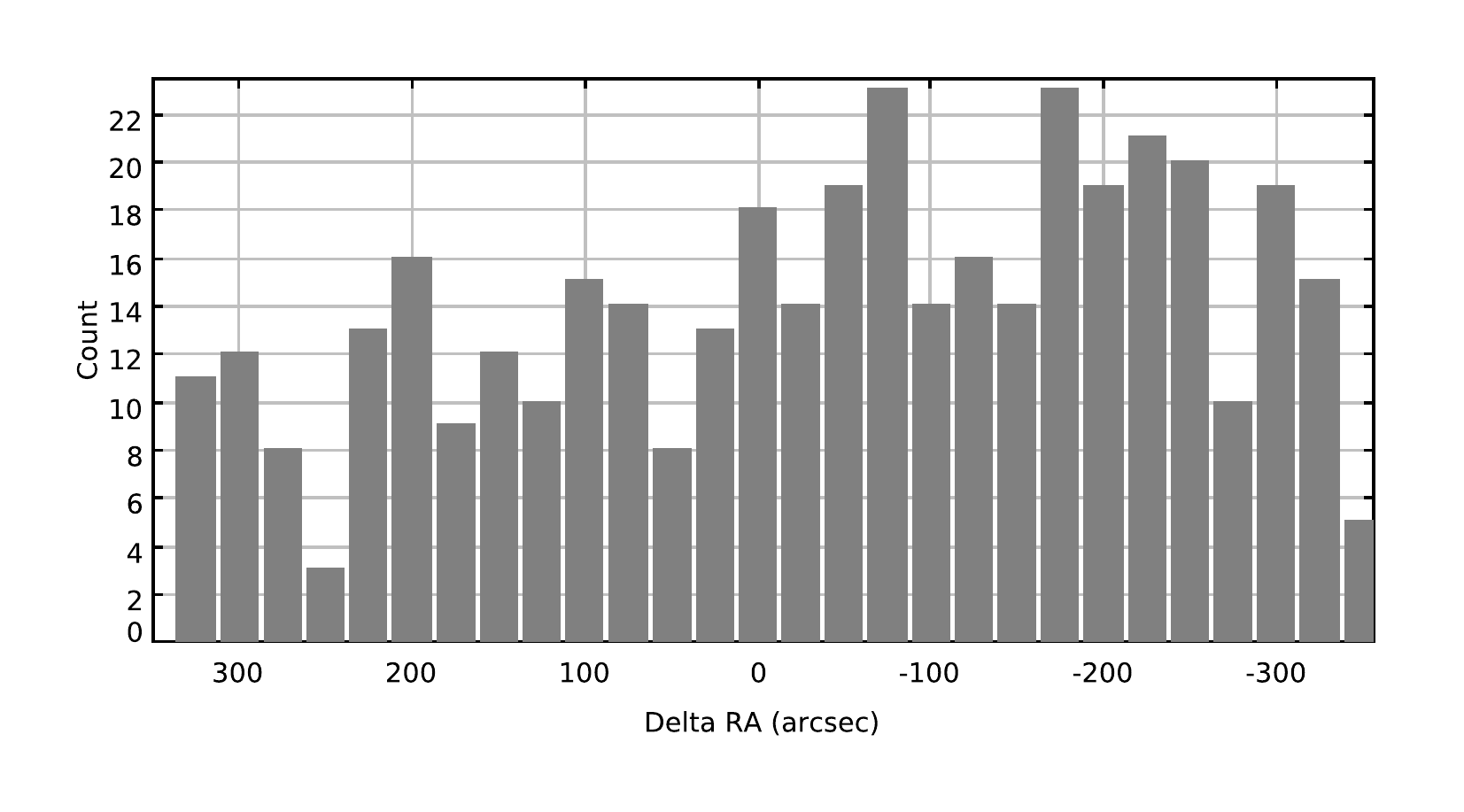} \\       
\includegraphics[width=80mm]{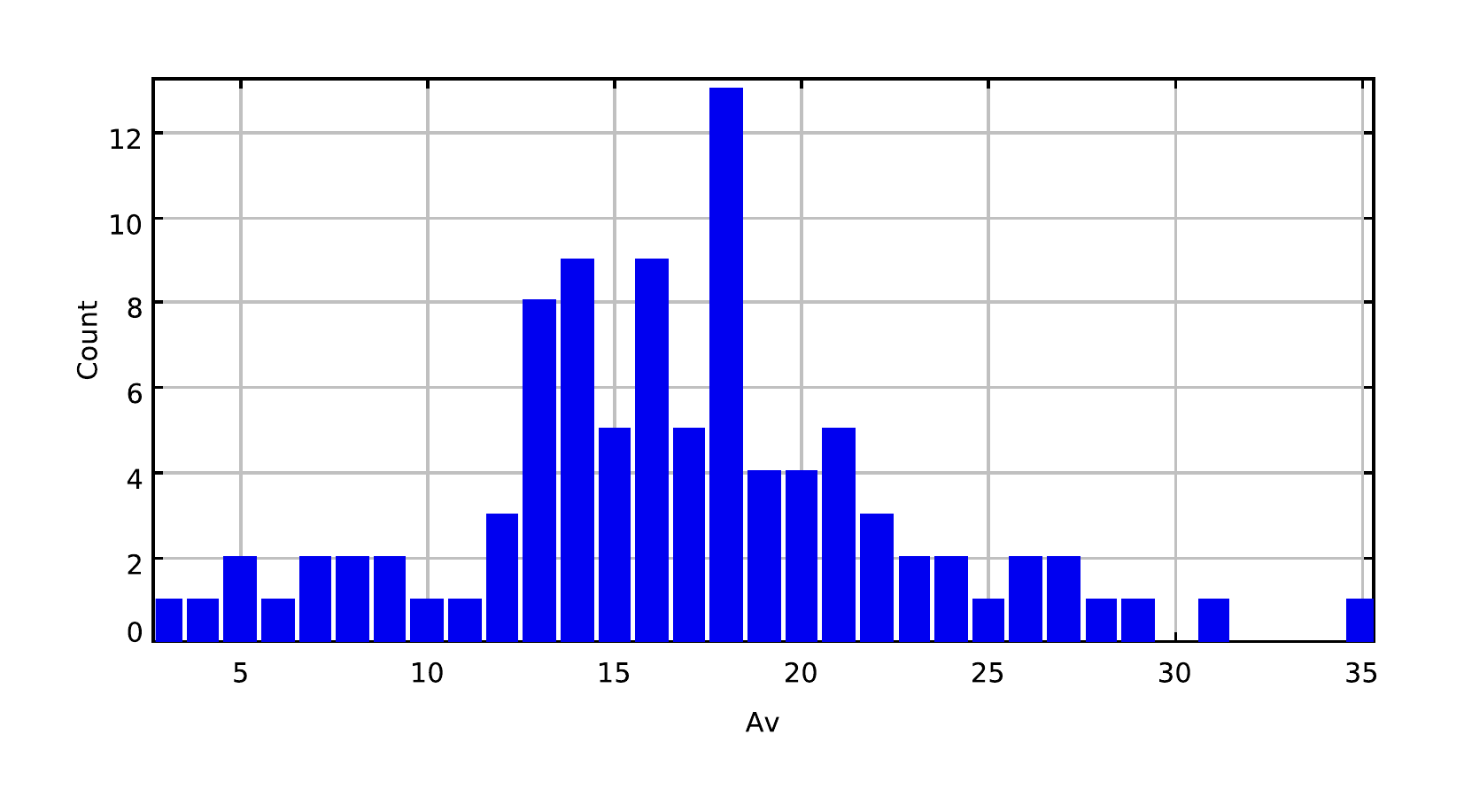}    
\includegraphics[width=80mm]{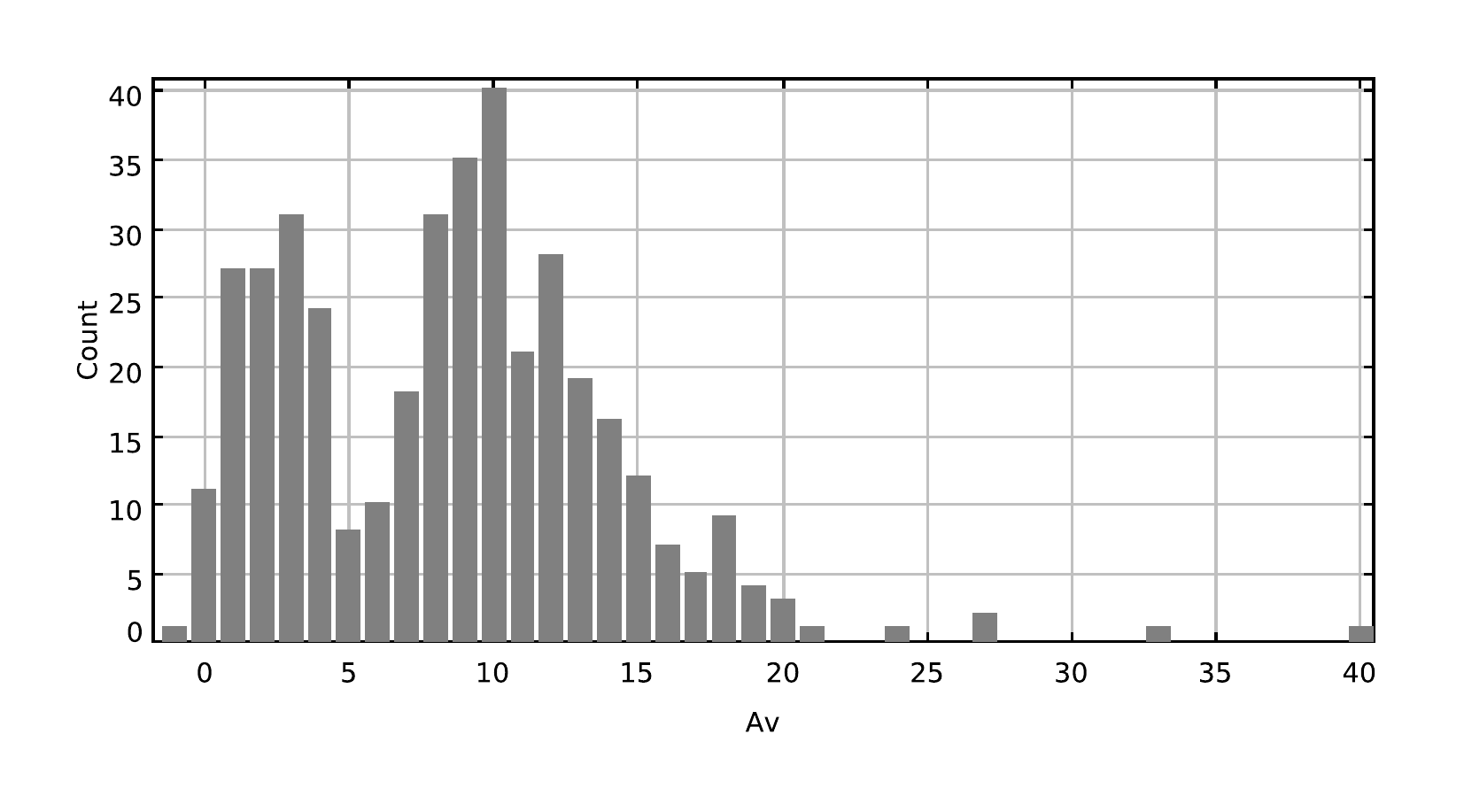}   \\                   
    \caption{{\it Top panel}: The separation in RA from the central Trapezium for the Class I/II ({\it left}) and the Class III/field stars ({\it right}). {\it Bottom panel}: The $A_{V}$ distribution for Class I/II ({\it left}) and Class III/field stars ({\it right}). }
      \label{class2}
 \end{figure*}

\begin{figure*}
\includegraphics[width=\linewidth]{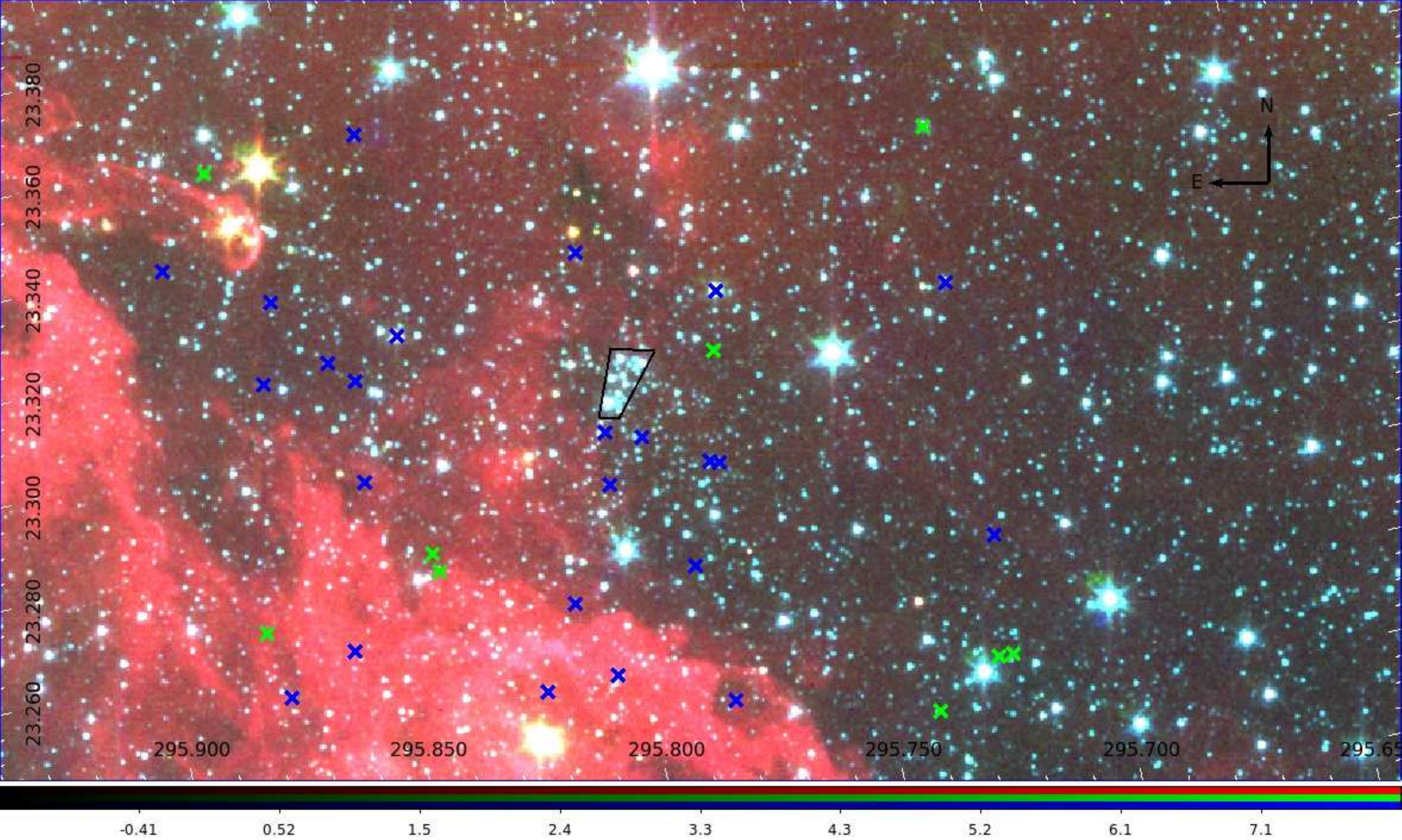}
    \caption{The spatial distribution for accretors (H$\alpha$ EW$>$10$\AA$). Class II systems are denoted by blue, Class III/field stars are denoted by green symbols. }
      \label{class-Ha}
 \end{figure*}

Figure~\ref{class} shows the spatial distribution for the different SED classes in NGC 6823. We find a higher concentration of Class II systems in the eastern region of the cluster and close to the central Trapezium. The two protostellar sources are located in the northern part, with one protostar located close to the {\it VulP12} pillar in the north-east. The western part of the cluster mostly contains Class III/field stars. The differences in the spatial distribution of the Class I/II and diskless population can be noted more clearly in Fig.~\ref{class2}. Most Class II systems lie within 100$\arcsec$ of the central Trapezium ({\it top panel, left}). Another peak is seen at a distance of $\sim$200$\arcsec$ east of the Trapezium stars. These are two groups of Class II sources, one located near the {\it VulP12} pillar observed in the north-east and another more dispersed group located in the south-east. The Class III/field stars are more evenly distributed in the cluster, though a higher number of these sources are found in the western region (Fig.~\ref{class2}; {\it top panel, right}). Fig.~\ref{class2} ({\it bottom panel}) shows the $A_{V}$ distribution for the Class I/II and the Class III/field stars. The disk population is mostly located in the high extinction regions, where $A_{V}$ is $\sim$10-20 magnitude. The diskless sources show a bimodal $A_{V}$ distribution, similar to that observed for the whole cluster (Fig.~\ref{Av}), which indicates their large spread over the whole region. The western region of NGC 6823 thus seems more devoid of disk sources. This western part could be more evolved, so that the disks have all been dissipated, or could have a high contamination of foreground/background field stars. Fig.~\ref{class-Ha} shows the spatial distribution of the possible accretors in NGC 6823 (H$\alpha$ EW $>$ 10$\AA$). Most of the accretors among the Class III sources are located in the southern region, while the Class II CTTS sources are more dispersed, but are located mainly near the Trapezium or the eastern part of the cluster. The main conclusion that can be made from Figs.~\ref{class} and \ref{class-Ha} is that star formation is not concentrated near the center in NGC 6823, and has spread to the outer regions, mainly towards the east of the central Trapezium.

\subsection{Age and Mass Estimates}
\label{age-mass}

Figure~\ref{age} shows the V versus (V-I) color-magnitude diagram (cmd) for NGC 6823. We have considered the pre-main sequence models from Siess et al. (2000; S00) (Fig.~\ref{age}a), Palla \& Stahler (1999; PS99) (Fig.~\ref{age}b) and D'Antona \& Mazitelli (1998; DAM98) (Fig.~\ref{age}c) to estimate the ages and masses for our sample. The evolutionary tracks and isochrones are for solar metallicity, and have been scaled to a distance of 2 kpc. A first analysis of the cmds from all three models indicates the possible presence of two different populations in the cluster. One is a sequence of older stars located along the zero-age main sequence. These sources are also the more massive in the cluster, with $M_{*} >$ 0.4 $M_{\sun}$. A majority of these objects are Class III/field stars, with photospheric ($r^{\prime}$-H$\alpha$) colors, and ($r^{\prime}$-$i^{\prime}$) colors consistent with FGK type stars (Fig.~\ref{Halpha}). The second group consists of younger sources with ages of $<$10 Myr, and at lower masses of $\sim$0.1-0.4 $M_{\sun}$. A large fraction ($\sim$75\%) of these are Class II systems. There is also a smaller group of mainly Class III/field stars that lie at the faint end, with V $\geq$ 20 and (V-I) $\geq$ 4.5 mag. A few of these could be the candidate young brown dwarfs in the cluster. In Fig.~\ref{age}, we have marked with circles the sources that may be possible accretors. Most of the H$\alpha$ emitters are among the younger group of sources, while the H$\alpha$ emitters among the older group are all Class II systems. The zero-age main sequence was assembled from the work of Bessell \& Brett (1988), Leggett (1992), Leggett, Allard \& Hauschildt (1998), and Kirpatrick et al. (2000). 

\begin{figure*}   
\includegraphics[width=80mm]{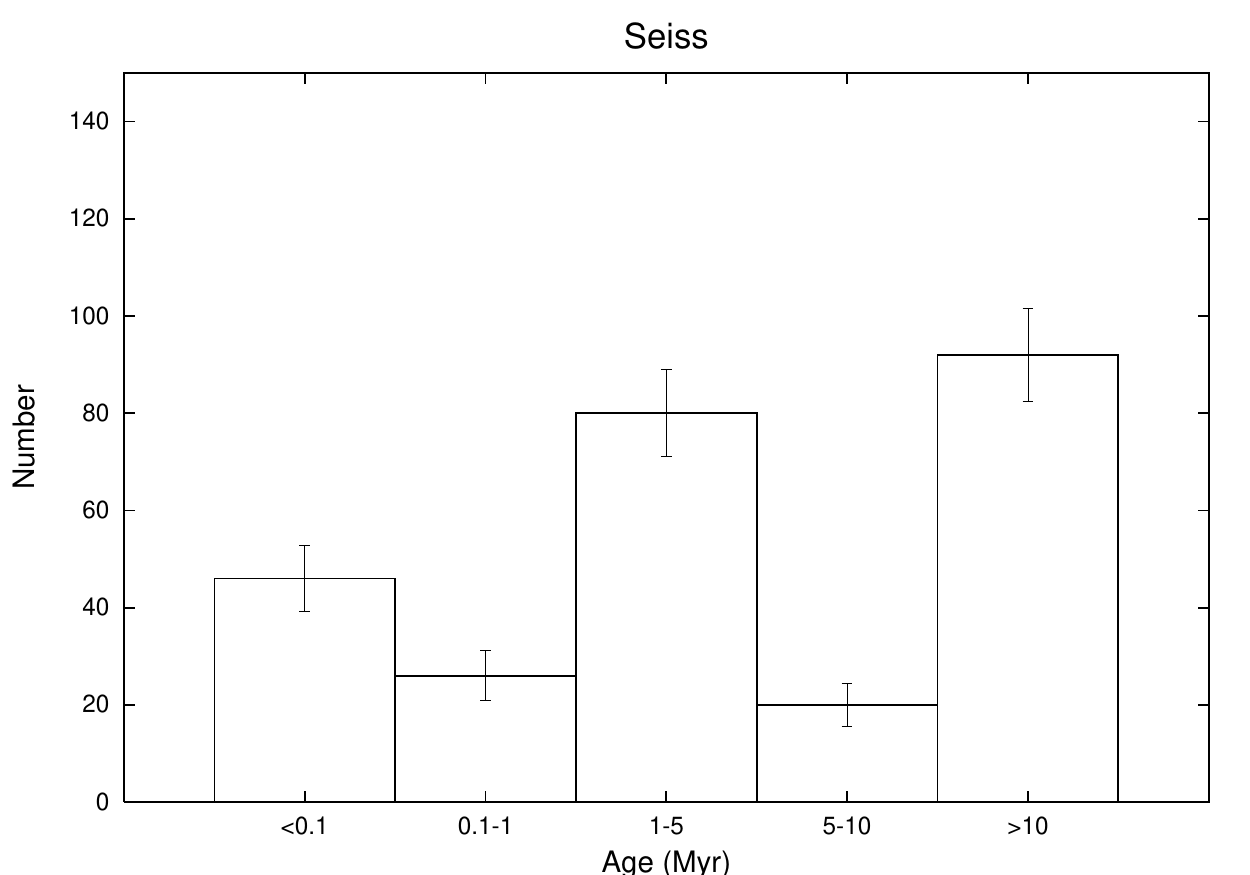} 
\includegraphics[width=80mm]{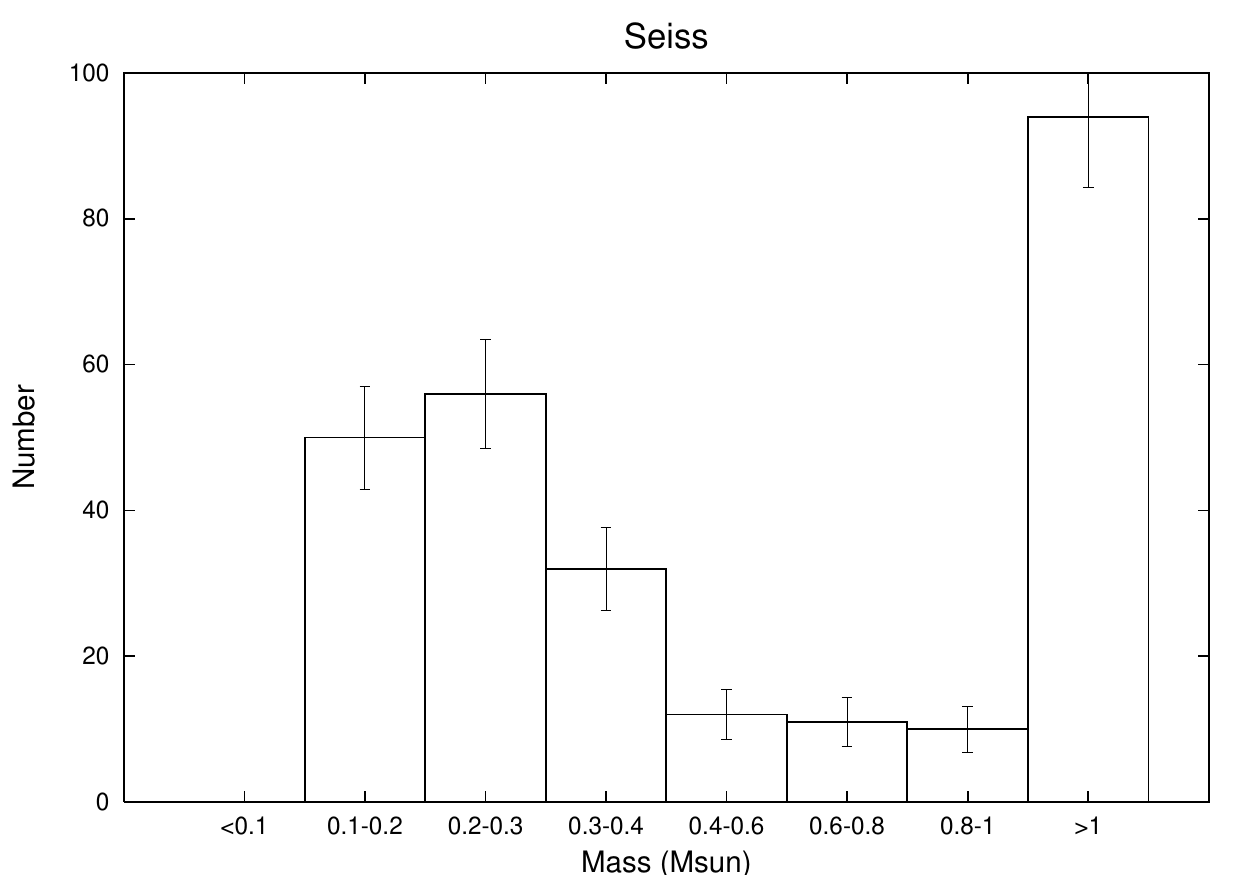} \\
\includegraphics[width=80mm]{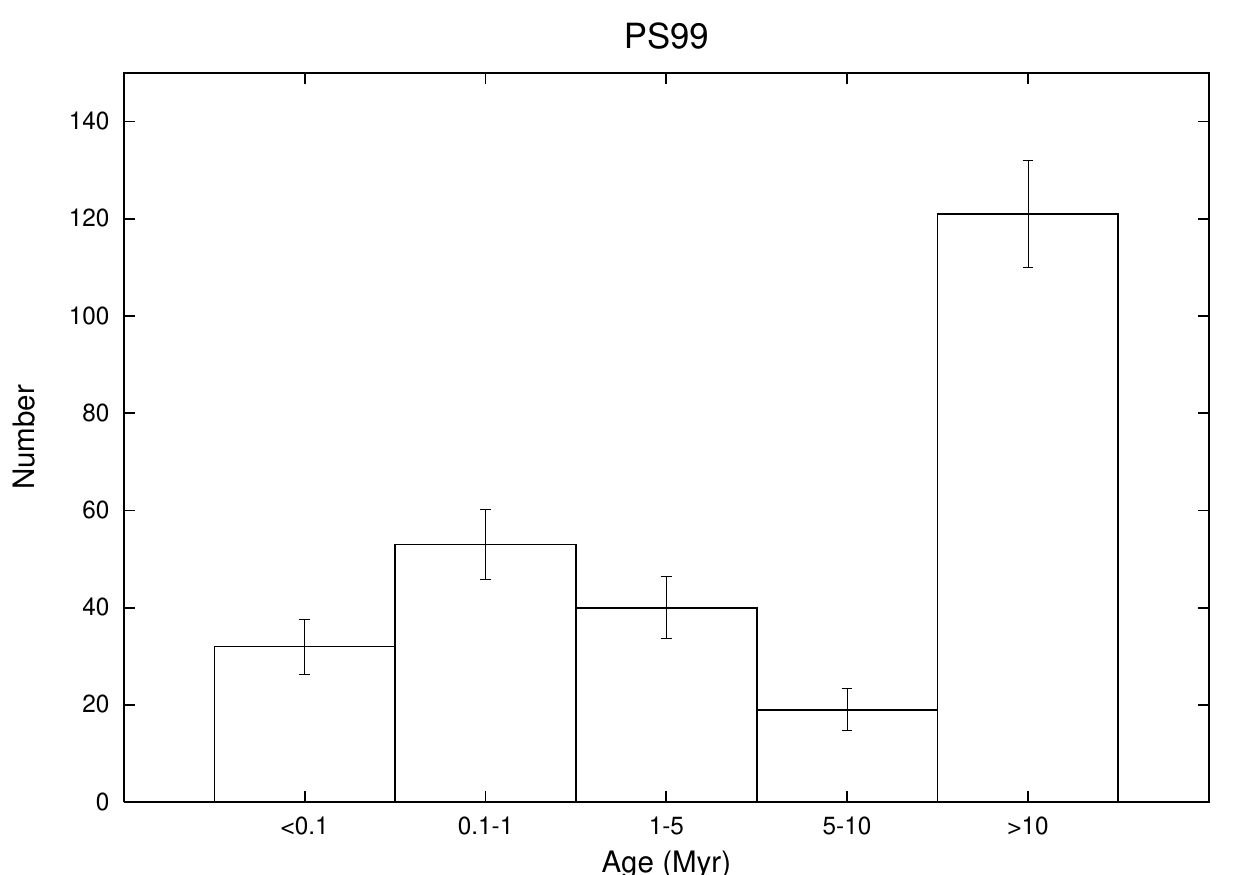}
\includegraphics[width=80mm]{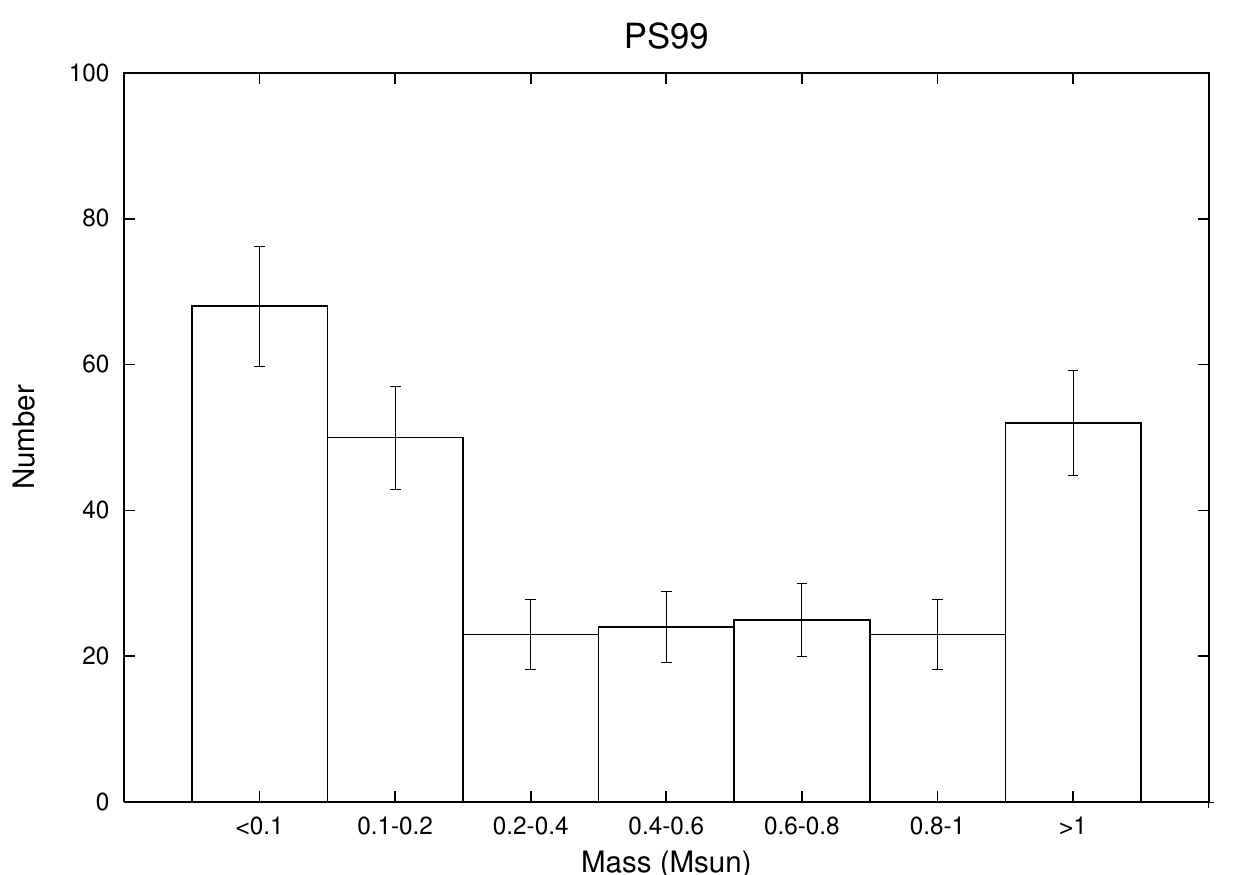} \\
\includegraphics[width=80mm]{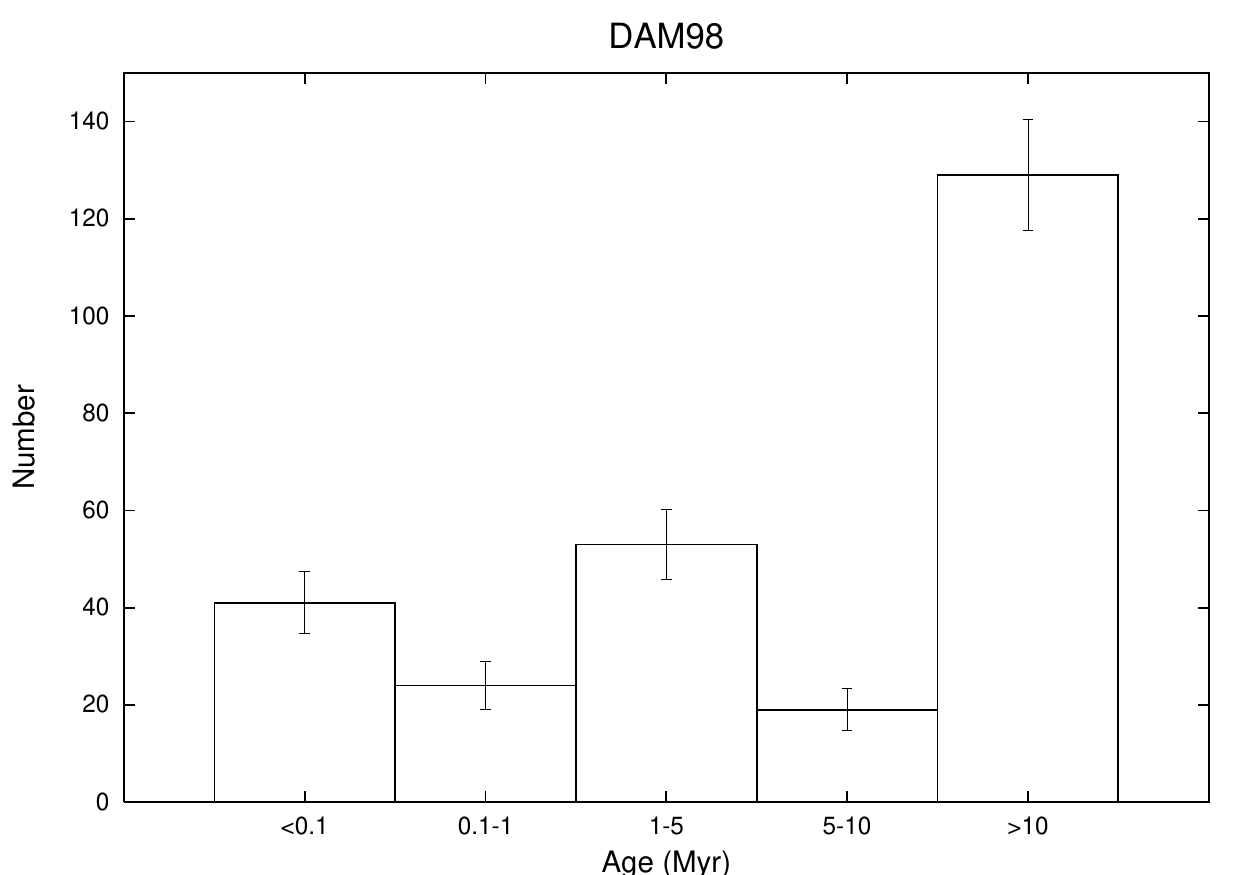}
\includegraphics[width=80mm]{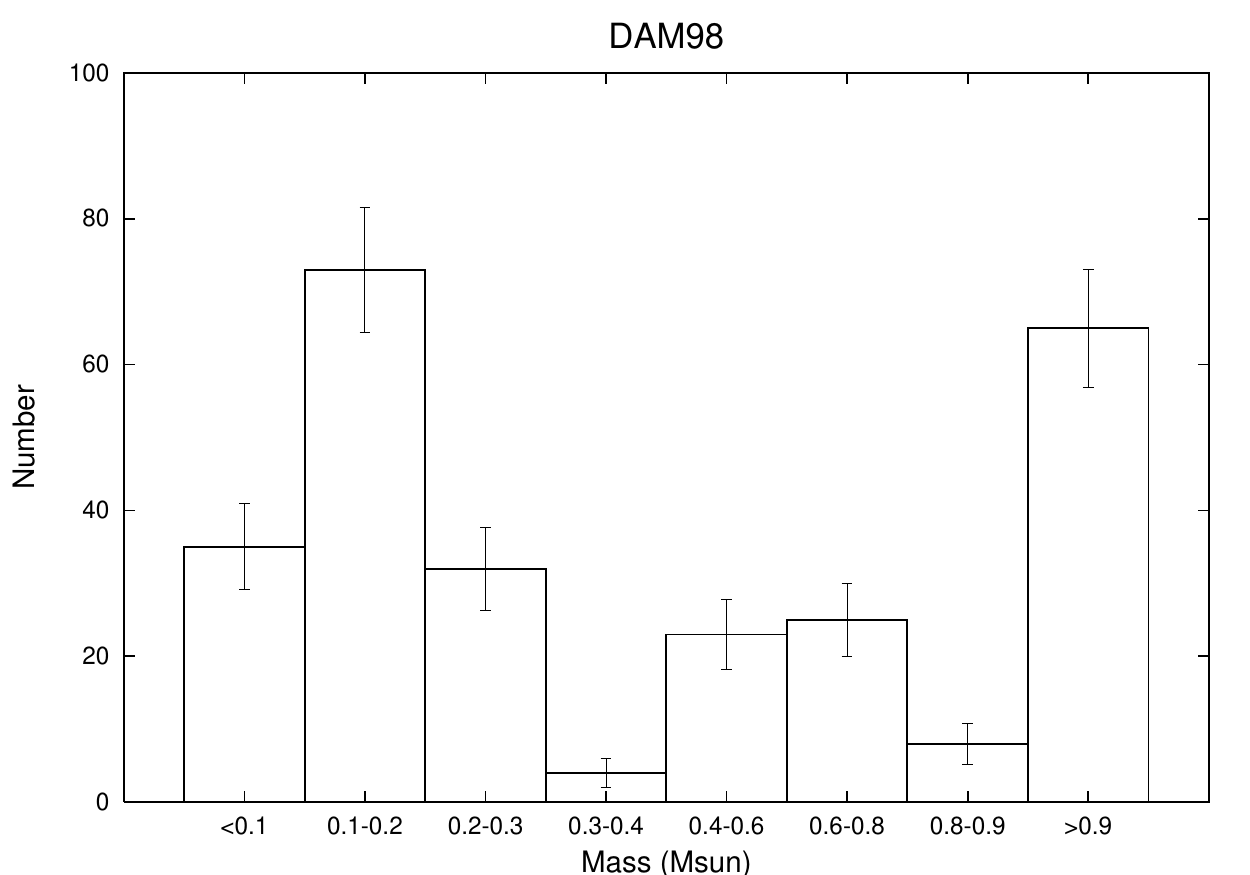} \\
    \caption{The age ({\it left}) and mass ({\it right}) distributions for NGC 6823, obtained using the models by Seiss et al. (2000) ({\it top panel}), Palla \& Stahler (1999) ({\it midle panel}), and D'Antona \& Mazitelli (1998) ({\it bottom panel}). The error bars are the 1-$\sigma$ Poisson uncertainties. }
      \label{distribution}
 \end{figure*}

Figure~\ref{distribution} shows the resulting age and mass distributions using the three different models. We have labelled the oldest age bin as ``$>$10 Myr''. Though we expect most of these sources to be the main sequence population in the cluster, this older bin also includes about 25 Class II systems that lie between the 10 Myr isochrone and the zero-age main sequence. Being disk sources, these are expected to be younger than the main-sequence population. These older disk sources are likely the secondary debris disk systems, with a regeneration of warm dust in the inner disk regions (e.g., Riaz et al. 2006b). The age distributions using the PS99 and DAM98 models indicate a strong peak for the older sources, along with a tail of objects at younger ages of $<$10 Myr. The shape of the age distribution using the S00 model is different, with two comparable peaks at 1-5 Myr and $>$10 Myr. There is a wider range in the V magnitudes between the 1 and 5 Myr isochrones for the S00 model compared to the other two cases (Fig.~\ref{age}a), which would explain the two comparable peaks observed for the S00 age distribution. From all three models, most of the young sources seem to be at ages of $\sim$0.1-5 Myr, and the least number of objects is found in the 5-10 Myr age bin. For the PS99 model, the 0.1 Myr isochrone is $\sim$2 magnitudes brighter than the 1 Myr isochrone at a given (V-I) color, while the difference is less than one magnitude for the S00 or the DAM98 models. This is particularly for objects redder than (V-I)$\sim$3. This results in a greater number of objects in the 0.1-1Myr bin for the PS99 model compared to the other two models. We note that for all three models, the $<$0.1 Myr age bin may not be reliable. This bin mainly contains the group of faint sources at V $\sim$ 22 and (V-I) $\geq$ 4.5 mag, most of which are Class III/field stars. It is unusual for a source which is presumably at an age of $<$0.1 Myr to be devoid of circumstellar material. These red objects could be extincted field stars rather than very young low-mass stars. Also, for the DAM98 model, the isochrones are not available for (V-I) bluer than $\sim$1.5. We have roughly extrapolated the 1, 5 and 10 Myr isochrones from this model to estimate the ages for these bluer sources, which are likely to be older, massive objects. Thus the 1-5 Myr and $>$10 Myr age bins for the DAM98 model could have fewer number of sources than plotted in Fig.~\ref{distribution}. 

\begin{figure*}   
\includegraphics[width=130mm]{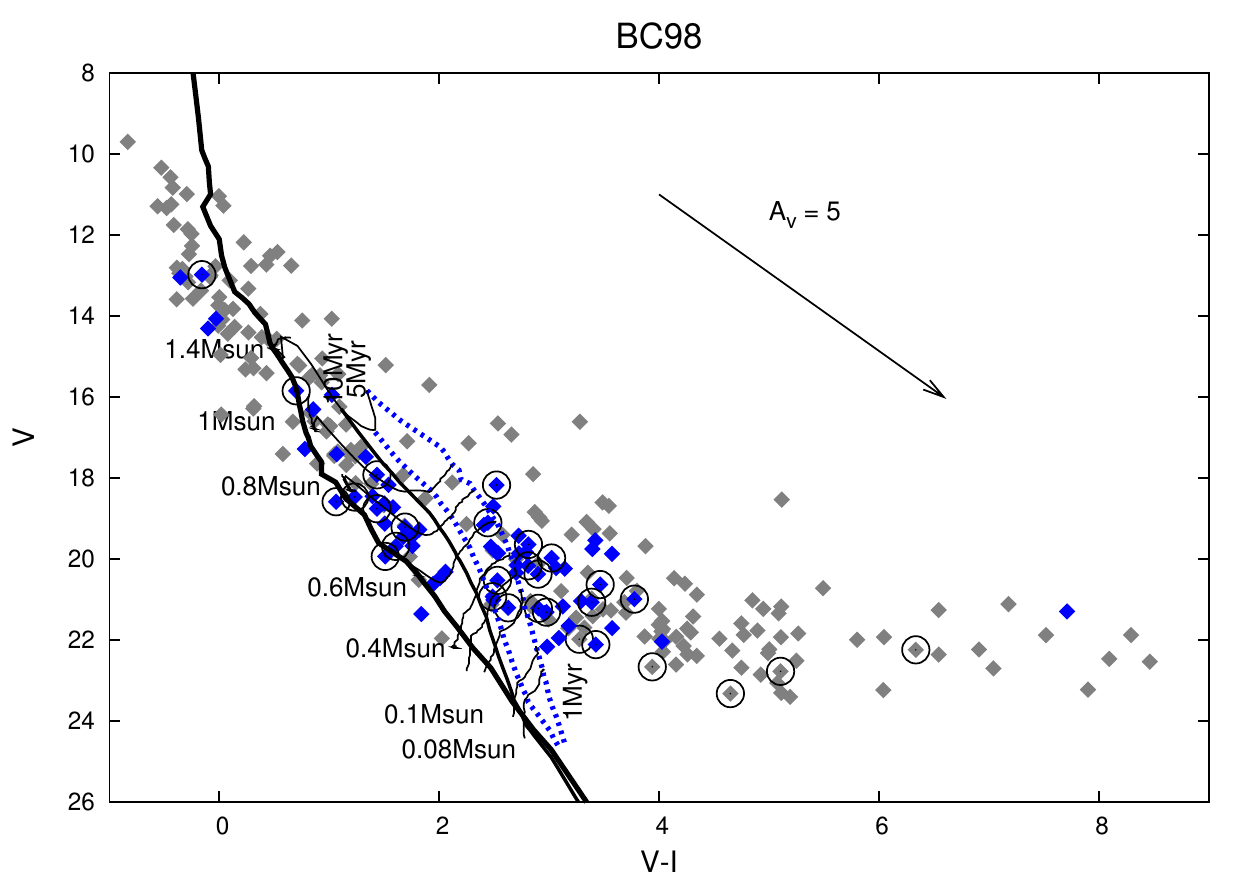} \\  \vspace{0.5in}
\includegraphics[width=90mm]{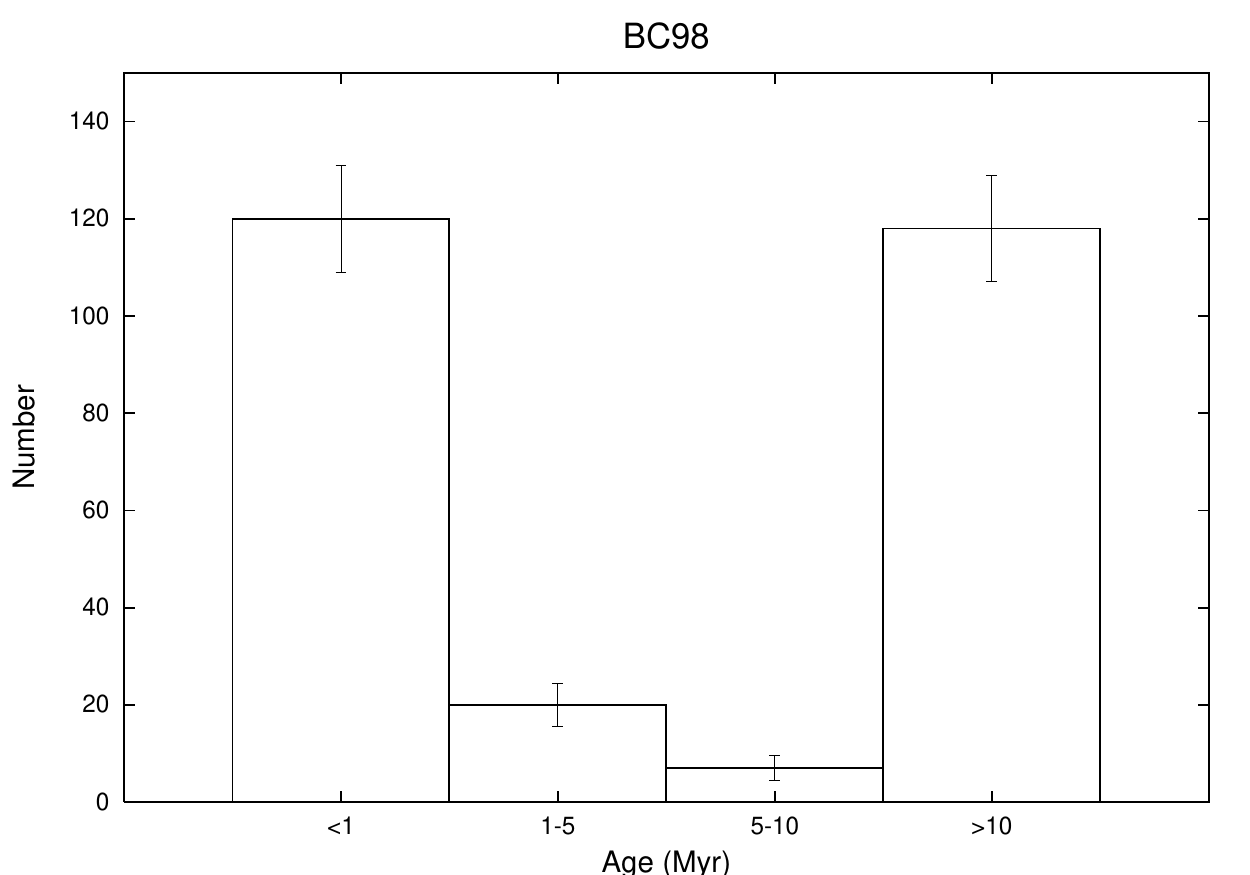} \\
    \caption{The V vs (V-I) cmd using Baraffe et al. (1998) models. Bottom panel shows the resulting age distribution. }
      \label{baraffe}
 \end{figure*}

Other than these three models, we have also considered the evolutionary models by Baraffe et al. (1998; BC98). The cmd and the resulting age distribution from this model are shown in Fig.~\ref{baraffe}. The BC98 model does not provide a 0.1 Myr isochrone. As an be seen, nearly all of the young sources lie above the 1 Myr isochrone. The resulting age distribution shows two strong peaks, one for $<$ 1 Myr age bin, and the other for the older likely main-sequence population. It is very difficult to estimate masses using this model, particularly for the younger sources extending from (V-I) $\geq$ 3. We have therefore not attempted to build the mass distribution using the BC98 model.



Despite the uncertainties and the differences among the models, we find a pre-main sequence population in NGC 6823, in addition to an upper main-sequence population. The age distributions for NGC 6823 indicate recent star formation activity in the last $\sim$1-5 Myr, and there may be candidate cluster members at ages as young as 0.1 Myr. The star formation history of NGC 6823 seems similar to that observed in young star-forming regions such as the Orion Nebula Cluster, Taurus-Auriga, Lupus and $\rho$ Ophiuchi, which also show an older population of $\sim$10$^{7}$ yr, along with more recent star formation activity in the last 1-2 Myr (e.g., Palla \& Gaali 1997; Mart\'{i}n et al. 1998; Muench et al. 2002). The similarities with NGC 6823 suggest that star formation may proceed in more than one epoch within a wide range of environments. 



The stellar mass tracks from these different evolutionary models show a more prominent difference than the isochrones (Fig.~\ref{age}). Based on the S00 model, all of the sources in the cluster have masses higher than 0.1 $M_{\sun}$ (Fig.~\ref{age}a), while using the DAM98 model results in a few sources having lower masses of $\sim$0.08$M_{\sun}$ (Fig.~\ref{age}c). In comparison, the lowest mass track from the PS99 model is for 0.1$M_{\sun}$, and it lies at a bluer color than the 0.1$M_{\sun}$ track from DAM98 or S00 model (Fig.~\ref{age}b). Thus the PS99 model results in a larger number of sources with masses less than 0.1$M_{\sun}$. Also notable are the nearly constant (V-I) colors at a given mass for the PS99 case, while the mass tracks for $M_{*}$$\geq$0.2 $M_{\sun}$ for the DAM98 model get redder with increasing age (Fig.~\ref{age}b,c). The resulting mass distributions from these models thus have very different shapes (Fig.~\ref{distribution}, right panel). The mass distribution from the S00 model shows a prominent peak for massive sources at $M_{*}$$>$1$M_{\sun}$, with a nearly flat distribution for lower masses and then a rise again for the 0.1-0.3 $M_{\sun}$ mass bins. The PS99 model (middle, right panel) also shows peaks at the two ends of the mass distribution, with a flat distribution for masses of 0.2-1 $M_{\sun}$. The distribution based on the DAM98 model also shows a peak for the very low-mass and the very high-mass sources, but there is almost a break in the distribution at 0.3-0.4$M_{\sun}$. The range in the (V-I) colors between the 0.3 and 0.4 $M_{\sun}$ tracks from this model is very narrow (Fig.~\ref{age}c), which results in a negligible number of sources in this mass bin. This is also the case for the 0.8-0.9 $M_{\sun}$ bin. On the other hand, the range in the (V-I) colors between 0.1 and 0.2 $M_{\sun}$ is much larger for DAM98 than the other two models (Fig.~\ref{age}c), which results in a higher number of sources in this mass bin compared to the other two models. 

We note that there are several inconsistencies in the mass tracks available. The boundaries between the evolutionary tracks for masses higher than $\sim$0.4$M_{\sun}$ are less clear for the older sources at ages $>$10 Myr (Fig.~\ref{age}). We have roughly extrapolated the mass tracks to estimate masses for these older sources. We would thus consider the mass bins of $>$0.4 $M_{\sun}$ to be unreliable. For the DAM98 model, the highest mass track is for 0.9$M_{\sun}$, and so the $>$0.9$M_{\sun}$ bin for this model includes all of the sources that lie above this track (Fig.~\ref{age}c). On the other hand, the $>$1 $M_{\sun}$ mass bin for the PS99 and S00 models includes sources with masses of 1-3 $M_{\sun}$ and 1-5 $M_{\sun}$, respectively. This makes it difficult to directly compare the high-mass peaks observed for the three models. It is also difficult to estimate the masses for the tail of faint sources observed at V$\sim$22 and (V-I) $\geq$ 4.5. We have included all of these faint objects in the $<$0.1 $M_{\sun}$ bin, which makes this bin unreliable as it may contain fewer sources than plotted. Another inconsistency is that the mass tracks from every model are not available at intervals of 0.1$M_{\sun}$, due to which the bin sizes in Fig.~\ref{distribution} are not the same.  

Overall, the mass distributions indicate a large spread in the stellar masses, with two main groups at $\sim$0.1-0.4 $M_{\sun}$, and at $>$1 $M_{\sun}$. There may be a possible mass-age relationship in the cluster, or a mass dependent age spread, with the older stars being more massive than the younger ones. None of these pre-main sequence models, however, can explain the small group of faint sources located at (V-I)$\geq$4.5. The position of a young star in a cmd is dependent on its mass, age, extinction, binarity, its intrinsic variability, and the presence of a circumstellar disk. These properties need to be taken into account in order to construct the IMF and study the star formation history of the cluster. Follow-up spectroscopic observations will allow us to confirm cluster membership for these potential low-mass YSOs, and to build up a more reliable age and mass distribution.

\begin{figure*}   
\includegraphics[width=140mm]{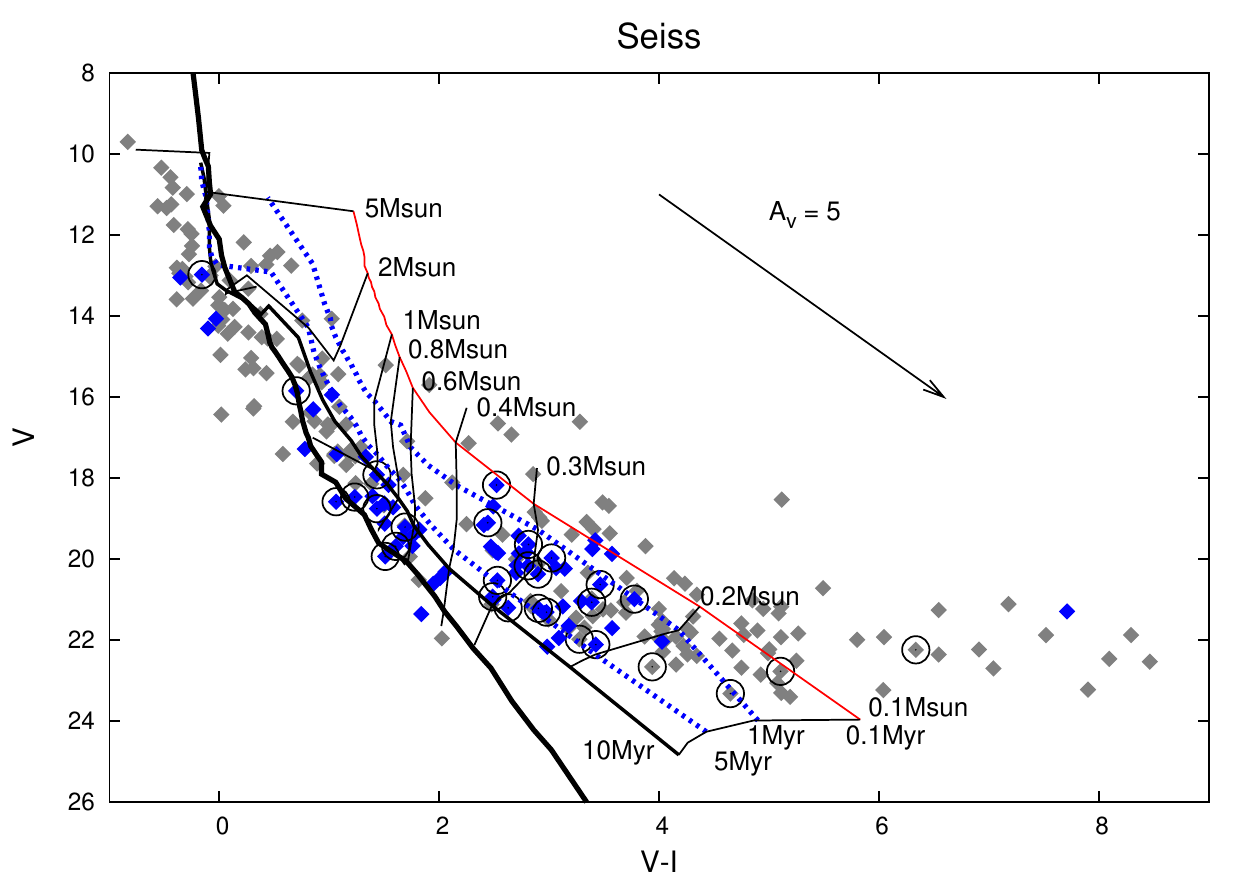}
    \caption{(a): The V vs (V-I) cmds for NGC 6823. The evolutionary tracks and isochrones are from Seiss et al. (2000) models. The reddening vector is plotted Reike \& Lebofsky (1985) reddening law. Grey symbols indicate Class III/field stars, blue symbols denote Class II systems. The objects marked with black circles show emission in H$\alpha$ (EW $>$10 $\AA$). Black thick line is the zero age main-sequence. }
      \label{age}
 \end{figure*}
 
 \begin{figure*}   
\setcounter{figure}{9}
\includegraphics[width=140mm]{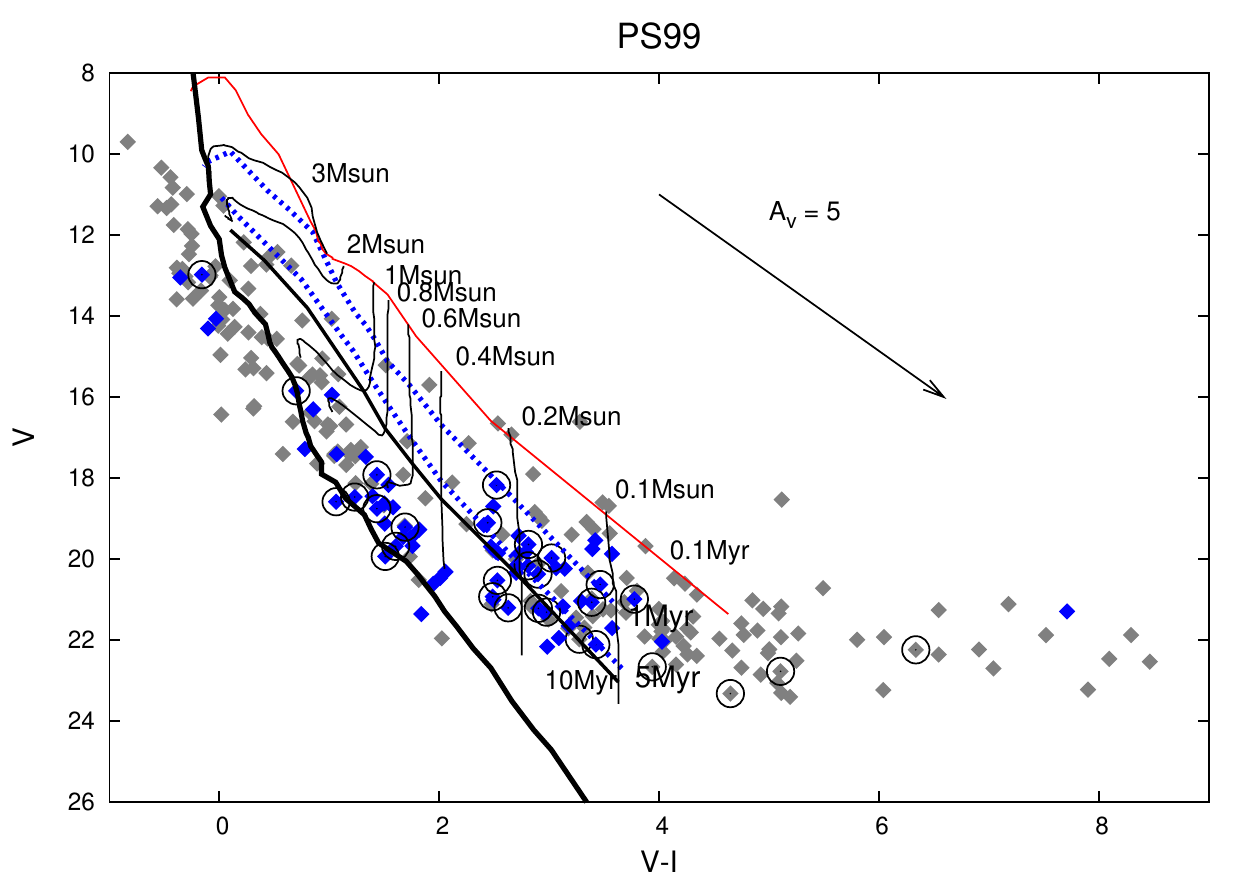}
    \caption{(b): The V vs (V-I) cmds for NGC 6823. The evolutionary models are from Palla \& Stahler (1999). Symbols are the same as in Fig.~\ref{age}a. }
      \label{age}
 \end{figure*}
 
 \begin{figure*}   
\setcounter{figure}{9}
\includegraphics[width=140mm]{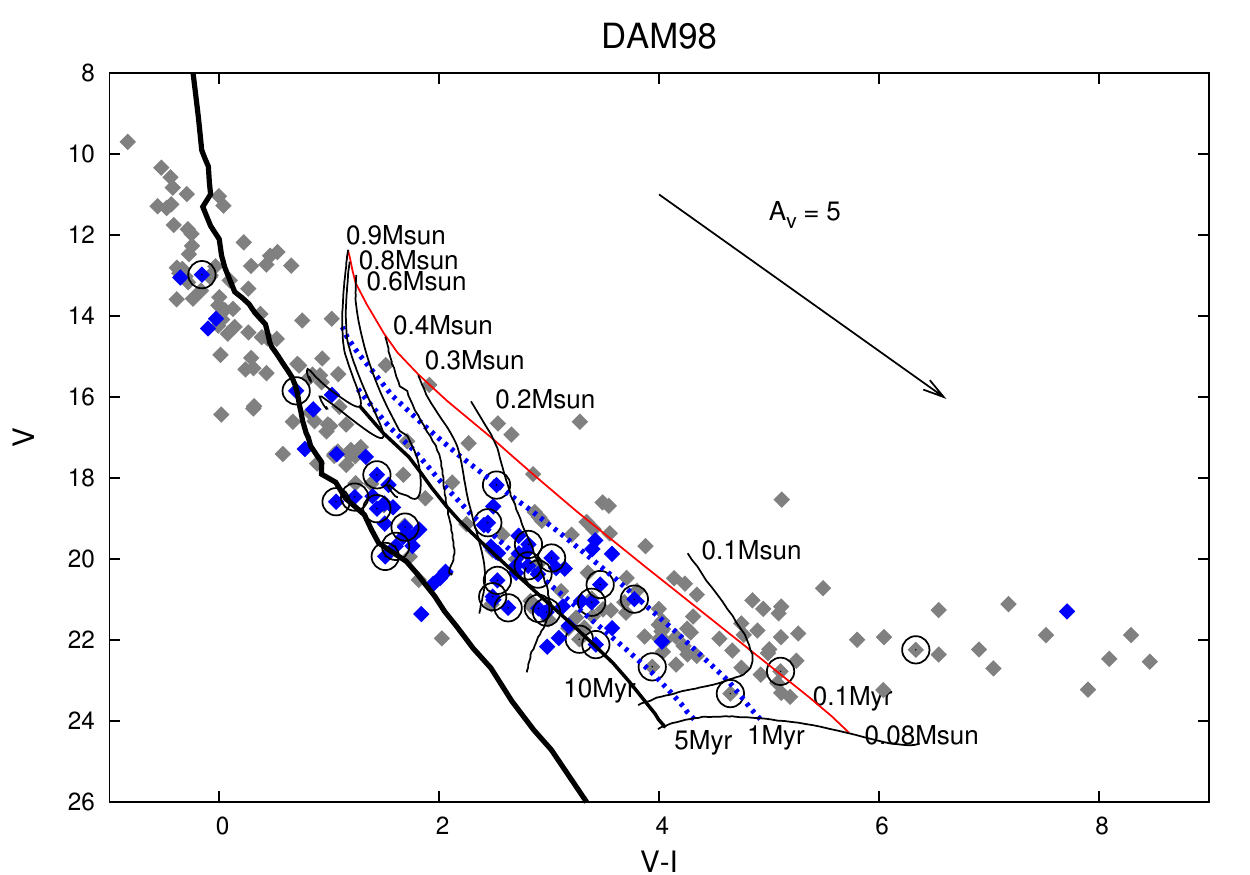}
    \caption{(c): The V vs (V-I) cmds for NGC 6823. The evolutionary models are from D'Antona \& Mazitelli (1998). Symbols are the same as in Fig.~\ref{age}a. }
      \label{age}
 \end{figure*}
 
\section{Discussion: A Possible Case of Triggered Star Formation}
\label{trigger}

Taylor et al. (1992) detected a supernova remnant, SNR G59.5+0.1, in the direction of the Vulpecula OB1 association. This SNR is located at a distance of $\sim$11 pc from the central Trapezium in NGC 6823 (Fig.~\ref{snr}). Taylor et al. described it as a shell-type SNR with a diameter of 15$\arcmin$. Billot et al. (2010) noted a high density of YSOs lined up along two arcs at the northern and southern rims of SNR G59.5+0.1, and suggested that the interaction of the expanding shell of SNR G59.5+0.1 and the neutral gas of Sh2-86 could have triggered the formation of young stars in the cluster. As explained in Preibisch \& Zinnecker (2006), close to a supernova, the shock wave will destroy the surrounding cloud, but at large distances of $\sim$20-100 pc, the shock velocities are below $\sim$50 km s$^{-1}$ and the shock properties are suitable for triggering cloud collapse. The proximity of SNR G59.5+0.1 to the central Trapezium suggests that the shock waves from this explosion may have triggered the star formation activity that we witness in the present epoch. An important aspect of supernova-triggered star formation is that the age spread in the new subgroup formed should be small, since the shock waves disperse through the cloud on a short timescale. The estimated spread in ages for the new generation of stars in NGC 6823 is $\sim$1-5 Myr, which is not significantly large. However, Billot et al. (2010) estimate an age of $\sim$10$^{4}$ yr for SNR G59.5+0.1. This event seems too recent to have triggered any of the star formation that we see now. 

NGC 6823 is also known to host some 20 spectroscopically confirmed OB stars (Fig.~\ref{class}), which have cluster membership probabilities of $>$95\% (e.g., Shi \& Hu 1999). The ionizing radiation and winds from massive stars can disperse the clouds in the immediate surroundings, and trigger the collapse of a molecular cloud which may otherwise not contract and fragment spontaneously. The classical model for triggered star formation from OB stars is the ``collect-and-collapse'' model (Elmgreen \& Lada 1977). In this model, as soon as the first massive star is born deep inside a cloud, the ionization fronts from the H II regions created by the massive star push the cloud from within, forming a cavity. The gas and dust in the cloud accumulate to a layer until a critical density is reached for gravitational collapse, triggering the formation of the next generation of stars. According to this model, low-mass stars should be systematically older than the massive OB stars, and there should be a large age spread, as large as the lifetime of the cloud (e.g., Preibisch \& Zinnecker 2006). Another observational evidence of triggering by this mechanism is the presence of large-scale cavities in the association, as observed in the W5 region (W5 West; Koenig et al. 2008). Such large-scale cavity-like structures have not been observed in the Vulpecula OB1 association (Billot et al. 2010) or in the NGC 6823 region surveyed. We do find evidence of the lower mass population to be younger than the massive stars in the cluster, but the age spread is not as large as the lifetime of the cloud. The collect-and-collapse model thus does not seem to be the applicable mechanism in this cluster.



Another mechanism of triggered star formation by OB stars is called the ``radiation-driven implosion'' (RDI) process (e.g., Hester \& Desch 2005). According to this model, an OB star drives an ionization shock front into the surrounding cloud, and cores within the cloud are thus triggered into collapse by the shock wave. The RDI  model predicts that the low-mass stars should be younger than the OB stars, since the OB stars initiated their formation. Chen et al. (2006) have discussed some more observational diagnostics of this model. The remnant cloud will have the shape of an extended pillar-like structure, and will extend towards or point towards the massive stars. The young stars thus formed will be roughly concentrated between the remnant cloud and the massive stars. In Fig.~\ref{class}, the spatial distribution of YSOs in NGC 6823 shows a concentration of Class II sources between the pillar {\it VulP12} in the north-east and the OB stars at the center. The pillar seems to be pointed towards the massive stars at the center. This suggests that the RDI mechanism may have triggered the formation of the YSOs seen in the eastern part of the cluster. In contrast, the western part does not contain any pillars or any such extended structures, and is also found to be nearly devoid of disk sources. 

Such triggered star formation is self-sustaining in time and self-propagating in space, as compared to spontaneous cloud collapse. The excess of YSOs mainly to the east of the Trapezium suggests the presence of some external stimulus, which could be the RDI mechanism.

\section{Summary}

We present results from an optical through MIR photometric survey of the young open cluster NGC 6823 that lies in the Vulpecula OB1 association at $\sim$2kpc. There is significant differential reddening in the cluster, with $A_{V}$ ranging between $\sim$3-20 mag. We find a $\sim$20\% fraction of Class I/II sources in the cluster. There is also a large population of photospheric Class III sources, a majority of which is likely the extincted background/foreground field star population. A large fraction of the disk sources are found to be among the younger group of low-mass stars in NGC 6823, at ages of $\sim$1-5 Myr and masses of $\sim$0.1-0.4$M_{\sun}$. The age and mass estimates are tentative and require cluster membership confirmation. A majority of the disk sources are located in the eastern part of the cluster, while the western region mostly consists of the Class III/field stars. Our survey has been deep enough to detect the low-mass population in this cluster. NGC 6823 provides a modest density comparison to the Orion Nebula Cluster, which also has an extreme environment with both a high- and a low-mass population extending from massive O-type stars down to sub-stellar objects (e.g., Slesnick et al. 2004). The presence of young disk sources in NGC 6823 indicates similar star formation properties in the outer regions of the Galaxy as observed for young clusters in the solar neighborhood. Follow-up studies including multi-object spectroscopy and X-ray imaging are needed to deepen our understanding of this cluster.

\section*{Acknowledgments}

This work was supported by the FP6 CONSTELLATION Marie Curie RTN which is governed by
contract number MRTN-CT-2006-035890 with the European Commission. N.P.-B has been supported by VietNam NAFOSTED grant 103.08-2010.07. BR would like to thank P. Lucas for many helpful discussions. This work is based on observations made with the William Herschel Telescope operated on the island of La Palma by the Isaac Newton Group in the Spanish Observatorio del Roque de los Muchachos of the Instituto de Astrof'sica de Canarias. This work is also based on observations obtained at the Cerro Tololo Inter-American Observatory, National Optical Astronomy Observatory, which are operated by the Association of Universities for Research in Astronomy, under contract with the National Science Foundation. This paper makes use of data obtained as part of the INT Photometric H$\alpha$ Survey of the Northern Galactic Plane (IPHAS) carried out at the Isaac Newton Telescope. All IPHAS data are processed by the Cambridge Astronomical Survey Unit, at the Institute of Astronomy in Cambridge. This work is based in part on observations made with the {\it Spitzer Space Telescope}, which is operated by the Jet Propulsion Laboratory, California Institute of Technology under a contract with NASA. Support for this work was provided by NASA through an award issued by JPL/Caltech. This work has made use of the Image Reduction and Analysis Facility (IRAF) software system. IRAF is distributed by the National Optical Astronomy Observatories, which are operated by the Association of Universities for Research in Astronomy, Inc., under cooperative agreement with the National Science Foundation.


\begin{table*}
\begin{minipage}{\linewidth}
\tiny
\caption{Optical, NIR, and MIR Photometry}
\label{phot}
\begin{tabular}{@{}l@{}ll@{}ll@{}ll@{}ll@{}ll@{}ll@{}ll@{}ll@{}ll@{}ll@{}ll}
\hline

RA (J2000)   &    Dec (J2000)    &  V  & & R  & & I   & &   J  & &  H   & &  K$_{s}$   & & [3.6]  & &[4.5]  && [5.8]  &&[8]  && SED Class   \\

\hline

19:42:47.21  &   +23:13:12.94  &  --  &   --   &  20.43  &   (0.08)   &  15.25  &   (0.03)   &  14.31  &   (0.04)   &  12.60  &   (0.02)   &  11.82  &   (0.01)   &  11.66  &   (0.00)   &  11.55  &   (0.01)   &  11.52  &   (0.03)   &  11.39  &   (0.04)   &  III  \\  
19:42:47.28  &   +23:19:35.00  &  18.69  &   (0.02)   &  16.93  &   (0.02)   &  15.15  &   (0.02)   &  11.68  &   (0.02)   &  10.36  &   (0.01)   &  8.92  &   (0.08)   &  9.50  &   (0.00)   &  9.53  &   (0.00)   &  9.37  &   (0.01)   &  9.34  &   (0.01)   &  III  \\  
19:42:47.39  &   +23:18:24.98  &  20.98  &   (0.02)   &  19.32  &   (0.05)   &  17.59  &   (0.03)   &  14.24  &   (0.03)   &  12.93  &   (0.01)   &  11.60  &   (0.19)   &  12.19  &   (0.01)   &  12.18  &   (0.01)   &  11.93  &   (0.05)   &  11.87  &   (0.07)   &  III  \\  
19:42:47.40  &   +23:17:18.85  &  --  &   --   &  22.92  &   (0.06)   &  20.21  &   (0.03)   &  16.37  &   (0.04)   &  14.75  &   (0.02)   &  13.93  &   (0.03)   &  13.32  &   (0.02)   &  13.36  &   (0.04)   &  13.03  &   (0.17)   &  12.91  &   (0.21)   &  III  \\  
19:42:47.49  &   +23:18:54.14  &  20.11  &   (0.02)   &  18.61  &   (0.05)   &  17.21  &   (0.04)   &  15.58  &   (0.03)   &  13.93  &   (0.04)   &  13.34  &   (0.04)   &  12.74  &   (0.01)   &  12.61  &   (0.02)   &  12.89  &   (0.15)   &  12.68  &   (0.17)   &  III  \\  
19:42:47.61  &   +23:17:1.32  &  18.54  &   (0.02)   &  --  &   --   &  13.42  &   (0.02)   &  12.17  &   (0.03)   &  10.95  &   (0.01)   &  10.38  &   (0.01)   &  10.18  &   (0.00)   &  10.17  &   (0.00)   &  10.12  &   (0.01)   &  10.00  &   (0.01)   &  III  \\
\hline
\end{tabular}
\end{minipage}
\end{table*}

\begin{table*}
\setcounter{table}{0}
\begin{minipage}{\linewidth}
\tiny
\caption{Continued}
\label{phot}
\begin{tabular}{@{}l@{}ll@{}ll@{}ll@{}ll@{}ll@{}ll@{}ll@{}ll@{}ll@{}ll@{}ll}
\hline

RA (J2000)   &    Dec (J2000)    &  V  & & R  & & I   & &   J  & &  H   & &  K$_{s}$   & & [3.6]  & &[4.5]  && [5.8]  &&[8]  && SED Class   \\

\hline

19:42:56.52  &   +23:19:45.48  &  22.04  &   (0.02)   &  19.99  &   (0.03)   &  18.01  &   (0.02)   &  15.40  &   (0.02)   &  14.50  &   (0.01)   &  13.92  &   (0.01)   &  13.58  &   (0.03)   &  13.41  &   (0.03)   &  13.06  &   (0.18)   &  12.83  &   (0.21)   &  II  \\  
19:42:56.52  &   +23:19:44.83  &  21.34  &   (0.02)   &  19.91  &   (0.03)   &  18.38  &   (0.03)   &  15.30  &   (0.01)   &  14.36  &   (0.01)   &  13.60  &   (0.02)   &  13.58  &   (0.02)   &  13.41  &   (0.04)   &  13.06  &   (0.13)   &  12.83  &   (0.15)   &  II  \\  
19:42:56.57  &   +23:14:43.40  &  --  &   --   &  --  &   --   &  15.42  &   (0.01)   &  13.47  &   (0.01)   &  12.96  &   (0.01)   &  12.66  &   (0.00)   &  12.64  &   (0.01)   &  12.60  &   (0.02)   &  12.25  &   (0.07)   &  12.31  &   (0.09)   &  III  \\  
19:42:56.62  &   +23:18:5.69  &  --  &   --   &  --  &   --   &  20.71  &   (0.02)   &  16.46  &   (0.04)   &  14.74  &   (0.01)   &  13.89  &   (0.01)   &  13.36  &   (0.02)   &  13.26  &   (0.03)   &  13.35  &   (0.23)   &  12.72  &   (0.18)   &  III  \\  
19:42:56.62  &   +23:17:59.06  &  --  &   --   &  20.97  &   (0.04)   &  18.76  &   (0.04)   &  14.72  &   (0.02)   &  13.26  &   (0.00)   &  12.10  &   (0.04)   &  12.19  &   (0.01)   &  12.01  &   (0.01)   &  11.82  &   (0.05)   &  11.87  &   (0.06)   &  III  \\  
19:42:56.70  &   +23:18:12.42  &  --  &   --   &  21.13  &   (0.04)   &  18.96  &   (0.05)   &  15.01  &   (0.02)   &  13.56  &   (0.00)   &  12.45  &   (0.03)   &  12.58  &   (0.01)   &  12.31  &   (0.01)   &  12.10  &   (0.07)   &  12.14  &   (0.08)   &  III  \\
\hline
\end{tabular}
\end{minipage}
\end{table*}

\begin{table*}
\setcounter{table}{0}
\begin{minipage}{\linewidth}
\tiny
\caption{Continued}
\label{phot}
\begin{tabular}{@{}l@{}ll@{}ll@{}ll@{}ll@{}ll@{}ll@{}ll@{}ll@{}ll@{}ll@{}ll}
\hline

RA (J2000)   &    Dec (J2000)    &  V  & & R  & & I   & &   J  & &  H   & &  K$_{s}$   & & [3.6]  & &[4.5]  && [5.8]  &&[8]  && SED Class   \\

\hline

19:43:5.13  &   +23:19:35.90  &  12.73  &   (0.02)   &  17.67  &   (0.14)   &  12.31  &   (0.02)   &  10.87  &   (0.00)   &  10.34  &   (0.00)   &  10.18  &   (0.01)   &  10.05  &   (0.00)   &  10.00  &   (0.00)   &  10.03  &   (0.01)   &  9.87  &   (0.01)   &  III  \\  
19:43:5.20  &   +23:17:8.41  &  13.59  &   (0.02)   &  13.92  &   (0.01)   &  13.97  &   (0.02)   &  13.00  &   (0.01)   &  12.83  &   (0.01)   &  12.40  &   (0.02)   &  12.73  &   (0.01)   &  12.53  &   (0.02)   &  12.36  &   (0.07)   &  12.47  &   (0.11)   &  III  \\  
19:43:5.23  &   +23:15:31.97  &  21.33  &   (0.05)   &  19.42  &   (0.06)   &  17.63  &   (0.03)   &  14.06  &   (0.00)   &  12.78  &   (0.01)   &  11.32  &   (0.10)   &  11.76  &   (0.01)   &  11.64  &   (0.01)   &  11.64  &   (0.04)   &  11.68  &   (0.06)   &  III  \\  
19:43:5.47  &   +23:15:24.66  &  21.73  &   (0.04)   &  19.61  &   (0.06)   &  17.70  &   (0.04)   &  14.00  &   (0.00)   &  12.63  &   (0.01)   &  10.75  &   (0.11)   &  11.74  &   (0.00)   &  11.61  &   (0.01)   &  11.43  &   (0.03)   &  11.61  &   (0.06)   &  III  \\  
19:43:5.47  &   +23:19:31.30  &  18.67  &   (0.01)   &  17.97  &   (0.03)   &  17.18  &   (0.02)   &  14.76  &   (0.01)   &  13.94  &   (0.00)   &  13.63  &   (0.00)   &  13.30  &   (0.02)   &  13.23  &   (0.03)   &  12.85  &   (0.11)   &  11.96  &   (0.07)   &  III  \\  
19:43:5.65  &   +23:15:47.88  &  --  &   --   &  --  &   --   &  21.41  &   (0.05)   &  16.51  &   (0.03)   &  14.21  &   (0.01)   &  13.18  &   (0.01)   &  12.30  &   (0.01)   &  12.20  &   (0.01)   &  12.02  &   (0.07)   &  11.81  &   (0.08)   &  III  \\
\hline
\end{tabular}
\end{minipage}
\end{table*}

\begin{table*}
\setcounter{table}{0}
\begin{minipage}{\linewidth}
\tiny
\caption{Continued}
\label{phot}
\begin{tabular}{@{}l@{}ll@{}ll@{}ll@{}ll@{}ll@{}ll@{}ll@{}ll@{}ll@{}ll@{}ll}
\hline

RA (J2000)   &    Dec (J2000)    &  V  & & R  & & I   & &   J  & &  H   & &  K$_{s}$   & & [3.6]  & &[4.5]  && [5.8]  &&[8]  && SED Class   \\

\hline

19:43:13.02  &   +23:16:53.18  &  --  &   --   &  22.76  &   (0.07)   &  21.28  &   (0.02)   &  18.14  &   (0.12)   &  16.21  &   (0.04)   &  14.69  &   (0.02)   &  13.18  &   (0.01)   &  12.62  &   (0.02)   &  11.88  &   (0.06)   &  11.11  &   (0.04)   &  II  \\  
19:43:13.10  &   +23:19:59.88  &  --  &   --   &  21.77  &   (0.03)   &  13.71  &   (0.01)   &  13.52  &   (0.01)   &  11.39  &   (0.00)   &  10.44  &   (0.00)   &  9.64  &   (0.00)   &  9.66  &   (0.00)   &  9.46  &   (0.01)   &  9.40  &   (0.01)   &  III  \\  
19:43:13.14  &   +23:15:38.09  &  --  &   --   &  23.96  &   (0.15)   &  19.88  &   (0.02)   &  14.19  &   (0.01)   &  11.80  &   (0.01)   &  10.61  &   (0.01)   &  9.69  &   (0.00)   &  9.66  &   (0.00)   &  9.37  &   (0.01)   &  9.38  &   (0.01)   &  III  \\  
19:43:13.15  &   +23:13:33.67  &  --  &   --   &  --  &   --   &  21.63  &   (0.05)   &  17.11  &   (0.06)   &  15.14  &   (0.02)   &  13.70  &   (0.02)   &  12.40  &   (0.01)   &  11.68  &   (0.01)   &  11.01  &   (0.03)   &  9.80  &   (0.01)   &  II  \\  
19:43:13.34  &   +23:14:34.48  &  20.38  &   (0.03)   &  18.83  &   (0.06)   &  17.48  &   (0.03)   &  15.10  &   (0.02)   &  14.00  &   (0.01)   &  13.19  &   (0.02)   &  11.95  &   (0.01)   &  11.44  &   (0.01)   &  10.88  &   (0.02)   &  10.38  &   (0.02)   &  II  \\  
19:43:13.34  &   +23:14:33.43  &  19.68  &   (0.03)   &  18.76  &   (0.06)   &  17.92  &   (0.04)   &  14.90  &   (0.01)   &  13.75  &   (0.02)   &  12.96  &   (0.38)   &  11.95  &   (0.01)   &  11.44  &   (0.01)   &  10.88  &   (0.02)   &  10.38  &   (0.02)   &  II  \\
\hline
\end{tabular}
\end{minipage}
\end{table*}

\begin{table*}
\setcounter{table}{0}
\begin{minipage}{\linewidth}
\tiny
\caption{Continued}
\label{phot}
\begin{tabular}{@{}l@{}ll@{}ll@{}ll@{}ll@{}ll@{}ll@{}ll@{}ll@{}ll@{}ll@{}ll}
\hline

RA (J2000)   &    Dec (J2000)    &  V  & & R  & & I   & &   J  & &  H   & &  K$_{s}$   & & [3.6]  & &[4.5]  && [5.8]  &&[8]  && SED Class   \\

\hline

19:43:24.12  &   +23:19:12.11  &  19.94  &   (0.01)   &  18.90  &   (0.01)   &  18.21  &   (0.02)   &  17.02  &   (0.06)   &  16.44  &   (0.07)   &  16.03  &   (0.08)   &  12.39  &   (0.01)   &  12.29  &   (0.01)   &  12.25  &   (0.08)   &  12.01  &   (0.10)   &  III  \\  
19:43:24.13  &   +23:19:23.59  &  21.17  &   (0.02)   &  19.52  &   (0.01)   &  18.04  &   (0.01)   &  15.97  &   (0.02)   &  14.94  &   (0.02)   &  14.37  &   (0.01)   &  13.91  &   (0.03)   &  13.42  &   (0.04)   &  12.75  &   (0.14)   &  12.26  &   (0.11)   &  II  \\  
19:43:24.16  &   +23:17:24.97  &  19.69  &   (0.02)   &  18.78  &   (0.03)   &  18.08  &   (0.02)   &  15.41  &   (0.01)   &  14.20  &   (0.02)   &  13.36  &   (0.02)   &  12.84  &   (0.01)   &  12.57  &   (0.02)   &  12.42  &   (0.08)   &  11.94  &   (0.07)   &  II  \\  
19:43:24.42  &   +23:20:12.16  &  17.28  &   (0.01)   &  17.49  &   (0.02)   &  16.50  &   (0.01)   &  14.37  &   (0.00)   &  13.43  &   (0.01)   &  12.57  &   (0.01)   &  11.77  &   (0.00)   &  11.38  &   (0.01)   &  11.16  &   (0.02)   &  10.21  &   (0.01)   &  II  \\  
19:43:24.52  &   +23:21:13.03  &  13.04  &   (0.01)   &  13.48  &   (0.02)   &  13.12  &   (0.01)   &  11.94  &   (0.01)   &  11.61  &   (0.01)   &  11.20  &   (0.02)   &  11.47  &   (0.00)   &  11.65  &   (0.01)   &  11.25  &   (0.03)   &  11.12  &   (0.03)   &  III  \\  
19:43:24.56  &   +23:18:12.78  &  --  &   --   &  20.02  &   (0.02)   &  18.14  &   (0.02)   &  13.99  &   (0.01)   &  12.46  &   (0.02)   &  11.64  &   (0.02)   &  11.51  &   (0.00)   &  11.40  &   (0.01)   &  11.33  &   (0.03)   &  11.26  &   (0.04)   &  III  \\
\hline
\end{tabular}
\end{minipage}
\end{table*}

\begin{table*}
\begin{minipage}{115mm}
\caption{MIPS Photometry}
\label{mips}
\begin{tabular}{cccccc}
\hline
RA (J2000) & Dec (J2000) & [24]\footnote{The uncertainty on the 24$\micron$ photometry is estimated to be $<$0.2 mag (Billot et al. 2010). } & SED Class \\ \hline
19:42:52.37 &  +23:13:04.87   & 6.05 & II \\ 
19:42:57.41 &  +23:14:41.46  & 5.07 & II \\ 
19:43:04.40   &  +23:18:47.92  & 4.75 & II \\ 
19:43:06.46  &  +23:14:50.96  & 7.31 & II \\ 
19:43:08.33  &  +23:19:13.01  & 3.49 & II \\ 
19:43:09.97  &  +23:18:01.08   & 7.3  & II \\ 
19:43:11.58 &  +23:17:32.42  & 7.37 & II \\ 
19:43:11.58 &  +23:18:20.45  & 7.37 & II \\ 
19:43:12.47 &  +23:18:17.03  & 8.03 & III \\
19:43:13.47 &  +23:19:15.78  & 6.97 & II \\ 
19:43:13.56 &  +23:17:58.74  & 7.76 & II \\ 
19:43:14.95 &  +23:13:56.6   & 6.99 & II \\ 
19:43:20.98 &  +23:19:02.39   & 3.73 & II \\ 
19:43:24.42 &  +23:20:12.16  & 7.23 & II \\ 
19:43:24.80  &  +23:18:53.35  & 7.57 & II \\ 
19:43:30.06 &  +23:15:05.65   & 6.75 & II \\ 
\hline
\end{tabular}
\end{minipage}
\end{table*}

\begin{table*}
\begin{minipage}{115mm}
\caption{IPHAS Photometry}
\label{iphas}
\begin{tabular}{cccccccc}
\hline

RA (J2000)   &    Dec (J2000)   &  [ r$^{\prime}$ ] & [ i$^{\prime}$ ] & [ H$\alpha$ ] &SED Class\footnote{Sources that lie above the H$\alpha$ EW=10 $\AA$ track are marked by an asterisk.} \\

\hline

19:42:47.28  &   +23:19:35.00  &  17.78 (0.008)   &  15.47 (0.005)   &  17.11 (0.014)   &   III   \\  
19:42:47.39  &   +23:18:24.98  &  20.15 (0.045)   &  17.84 (0.023)   &  19.66 (0.119)   &   III   \\  
19:42:47.49  &   +23:18:54.14  &  19.05 (0.019)   &  17.47 (0.018)   &  18.53 (0.045)   &   III   \\  
19:42:47.61  &   +23:17:1.32  &  17.80 (0.008)   &  15.67 (0.006)   &  17.14 (0.013)   &   III   \\  
19:42:47.64  &   +23:19:51.82  &  20.26 (0.049)   &  17.97 (0.025)   &  19.41 (0.096)   &   III   \\  
19:42:47.65  &   +23:17:9.24  &  21.01 (0.089)   &  18.52 (0.037)   &  20.21 (0.157)   &   III   \\  
19:42:48.45  &   +23:17:30.01  &  19.94 (0.036)   &  17.69 (0.020)   &  19.28 (0.069)   &   III   \\
\hline
\end{tabular}
\end{minipage}
\end{table*}

\begin{table*}
\setcounter{table}{2}
\begin{minipage}{115mm}
\caption{Continued}
\label{iphas}
\begin{tabular}{cccccccc}
\hline

RA (J2000)   &    Dec (J2000)   &  [ r$^{\prime}$ ] & [ i$^{\prime}$ ] & [ H$\alpha$ ] &SED Class  \\

\hline

19:42:58.85  &   +23:17:3.23  &  21.09 (0.096)   &  18.16 (0.028)   &  20.07 (0.138)   &   III   \\  
19:43:0.40  &   +23:20:57.98  &  19.82 (0.037)   &  16.56 (0.011)   &  18.93 (0.063)   &   III   \\  
19:43:0.47  &   +23:14:51.61  &  17.17 (0.005)   &  15.93 (0.007)   &  16.78 (0.011)   &   III   \\  
19:43:0.52  &   +23:18:54.43  &  16.37 (0.003)   &  15.39 (0.005)   &  15.88 (0.006)   &   III   \\  
19:43:0.64  &   +23:18:36.36  &  16.62 (0.004)   &  15.62 (0.006)   &  16.12 (0.007)   &   III   \\  
19:43:1.20  &   +23:14:51.40  &  17.24 (0.006)   &  16.21 (0.008)   &  16.74 (0.011)   &   III   \\  
19:43:1.43  &   +23:14:40.96  &  14.24 (0.001)   &  13.77 (0.002)   &  14.02 (0.002)   &   III   \\
\hline
\end{tabular}
\end{minipage}
\end{table*}

\begin{table*}
\setcounter{table}{2}
\begin{minipage}{115mm}
\caption{Continued}
\label{iphas}
\begin{tabular}{cccccccc}
\hline

RA (J2000)   &    Dec (J2000)   &  [ r$^{\prime}$ ] & [ i$^{\prime}$ ] & [ H$\alpha$ ] &SED Class \\

\hline

19:43:9.85  &   +23:18:14.33  &  16.74 (0.004)   &  16.12 (0.007)   &  16.43 (0.009)   &   III   \\  
19:43:9.96  &   +23:17:7.55  &  19.55 (0.027)   &  17.87 (0.023)   &  18.71 (0.053)   &   III   \\  
19:43:9.97  &   +23:18:1.08  &  13.44 (0.001)   &  13.00 (0.002)   &  13.19 (0.002)   &   II   \\  
19:43:9.99  &   +23:17:38.62  &  19.59 (0.028)   &  18.38 (0.034)   &  19.11 (0.075)   &   II   \\  
19:43:10.00  &   +23:17:17.56  &  13.28 (0.001)   &  12.87 (0.001)   &  13.04 (0.001)   &   III   \\  
19:43:10.17  &   +23:19:16.93  &  17.32 (0.006)   &  16.17 (0.008)   &  16.65 (0.011)   &   II   \\  
19:43:10.27  &   +23:20:48.62  &  13.22 (0.001)   &  12.85 (0.001)   &  12.95 (0.001)   &   III   \\
\hline
\end{tabular}
\end{minipage}
\end{table*}

\begin{table*}
\setcounter{table}{2}
\begin{minipage}{115mm}
\caption{Continued}
\label{iphas}
\begin{tabular}{cccccccc}
\hline

RA (J2000)   &    Dec (J2000)   &  [ r$^{\prime}$ ] & [ i$^{\prime}$ ] & [ H$\alpha$ ] &SED Class \\

\hline

19:43:23.57  &   +23:18:37.19  &  18.27 (0.011)   &  17.09 (0.013)   &  17.49 (0.019)   &   II *   \\  
19:43:23.98  &   +23:20:7.66  &  16.58 (0.004)   &  15.50 (0.005)   &  16.05 (0.007)   &   III   \\  
19:43:24.12  &   +23:19:12.11  &  19.09 (0.019)   &  18.27 (0.032)   &  18.62 (0.048)   &   III   \\  
19:43:24.13  &   +23:19:23.59  &  20.20 (0.047)   &  18.50 (0.038)   &  19.46 (0.100)   &   II   \\  
19:43:24.16  &   +23:17:24.97  &  19.30 (0.022)   &  17.93 (0.024)   &  18.40 (0.041)   &   II *   \\  
19:43:24.42  &   +23:20:12.16  &  17.27 (0.006)   &  16.30 (0.008)   &  16.64 (0.011)   &   II   \\  
19:43:24.52  &   +23:21:13.03  &  13.05 (0.001)   &  12.67 (0.001)   &  12.76 (0.001)   &   III   \\
\hline
\end{tabular}
\end{minipage}
\end{table*}

\label{lastpage}

\end{document}